\documentclass[structabstract]{aa}
\usepackage[switch,pagewise]{lineno}
\usepackage{graphicx}
\usepackage{rotating}
\usepackage{txfonts}
\usepackage{natbib}
\usepackage{lscape}
\usepackage{booktabs}
\usepackage{dcolumn}
\usepackage{float}
\usepackage{lscape}

\bibpunct{(}{)}{;}{a}{}{,} 

\begin{document}

\title{Catching the radio flare in \object{CTA\,102}}
\subtitle{II. VLBI kinematic analysis}

\author{C. M. Fromm
\inst{1}, E. Ros\inst{2,1}, M. Perucho\inst{2}, T. Savolainen\inst{1}, P. Mimica\inst{2}, M. Kadler\inst{3}, A. P. Lobanov\inst{1}, M. L. Lister\inst{4}, Y.~Y.~Kovalev\inst{5,1} \and J. A. Zensus\inst{1}}
\institute{Max-Planck-Institut f\"ur Radioastronomie, Auf dem H\"ugel 69, D-53121 Bonn, Germany\
\email{cfromm@mpifr.de}
\and Departament d'Astronomia i Astrof\'\i sica, Universitat de Val\`encia, Dr. Moliner 50, E-46100 Burjassot, Val\`encia, Spain\
\and Institut f\"ur Theoretische Physik und Astrophysik, Universit\"at W\"urzburg, Am Hubland, 97074 W\"urzburg, Germany\
\and Department of Physics, Purdue University, 525 Northwestern Avenue, West Lafayette, IN, 47907 USA\
\and Astro Space Center of Lebedev Physical Institute, Profsoyuznaya 84/32, 117997 Moscow, Russia}

\abstract
   {Very Long Baseline Interferometry (VLBI) observations can resolve the radio structure of active galactic nuclei (AGN) and provide estimates of the structural and kinematic characteristics on parsec-scales in their jets. The changes in the kinematics of the observed jet features can be used to study the physical conditions in the innermost regions of these sources. We performed multifrequency multiepoch Very Long Baseline Array (VLBA) observations of the blazar \object{CTA\,102} during its 2006 radio flare, the strongest ever reported for this source. These observations provide an excellent opportunity to investigate the evolution of the physical properties of blazars, especially during these flaring events}
   {We want to study the kinematic changes in the source during the strong radio outburst in April 2006 and test the assumption of a shock-shock interaction. This assumption is based on the analysis and modeling of the single-dish observations of \object{CTA\,102} (Paper I).}  
   {In this paper we study the kinematics of \object{CTA\,102} at several frequencies using VLBI observations. From the modeled jet features we derived estimates for the evolution of the physical parameters, such as the particle density and the magnetic field. Furthermore, we combined our observations during the 2006 flare with long-term VLBA monitoring of the source at $15\,\mathrm{GHz}$ and $43\,\mathrm{GHz}$.}
   {We cross-identified seven features throughout our entire multifrequency observations and find evidence of two possible recollimation shocks around $0.1\,\mathrm{mas}$ (deprojected 18\,$\mathrm{pc}$ at a viewing angle $\vartheta=2.6^\circ$) and $6.0\,\mathrm{mas}$ (deprojected $1\,\mathrm{kpc}$) from the core. The $43\,\mathrm{GHz}$ observations reveal a feature ejected at epoch $t_\mathrm{ej}=2005.9\pm0.2$, which could be connected to the 2006 April radio flare. Furthermore, this feature might be associated with the traveling component involved in the possible shock-shock interaction, which gives rise to the observed double peak structure in the single-dish light curves reported in Paper I.}
   {}
\keywords{galaxies: active, -- galaxies: jets, -- radio continuum: galaxies, -- radiation mechanisms: non-thermal, -- galaxies: quasars: individual: CTA\,102}

\titlerunning{CTA\,102 VLBI Kinematic Analysis}
\authorrunning{C. M. Fromm et al.}

\maketitle
\section{Introduction} 
High-resolution VLBI observations offer the unique possibility of studing 
the structural and spectral evolution in AGN on parsec-scales \citep[e.g.,][]{Lobanov:1999p2299,Savolainen:2002p2410}. These observations provide a tool to test several jet models and to probe how variations within the first few parsecs contribute to the high-energy emission. The brightness temperature, $T_b$, can be used to derive estimates of the dominant energy-loss mechanism and the evolution of intrinsic properties, such as the jet speed, the particle density, and the magnetic field \citep[e.g.,][]{Kadler:2004p2884,Schinzel:2012p5740}.

The blazar \object{CTA\,102}  (\object{B2230+114}) has been the target of several VLBI observations. Within the Radio Reference Frame Image Database, the source was observed five times at $8\,\mathrm{GHz}$ between 1994 and 1998 \citep{Piner:2007p5743}. The authors report apparent velocities from $(-7\pm14)\,\mathrm{c}$ up to $(24\pm29)\,\mathrm{c}$. Higher quality images have been obtained within the $15\,\mathrm{GHz}$ VLBI observations within the VLBA 2\,cm-Survey \citep[e.g.,][]{Kellermann:1998p5022,Zensus:2002p2202} and its successor, the MOJAVE\footnote{Monitoring of Jets in Active galactic nuclei with VLBA Experiments http://www.physics.purdue.edu/MOJAVE} program \citep{Lister:2009p90}. Those results show an extended jet region towards the southeast spanning a distance up to $25\,\mathrm{mas}$ (deprojected $4.5\,\mathrm{kpc}$ {using a viewing angle of $2.6^\circ$ \citep{Jorstad:2005p4121}}). This region contains several kinks and a kinematic analysis yields apparent speeds between $(-3\pm1)\,\mathrm{c}$ and $(19\pm1)\,\mathrm{c}$  \citep{Lister:2009p8}. The structure and kinematics of the innermost jet region between March 1998 and April 2001 is revealed by the $43\,\mathrm{GHz}$ VLBA observations of \object{CTA\,102}  within the Boston University monitoring program\footnote{http://www.bu.edu/blazars/research.html}. These observations show two stationary features, one close to the core $\left(r\approx0.1 - 0.2\,\mathrm{mas}\right)$ and the other further downstream at $r\approx2\,\mathrm{mas}$. Furthermore, two newly ejected features ($t_\mathrm{ej}=1997.9\pm0.2$ and $t_\mathrm{ej}=1999.5\pm0.1$) that are separating from the core split into two components further downstream. The flux density associated to these moving features outshine the core for a short time in both total and polarized intensity \citep{Jorstad:2005p4121}. The component ejected at $t_\mathrm{ej}=1997.9\pm0.2$ was connected to the 1997 flare in \object{CTA\,102} \citep{Savolainen:2002p2410}.

\citet[][hereafter Paper I]{Fromm:2011p4088}, we analyzed the centimeter- and millimeter- wavelength light curves of the historical flux density outburst of this source that took place in 2006. Our spectral analysis revealed a double-hump structure in the turnover frequency -- turnover flux density $\left(\nu_{m}-S_{m}\right)$ plane. The evolution of the physical parameters (e.g., the evolution of the magnetic field, $b$ $(B\propto r^{-b})$, in jet during the flare) were extracted from the obtained spectral parameters, $\left(\nu_{m}-S_{m}\right)$ by modeling their evolution within the shock-in-jet model \citep{Marscher:1985p50,Bjornsson:2000p32}. As a possible scenario to explain the observations, we suggested an interaction between a traveling and a standing shock wave.

In this second paper of the series, we test our hypothesis of a shock-shock interaction by studying the kinematical evolution on a parsec scale. Since single-dish observations cannot resolve the jet structure, we used several high-resolution multifrequency VLBA observations with a frequency range between $5\,\mathrm{GHz}$ and $86\,\mathrm{GHz}$ during the flaring time to study the kinematical variations. We combined our observations with the results of the $15\,\mathrm{GHz}$ MOJAVE observations and additionally analyzed the available $43\,\mathrm{GHz}$ VLBA observations of \object{CTA\,102} within the Boston Blazar Monitoring program. This extended data set allows us to study the morphological evolution of the source over nearly two decades (early 1995 until late 2011). The analysis of the core shift and the  temporal and spatial spectral evolution will be presented in the forthcoming Paper III (Fromm et. al in prep.).

The organization of the paper is as follows. In Sect.~\ref{obs}, we describe the VLBA observations and present the performed data analysis. The results of the analysis are presented in Sect.~\ref{results} and are discussed in Sect~\ref{disc}.  A summary and a conclusions are provided in Sect.~\ref{sum}. 

Throughout the paper we define the spectral index, $\alpha$, using the relation $S_\nu \propto \nu^{\alpha}$. The optically thin spectral index, $\alpha_0$, can be derived from the spectral slope, $s$ of the relativistic electron distribution ($N\propto E^{-s}$), via the relation $\alpha_{0}=-(s-1)/2$. We adopt a cosmology with $\Omega_m=0.27$, $\Omega_\Lambda=0.73$ and $H_0=71\,\mathrm{km\,s^{-1}\, Mpc^{-1}}$. This results in a linear scale of $8.11\,\mathrm{pc\,mas^{-1}}$ or $26.45\,\mathrm{ly\,mas^{-1}}$ for \object{CTA\,102} (z=1.037). With these conventions, $1\,\mathrm{mas}\,\mathrm{yr}^{-1}$ corresponds to $52.9\,\mathrm{c}$ . 

\section{VLBA observations and data analysis}
\label{obs}
\subsection{Multi-frequency VLBA observations}
We analyzed eight multifrequency VLBA observations with a frequency range between $2\,\mathrm{GHz}$ and $86\,\mathrm{GHz}$ centered around the 2006 radio flare. This coverage allowed us to study the flaring process during its different stages; the rise in the flux density from the quiescent stage, the flare, and the decay of the flare. The source was observed with all ten antennas of the VLBA, where \object{CTA\,102} was the main target within the experiment BR122. In the other observations \object{CTA\,102} was used as a D-term calibrator (see Table \ref{vlbacode}). In Table \ref{obspara} we present the characteristic image parameters for the multifrequency VLBA observations. {Out of the six $86\,\mathrm{GHz}$ observations only the May 2005 run has a high enough dynamic range to allow us to detect the extended structure of the source. Therefore, we cannot provide any kinematic results for images at this frequency.} 

\begin{figure}[h!]
\resizebox{\hsize}{!}{\includegraphics{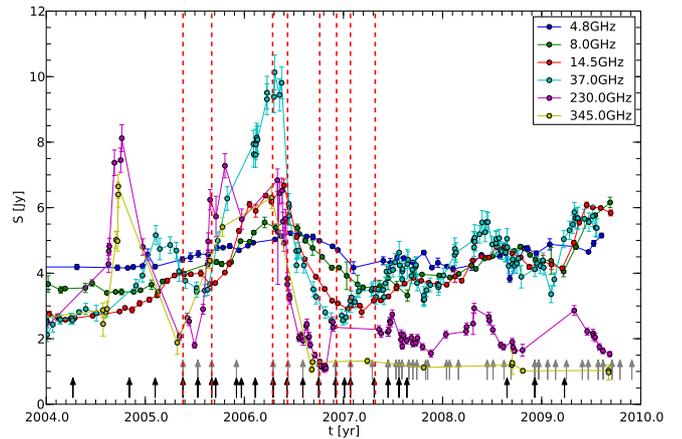}} 
\caption{$\mathrm{cm}$-$\mathrm{mm}$ single-dish light curves for \object{CTA\,102} covering the 2006 radio flare (see Paper I). The red dashed lines correspond to the epochs of multifrequency VLBI observations presented in this paper, the black arrows to the $15\,\mathrm{GHz}$, and the gray arrows to the $43\,\mathrm{GHz}$ VLBA observations.} 
\label{lc} 
\end{figure}

\begin{table}[h!]
\caption{Used VLBI observations and frequency range}  
\label{vlbacode}
\centering
\small
\begin{tabular}{@{}l c l l @{}} 
\hline\hline
VLBA-ID	&	Date  		&	Radio bands $^\mathrm{a}$	&	Notes\\
{}			& [yyyy-mm-dd]	& 		&	\\
\hline
BS157A	&	2005-05-19	&	CXUKQW		&  HN receiver warm	\\
BS157C	&	2005-09-01	&	CXUKQW		&	--		\\
BR122A	&	2006-04-14	&	SCXUKQ		& BR  receiver	warm\\
BR122B	&	2006-06-08	&	SCXUKQ		& HN 3 hours lost	\\ 
BW086B	&	2006-10-02	&	CXUKQW		& BR focus problems 	\\
BW086C	&	2006-12-04	&	CXUKQW		& --			\\
BW086D	&	2007-01-26	&	CXUKQW		& --			\\
BW086E	&	2007-04-26	&	CXUKQW		& SC weather problems	\\	
\hline
\multicolumn{4}{l}{\fontsize{6}{2}{$^\mathrm{a}$ S=$2\,\mathrm{GHz}$, C=$5\,\mathrm{GHz}$, X=$8\,\mathrm{GHz}$, U=$15\,\mathrm{GHz}$, K=$22\,\mathrm{GHz}$},Q=$43\,\mathrm{GHz}$,W=$86\,\mathrm{GHz}$}\\
\end{tabular} 
\end{table}

\begin{table*}
\centering
\begin{minipage}[t]{\columnwidth}
\caption{Image parameters of the multifrequency VLBA observations (see Table \ref{vlbacode})}  
\label{obspara}
\begin{tabular}{@{}c c c c c c c c @{}} 
\hline\hline
Frequency & Epoch & RMS$^{a}$& S$_\mathrm{peak}$ &S$_\mathrm{total}$& $\Theta_\mathrm{maj}$ & $\Theta_\mathrm{min}$ & P.A.\\
$[\mathrm{GHz}]$& [yyyy-mm-dd] & [mJy/beam] & [Jy/beam] & [Jy] & [mas] & [mas] & [$\deg$]\\
\hline 
5.0 & 2005-05-19 & 0.21 & 1.61 & 4.05 & 3.65 & 1.62 & $-$8.0\\
8.3 & 2005-05-19 & 0.25 & 1.68 & 3.91 & 2.32 & 1.02 & $-$6.7\\
15.3 & 2005-05-19 & 0.26 & 2.22 & 3.71 & 1.33 & 0.55 & $-$8.8\\
22.2 & 2005-05-19 & 0.55 & 2.33 & 3.21 & 0.95 & 0.40 & $-$15.4\\
43.1 & 2005-05-19 & 0.60 & 2.63 & 3.16 & 0.53 & 0.20 & $-$15.9\\
86.2 & 2005-05-19 & 2.76 & 1.94 & 2.34 & 0.25 & 0.11 & $-$19.3\\
\hline
5.0 &  2005-09-01 & 0.32 & 1.73 & 4.13 & 3.36 & 1.57 & $-$3.7\\
8.3 & 2005-09-01 & 0.39 & 1.48 & 3.23 & 1.98 & 0.94 & 1.4 \\
15.3 & 2005-09-01 & 0.31 & 1.96 & 3.37 & 1.21 & 0.52 & $-$6.1\\
22.2 & 2005-09-01 & 0.51 & 1.67 & 2.49 & 0.78 & 0.34 & $-$7.4\\
43.1 & 2005-09-01 & 0.58 & 3.48 & 4.18 & 0.45 & 0.18 & $-$12.3\\
\hline
2.3 & 2006-04-14 & 0.51 & 2.34 & 4.66 & 8.60 & 3.91 & $-$7.7\\
5.0 & 2006-04-14 & 0.18 & 2.05 & 4.48 & 3.90 & 1.45 & $-$11.2\\
8.4 & 2006-04-14 & 0.31 & 2.94 & 4.91 & 2.74 & 0.97 & $-$13.2\\
15.4 & 2006-04-14 & 0.26 & 4.88 & 6.32 & 1.46 & 0.51 & $-$14.1\\
22.2 & 2006-04-14 & 0.43 & 5.85 & 6.84 & 1.04 & 0.34 & $-$16.6\\
43.2 & 2006-04-14 & 0.75 & 6.43 & 7.35 & 0.65 & 0.15 & $-$15.8\\
\hline
2.3 & 2006-06-08 & 0.29 & 2.30 & 5.47 & 7.75 & 3.54 & $-$2.4\\
5.0 & 2006-06-08 & 0.17 & 2.04 & 4.40 & 3.52 & 1.52 & $-$6.7\\
8.4 & 2006-06-08 & 0.22 & 2.78 & 4.71 & 2.11 & 0.95 & $-$1.1\\
15.4 & 2006-06-08 & 0.25 & 4.13 & 5.37 & 1.26 & 0.51 & $-$8.7\\
22.2 & 2006-06-08 & 0.42 & 3.83 & 4.76 & 0.76 & 0.33 & $-$6.2\\
43.2 & 2006-06-08 & 0.43 & 3.62 & 4.74 & 0.39 & 0.18 & $-$5.0\\
\hline
5.0 & 2006-10-02 & 0.22 & 2.04 & 4.44 & 3.51 & 1.69 & $-$9.5\\
8.3 & 2006-10-02 & 0.59 & 1.72 & 3.45 & 2.16 & 1.11 & $-$4.9\\
15.3 & 2006-10-02 & 0.23 & 1.96 & 2.94 & 1.35 & 0.60 & $-$14.7\\
22.2 & 2006-10-02 & 0.46 & 1.49 & 2.20 & 0.76 & 0.37 & $-$7.5\\
43.1 & 2006-10-02 & 0.73 & 1.89 & 2.78 & 0.43 & 0.19 & $-$10.6\\
\hline
5.0 & 2006-12-04 & 0.25 & 1.84 & 4.16 & 3.65 & 1.75 & $-$9.7\\
8.3 &  2006-12-04& 0.23 & 1.48 & 3.32 & 2.11 & 1.17 & $-$4.0\\
15.3 &  2006-12-04 & 0.20 & 1.26 & 2.23 & 1.21 & 0.61 & $-$10.7\\
22.2 &  2006-12-04 & 0.39 & 1.08 & 1.81 & 0.96 & 0.55 & 9.9\\
43.1 &  2006-12-04 & 0.51 & 1.76 & 2.40 & 0.45 & 0.21 & $-$14.4\\
\hline
5.0   & 2007-01-26 & 0.72 & 1.43 & 3.32 & 3.36 & 1.71 & $-$1.6\\
8.3   & 2007-01-26 & 0.26 & 1.12 & 2.55 & 2.06 & 1.11 & $-$0.8\\
15.3 & 2007-01-26 & 0.22 & 1.13 & 1.99 & 1.20 & 0.58 & $-$10.4\\
22.2 & 2007-01-26 & 0.39 & 1.22 & 1.94 & 0.72 & 0.39 & $-$7.7\\
43.1 & 2007-01-26 & 0.66 & 2.61 & 3.31 & 0.38 & 0.19 & $-$9.9\\
\hline
5.0 & 2007-04-26 & 0.24 & 1.39 & 3.59 & 3.46 & 1.81 & $-$6.8\\
8.3 & 2007-04-26 & 0.28 & 1.16 & 2.99 & 2.10 & 1.14 & $-$1.2\\
15.3 & 2007-04-26 & 0.20 & 1.55 & 2.50 & 1.23 & 0.57 & $-$9.4\\
22.2 & 2007-04-26 & 0.34 & 1.58 & 2.24 & 0.88 & 0.39 & $-$13.0\\
43.1 & 2007-04-26 & 0.54 & 2.54 & 3.04 & 0.44 & 0.20 & $-$14.7\\
\hline
\multicolumn{8}{l}{$^a$ RMS values are determined in a region of the final map without significant emission}\\
\end{tabular}
\end{minipage}
\end{table*}

\subsection{Additional $15\,\mathrm{GHz}$ and $43\,\mathrm{GHz}$ VLBA observations from survey programs}
The $15\,\mathrm{GHz}$ VLBA observations were taken from the MOJAVE archive \citep{Lister:2009p8} and combined with our $15\,\mathrm{GHz}$ observations. We analyzed 36 epochs starting from June 1995 until October 2010 $\left(\left<\Delta t\right>=157\,\mathrm{days}\right)$.

The $43\,\mathrm{GHz}$ VLBA observations were partially taken from the Boston blazar monitoring program database$^2$. The source was on average observed every $58\,\mathrm{days}$ between May 2005 and January 2011, which led to a total number of 46 epochs.\\
In Figure \ref{lc} we present the temporal correspondence between the $\mathrm{cm}$-$\mathrm{mm}$ light curves and the VLBA observations of  \object{CTA\,102}. 

\subsection{Data reduction}
\label{datareduction}
We used National Radio Astronomy Observatory's (NRAO) Astronomical Image Processing System (AIPS) for calibrating the data. We performed an amplitude calibration and applied a correction to the atmospheric opacity for the high-frequency data $\left(\nu>15\,\mathrm{GHz}\right)$. The parallactic angle correction was taken into account before we calibrated the phases using the pulscal signal and a final fringe fit. The time- and frequency-averaged data were imported to DIFMAP \citep{Shepherd:1997p2298}, where we used the CLEAN algorithm combined with phase and amplitude self-calibration and MODELFIT algorithms for imaging and model fitting, respectively (see Sect.~\ref{modelfit}).

The uncertainties on the obtained fluxes were estimated by comparing the total VLBA flux densities with the values measured by the University of Michigan Radio Astronomy Observatory (UMRAO) program (see Paper I). For the $5\,\mathrm{GHz}$ and $8\,\mathrm{GHz}$ VLBA observations we calculated a flux density difference between 10\% and 15\%, where the difference was between 5\% and 10\% for the $15\,\mathrm{GHz}$ ones. The estimates for frequencies above $15\,\mathrm{GHz}$ were more difficult to derive since the observations could be easily effected by variations in the wet component of the troposphere, but a conservative estimate led to an uncertainty value of 10\% of the UMRAO flux density \citep{Savolainen:2008p2958}.

\subsection{Model fitting}
\label{modelfit}
To parametrize the jet and trace the evolution of local flux density peaks along the jet, we fitted 2D circular Gaussian components to the fully calibrated visibilities using the MODELFIT algorithm in DIFMAP. These components were characterized by their flux density, $S_\mathrm{mod}$, position $r_\mathrm{mod}$, position angle (P.A.), $\theta_\mathrm{mod}$ (measured from north through east) and full width half maximum (FWHM), i.e., the deconvolved size of the components. We modeled the data at each frequency separately to avoid biasing effects. Since the number of fitted Gaussians was initially not limited, we added only a new component at a given frequency if this resulted in a significant decrease in the $\chi^{2}$ value. This approach led to a minimum number of Gaussians that can be regarded as a reliable representation of the source. As mentioned in \citet{Lister:2009p8} the fitted components do not necessary correspond to physical features in the jet and may reflect a mathematical requirement to properly describe the complex structure of the source. To minimize the above-mentioned effect of a fitted Gaussian having a purely mathematical nature, we applied the following criterion: A fitted feature at a given frequency should be traceable at least within five consecutive epochs, and its evolution in position and flux density should be smooth without any strong jumps. If a component satisfies this principle we may interpret it as a region of locally enhanced emission generated by perturbations (shocks or instabilities) in the underlying jet flow.\\
 the following we identify the most northern component at each frequency as the core, e.g., the foot point of the jet and compute the position of the additional components relative to the core. The uncertainties of the fitted components were typically around 5\% on the total intensities\citep{Lister:2005p5497}. In this paper we follow the suggestion of \citet{Lister:2009p8} that the uncertainties of the fitted components are around 10\% of the component size convolved with the beam size. A more detailed determination of the component uncertainties is beyond the scope of this paper.


\section{Results}
\label{results}
Here we present the results of modeling the source at different frequencies. Following the criteria presented by \citet{Lister:2009p8}, we derive the kinematical parameters, e.g., the apparent speed, $\beta_\mathrm{app}$, only for those that are identified in five {consecutive} epochs or more, and we regard a component as stationary if the angular speed obtained is consistent with $\mu=0$ within the observational accuracy. All analyzed epochs and modeled components at a given frequency are presented in the Appendix.

\subsection{Cross-frequency component identification}
With multifrequency VLBI observations we can study the same physical region of a jet at different angular resolutions. In addition to the {increase} in the angular resolution with increasing frequency, there is the physical effect of the self-absorption of the emission. The shape of spectrum of the observed radiation, assuming that the detected emission is mainly generated by an ensemble of relativistic electrons, can be characterized by its spectral turnover, e.g., the frequency and flux density at the peak ($\nu_m$ and $S_m$). If the observing frequency is lower than the turnover frequency, $\nu_m$, the emission will be self-absorbed (optically thick), in contrast, if the observing frequency is higher than the turnover frequency, the emission will be optically thin. This transition between optically thick and optically thin emission occurs in general in the core region.These two effects are reflected in the modeling of the source by splitting the core region into several features with increasing frequency.

\begin{figure*}[t!]
\centering 
\includegraphics[width=17cm]{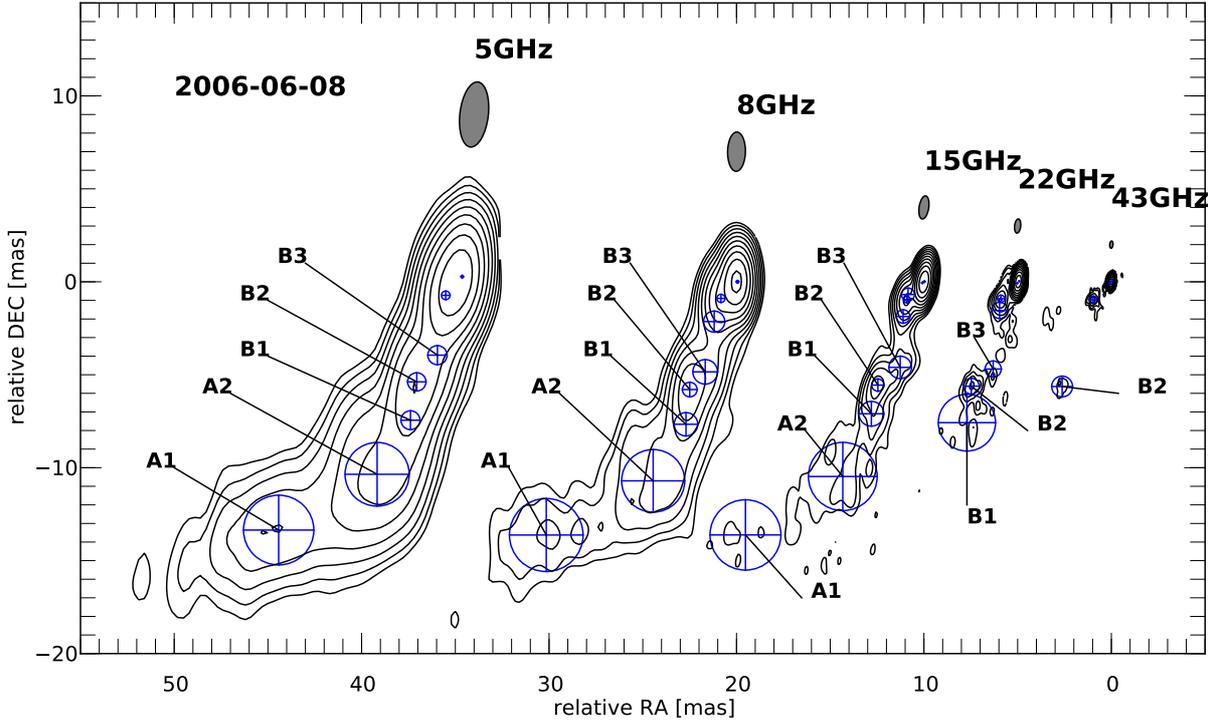} 
\caption{Uniform weighted VLBA CLEAN images with fitted circular Gaussian components at different frequencies for the 2006 July observation of \object{CTA\,102}. The lowest contour is plotted at $10\times$ the off-source $\mathrm{rms}$ at $43\,\mathrm{GHz}$ and increases in steps of 2. The observing frequency and the restoring beam size are plotted above each map. For the labeling we used capital letters for the same physical region in the jet and the numbers increase with inverse distance from the core. For a more detailed picture of the core region see Fig.~\ref{allcontcore}} 
\label{allcont} 
\end{figure*}

\begin{figure*}[b!]
\centering 
\includegraphics[width=17cm]{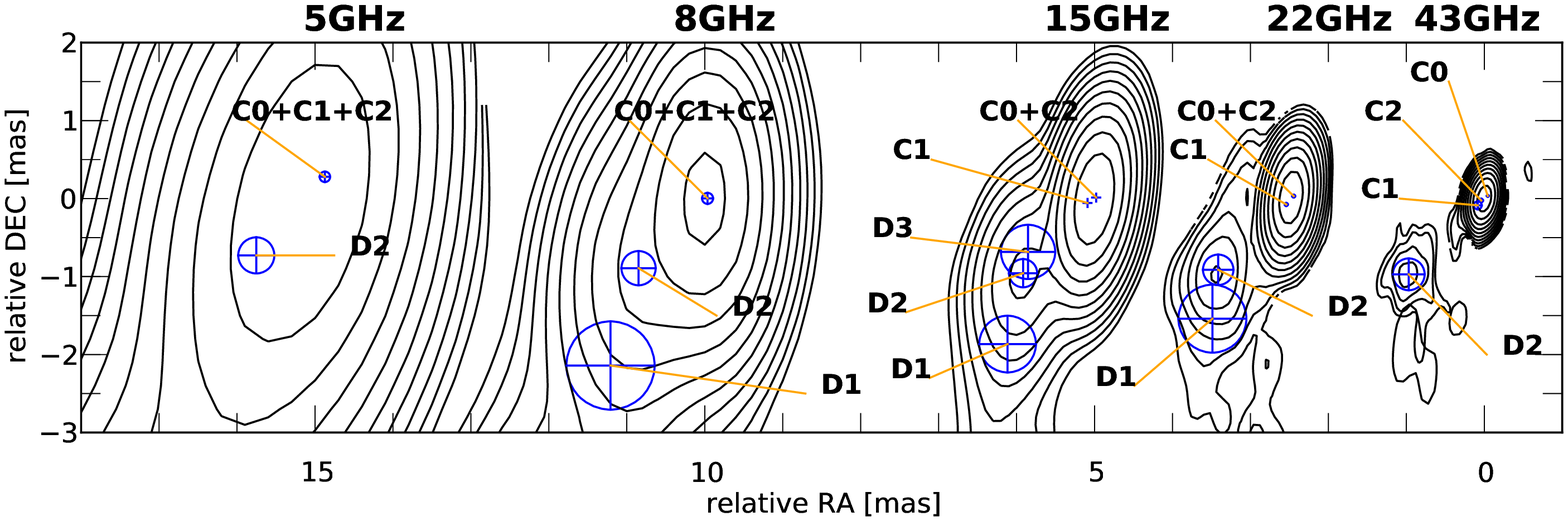} 
\caption{Zoom into the core region for the images presented in Fig.~\ref{allcont}, showing the splitting of the C-components with increasing frequency. For map details see Fig.~\ref{allcont}. } 
\label{allcontcore} 
\end{figure*}

For labeling the fitted components we use capital letters for the same physical regions and the numbers {increase with decreasing core distance}. The first criterion for the cross-frequency identification of fitted Gaussians is based on their position. Since we have not aligned the maps, e.g., corrected for the opacity shift, there is a slight shift in the position of these components. (The analysis of the opacity shift will be presented in Paper III.) However, if a fitted component at a given frequency parametrizes a physical region in the jet, the observed angular speed obtained for this feature should be consistent with the angular speed derived from its counterpart at the other frequencies. Based on these two criteria we can clearly identify components across the different frequencies involved in this study. From our modeling we could cross-identify (in frequency and time) four different regions, labeled from A to D (see Figs. \ref{allcont} and \ref{allcontcore}). The influence of the increased resolution on frequency is clearly visible by the amount of detail on the jet structure, which is most obvious on the kink around $4\,\mathrm{mas}$ away from the core (compare $5\,\mathrm{GHz}$ and $15\,\mathrm{GHz}$ image in Fig. \ref{allcont}). The combined effect of increased resolution and the transition from optically thick and optically thin emission can be seen in Fig. \ref{allcontcore}.

\subsection{Calculation of kinematic parameters}
The motions for the fitted components in \object{CTA\,102} can be divided into three different categories: i) stationary, ii) non-accelearting and iii) accelerating. To derive the kinematics of the source, we fit polynomials to the observed $\mathrm{x}$- and $\mathrm{y}$-positions of the fitted components. We follow the procedure presented in e.g., \citet{Lister:2009p8} and \citet{Schinzel:2012p5740}, using first- and second-order polynomials if the components was detected in more than ten epochs. (The latter can only be applied to the long-term monitoring of the source at $15\,\mathrm{GHz}$ and $43\,\mathrm{GHz}$):
\begin{eqnarray}
x(t)=\mu_x\left(t-t_\mathrm{x,0}\right)+\frac{\dot{\mu_x}}{2}\left(t-t_\mathrm{mid}\right)^2\\
y(t)=\mu_y\left(t-t_\mathrm{y,0}\right)+\frac{\dot{\mu_y}}{2}\left(t-t_\mathrm{mid}\right)^2,
\end{eqnarray}
where $t_\mathrm{mid}=(t_\mathrm{min}+t_\mathrm{max})/2$, $t_\mathrm{x,0}=t_\mathrm{mid}-x(t_\mathrm{mid})/\mu_x$ and $t_\mathrm{y,0}=t_\mathrm{mid}-y(t_\mathrm{mid})/\mu_y$ are the component ejection times. Owing to the bent structure of \object{CTA\,102}, the ejection cannot be computed from a simple back extrapolation if the trajectory is not clearly radial outwards. We only use the results of the second-order polynomial fit, if there is a significant decrease in the $\chi^2$ value as compared to the linear fit. Using the average angular speed $\left<\mu\right>$ obtained, we compute the kinematic parameters, e.g., apparent speed, $\beta_\mathrm{app}$, and the Doppler factor, $\delta$. The apparent speed of the components, $\beta_\mathrm{app}$, is derived from the angular speed, $\mu$:
\begin{equation}
\beta_\mathrm{app}=\frac{\left<\mu\right> D_\mathrm{L}}{1+z},
\end{equation}
where $D_\mathrm{L}$ is the luminosity distance and $z$ the redshift.

{The Doppler factor is defined as $\delta=\Gamma^{-1}(1-\beta\cos\vartheta)^{-1}$, where $\Gamma=(1-\beta^2)^{-1/2}$ is the bulk Lorentz factor and $\beta=v/c$. Since the $\beta$ and $\vartheta$ are not measurable directly from the observations, two approaches have been used in the recent past to estimate the the Doppler factor, $\delta$:\\
i) The first one assumes that the observed features are traveling at the critical angle, the angle that maximizes the speed of the feature. The critical viewing angle, $\vartheta_\mathrm{crit}$ can be derived by calculating the derivative of the apparent speed $\beta_\mathrm{app}=\beta\sin\vartheta/(1-\beta\cos\vartheta)$ with respect to $\vartheta$ for any given $\beta$. This leads to the result that the maximum apparent speed is obtained if $\cos\vartheta_\mathrm{crit}=\beta=\left[\beta^2_\mathrm{app}/(1+\beta^2_\mathrm{app})\right]^{1/2}$. For jets seen at this critical angle, the Doppler factor is $\delta_\mathrm{crit}=\sqrt{1+\beta_\mathrm{app}^2}$.}

{ii) We can also use causality arguments and estimate $\delta$ from the variability of the flux density
and the size of the component \citep[see, e.g.,][]{Jorstad:2005p4121}, to get a 'variability' Doppler factor}
\begin{equation}
\delta_\mathrm{var}=\frac{d_\mathrm{eff} D_\mathrm{L}}{c\Delta t_\mathrm{var}(1+z)},
\end{equation}
{with $d_\mathrm{eff}=1.8\times \mathrm{FWHM}$, the effective size of the component and
$\Delta t_\mathrm{var}=dt/\ln(S1/S2)$, where $S1$ and $S2$ are the maximum and minimum of the component
flux density and $dt$ the time difference between these two flux density values. The variability Doppler factor and
the apparent speed can be used to calculate estimates for the bulk Lorentz factor, $\Gamma$, and the viewing angle, $\vartheta$, according to}
\begin{eqnarray}
\Gamma_\mathrm{var}&=&\frac{\beta_\mathrm{app}^2+\delta_\mathrm{var}+1}{2\delta_\mathrm{var}}\\
\vartheta_\mathrm{var}&=&\arctan\frac{2\beta_\mathrm{app}}{\beta_\mathrm{app}^2+\delta_\mathrm{var}^2-1}.
\end{eqnarray}
The limited resolution at low frequencies means we cannot resolve some of the components that leads to blending effects between two or more features. Under these circumstances the values obtained for the apparent speed, $\beta_\mathrm{app}$ are too low and the calculated values for the viewing angle, $\vartheta$ are too high. The kinematic parameters for the fitted components are presented in Sects.~\ref{dopsec} and \ref{sumkin}. Since we observe \object{CTA\,102} at $2\,\mathrm{GHz}$ only for two epochs, April 2006 and June 2006, we neither derived kinematic parameters nor present model fitting results in the Appendix.

\subsection{Jet region A}
The region labeled as A spreads from $r\sim10\,\mathrm{mas}$ to $r\sim18\,\mathrm{mas}$. This region could be parametrized with two Gaussian components, labeled as A1 and A2, which we traced and cross-identified within our $5\,\mathrm{GHz}$, $8\,\mathrm{GHz}$, and $15\,\mathrm{GHz}$ observations (see Fig.~\ref{allcont}).  From our multifrequency observations we derived an angular speed for A1 between $(0.15\pm0.09)\,\mathrm{mas/yr}$ at $5\,\mathrm{GHz}$ and $(0.29\pm0.16)\,\mathrm{mas/yr}$ at $8\,\mathrm{GHz}$. With the long-term monitoring of \object{CTA\,102} at $15\,\mathrm{GHz}$, the angular speed we obtained is $(0.18\pm0.03)\,\mathrm{mas/yr}$.  
 
The fit of the trajectory of A1 by a second-order polynomial led to $\chi^2_\mathrm{red}=5.08$, which is no significant increase as compared to a first-order polynomial ($\chi^2_\mathrm{red}=5.14$), from which we conclude that there is no detectable acceleration for this component. Based on the $15\,\mathrm{GHz}$ observations we derived an apparent speed, $\beta_\mathrm{app}=(9.4\pm1.9)$\textit{c} (for additional kinematic parameters see Table \ref{kinres}).The angular speed for the second feature in this region, A2, spans from $(0.11\pm0.06)\,\mathrm{mas/yr}$ to $(0.21\pm0.045)\,\mathrm{mas/yr}$. We could not detect any clear indication of deceleration or acceleration of this feature from the long-term monitoring at $15\,\mathrm{GHz}$ and the calculated apparent speeds cover a range between $(6\pm3)$\,\textit{c} and $(11\pm2)$\,\textit{c}.

\begin{figure}[h!]
\resizebox{\hsize}{!}{\includegraphics{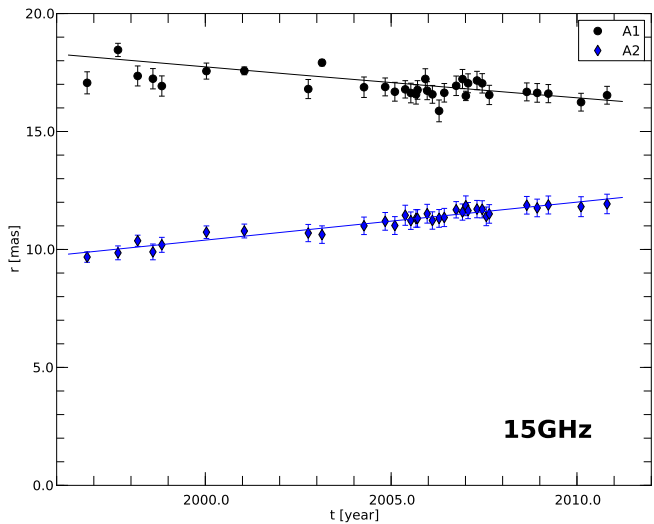}} 
\caption{Temporal separation from the core for the components A1 and A2 at $15\,\mathrm{GHz}$ taken as the representative frequency for the evolution of the cross-identified components within this region. The solid lines correspond to a first-order polynomial fit to the $x$- and $y$-position separately.} 
\label{A1A2sep} 
\end{figure}

\begin{figure*}[h!]
\centering 
\includegraphics[width=17cm]{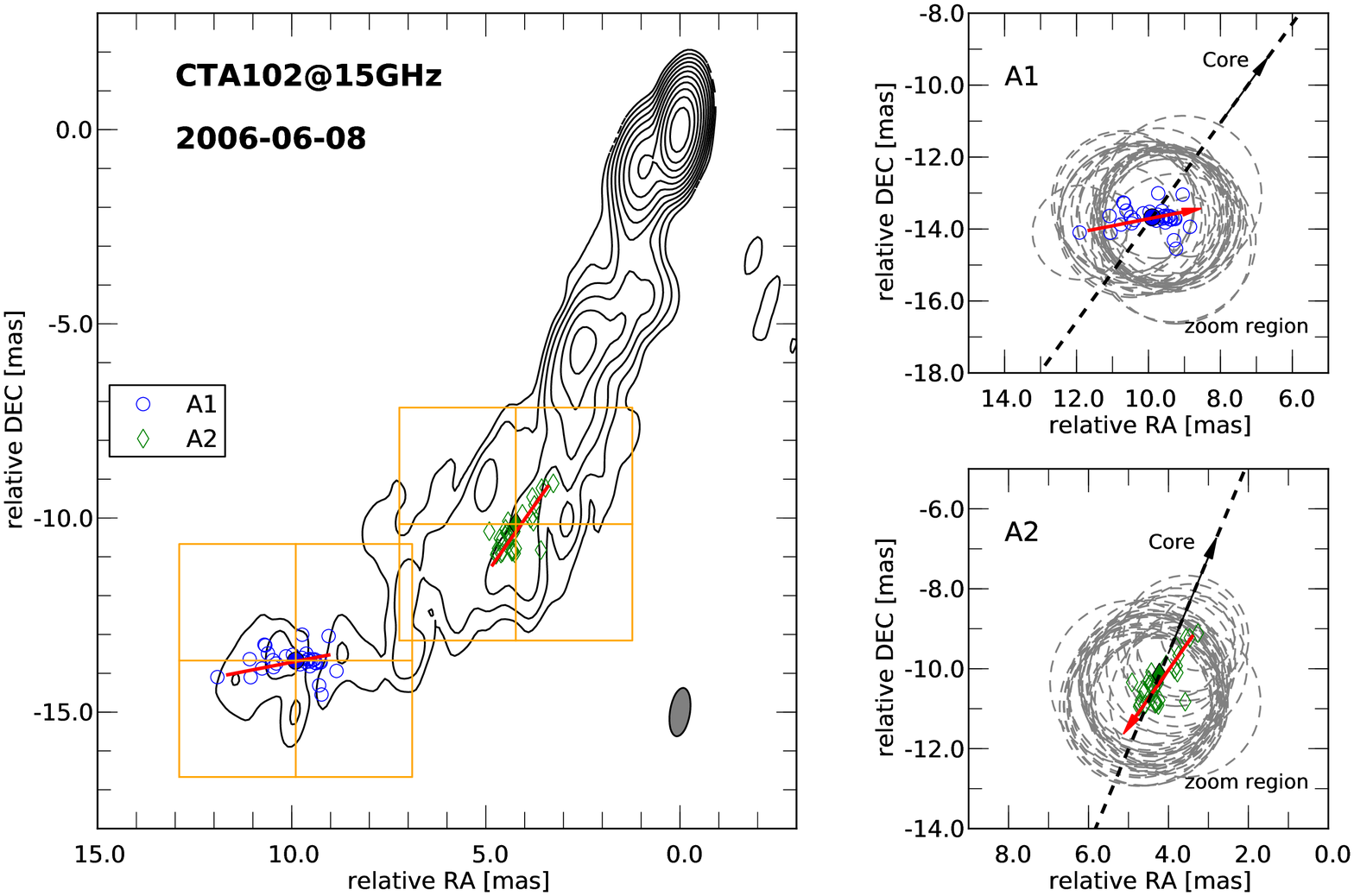} 
\caption{Vector motion fits and position of the fitted component in the sky. Left: $15\,\mathrm{GHz}$ CLEAN VLBI map of the 2006 June observations of \object{CTA\,102}. The lowest contour level is drawn at 10$\times$ the off-source rms, and the contours increase with steps of 2. The observational beam is plotted in the bottom right corner. The blue open symbols indicate the position of the components labeled as A1 and A2. The filled symbol corresponds to the component position at the middle time, $t_0$, and the red solid line to the trajectory of the feature. The orange squares show the zoom region. Right: Zoom region for the individual components. The dashed gray circle are the components size (FWHM) and the dashed black line and black arrow correspond the direction to the core drawn from the position of the component at $t_0$. The red solid line illustrates the trajectory of the component, and the direction of the component is indicated by the arrow.} 
\label{A1A2vec} 
\end{figure*}

For the presentation of the component trajectories we selected $15\,\mathrm{GHz}$ as the representative frequency.
In Fig.~\ref{A1A2sep} we show the temporal separation of A1 and A2 from the core and the vector motion fits are presented in Fig.~\ref{A1A2vec}, where the lefthand panel shows the position of the components in the sky and the righthand side a zoom of the individual regions with overplotted trajectory 
in the contour plot. 
The components are plotted relative to the core position and the contour map corresponds to the 2006 June observations \object{CTA\,102}. The lowest contour level is drawn at ten times the off-source rms ($1\,\mathrm{mJy}$ and increase with a factor of 2). The apparent inward motion of A1 (see Fig.~\ref{A1A2sep}) is partly due to the nonradial, nearly transversal motion of this feature and in contrast, component A2 is mainly moving radially outwards (see inlets in Fig.~\ref{A1A2vec}).  

Based on the derived angular speeds, $\mu$, we could, within the accuracy and limitation of our time sampling in the multifrequency observations, identifiy A1 and A2 across our observations.

\subsection{Jet region B}
We parametrized the region further upstream the jet, between $4\,\mathrm{mas}$ and $8\,\mathrm{mas}$ with three circular Gaussian components. One of these features, labeled as B2, could be identified throughout our entire multifrequency data set ($5\,\mathrm{GHz}$ to $43\,\mathrm{GHz}$). All three components, B1, B2, and B3, show very low pattern speeds. In the following we discuss the properties of those components separately and in detail.

The angular speed for B1 is on average comparable to $\mu=0$. However, we found from the $15\,\mathrm{GHz}$ observations an angular speed of $\mu=0.016\pm0.005$ that corresponds to an apparent speed $\beta_\mathrm{app}=(0.85\pm0.25)$\textit{c}. Furthermore, the second-order polynomial fit of the $y$-component of the B1 trajectory leads to a significant reduction of the $\chi^2$ value compared to a linear fit, from $\chi^2_\mathrm{red}=2.0$ to $\chi^2_\mathrm{red}=1.0$. The value obtained for acceleration $\dot{\mu_y}=(0.01\pm0.002)\,\mathrm{mas/yr^2}$

\begin{figure*}[h!]
\centering 
\includegraphics[width=17cm]{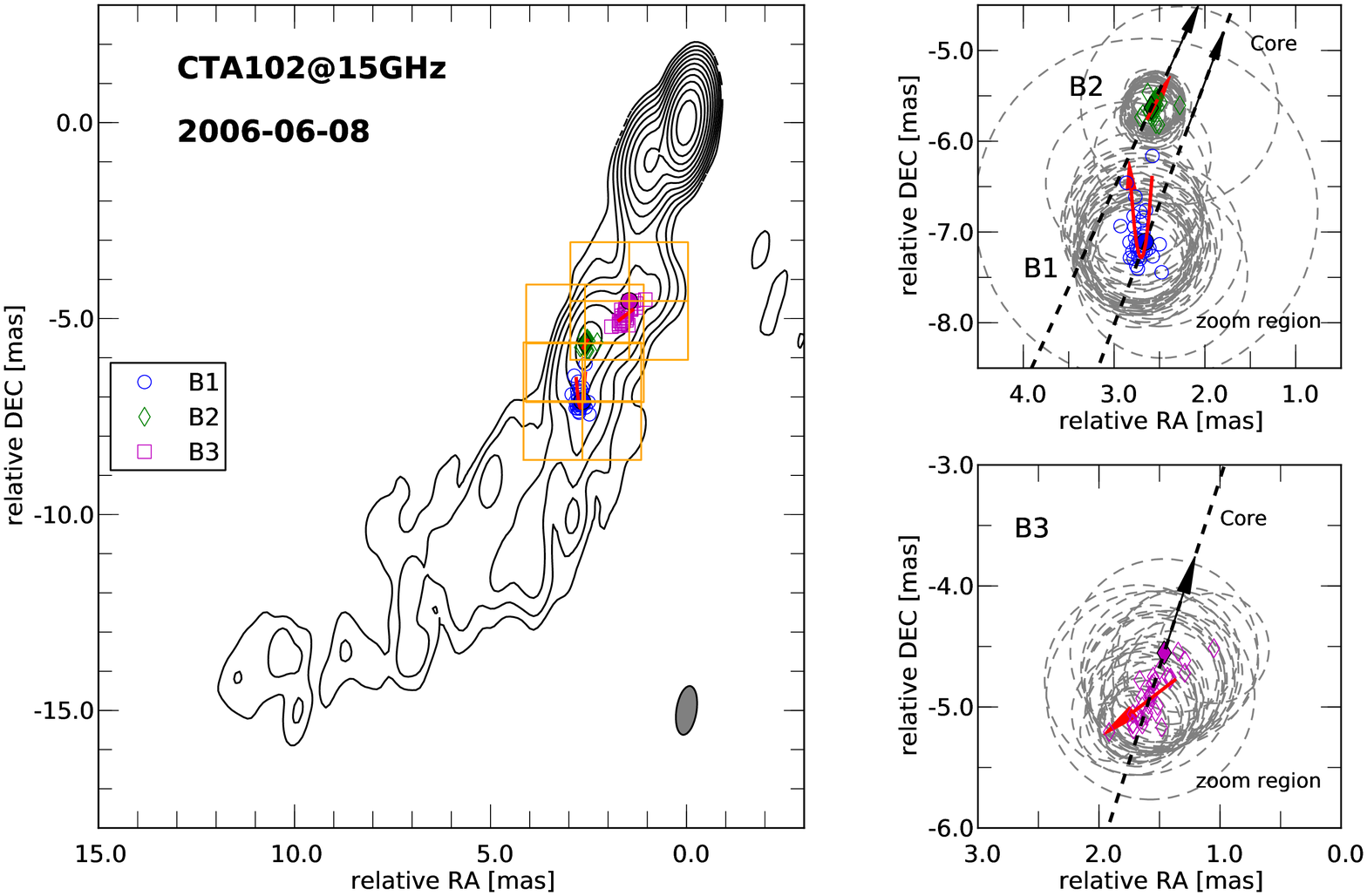} 
\caption{Same as Fig.~\ref{A1A2vec} for components B1, B2, and B3.} 
\label{Bvec} 
\end{figure*}

Similar behavior to B1 is found for B2. The angular speed takes values between $0.05\pm0.04\,\mathrm{mas/yr}$ and $0.015\pm0.003\,\mathrm{mas/yr}$, and the corresponding apparent speeds span $0.77\pm0.14$\,\textit{c} to $3\pm2$\,\textit{c}. From the long-term monitoring programs at $15\,\mathrm{GHz}$ and $43\,\mathrm{GHz}$ we compute angular velocities around $0.015\,\mathrm{mas/yr}$ and subluminal apparent speeds $\beta_\mathrm{app}\sim0.77$\,\textit{c}. Based on the high-resolution $43\,\mathrm{GHz}$ observations, we could identify B2 within our observational limitations as quasi-stationary with a slight radial inward motion (see inlet in Fig.~\ref{Bvec}).

The innermost component of this triplet is B3. Similar to the features farther downstream in region B, it is characterized by a small angular speed, where the scatter in $\mu$ obtained from the multifrequency campaign is larger due to the limited time span but is, within the uncertainties comparable to the derived angular speed from the $15\,\mathrm{GHz}$ observation. From these observation we calculated an angular speed of $\mu=(0.027\pm0.006)\,\mathrm{mas/yr}$ and an apparent speed of $\beta_\mathrm{app}=(1.4\pm0.3)$\,\textit{c}. The trajectory of B3 shows a nonradial outward motion with slow pattern speed. 

\begin{figure}[h!]
\resizebox{\hsize}{!}{\includegraphics{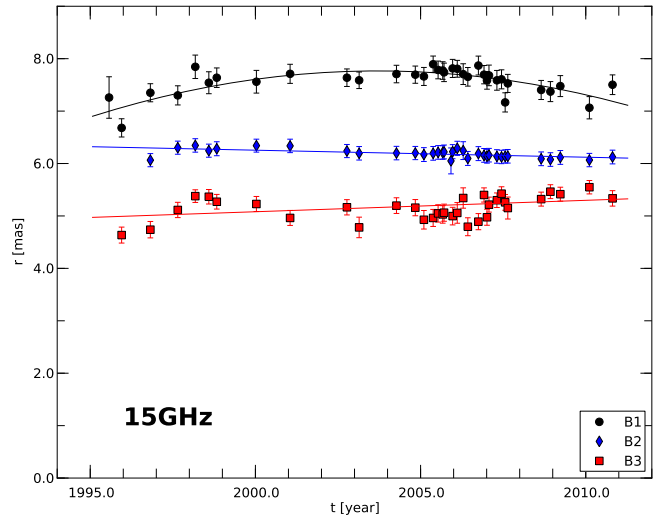}} 
\caption{Temporal separation from the core for the components B1, B2, and B3 at $15\,\mathrm{GHz}$ taken as the representative frequency for the evolution of the cross-identified components within this region. The solid lines correspond to polynomial fits to the $x$- and $y$-positions separately.} 
\label{Bsep} 
\end{figure}

As mentioned before, we used the $15\,\mathrm{GHz}$ modelfit results of the fitted components in region B as representative of the kinematics within this area. The evolution of the separation from the core for all components in region B is plotted in Fig.~\ref{Bsep}.
The vector motion of the components B1, B2, and B3 is presented in Fig.~\ref{Bvec}. 

\subsection{Jet region D}
The third region of the jet spans $1\,\mathrm{mas}$ to $3\,\mathrm{mas}$ away from the core and we parametrized this area with 2 to 3 Gaussians, labeled as D1, D2, and D3. At these distances from the core, the mentioned effects of increasing resolution with frequency lead to splitting fitted Gaussians into additional components. Therefore, it is difficult to distinguish and clearly cross-identify them.

\begin{figure*}[h!]
\centering 
\includegraphics[width=17cm]{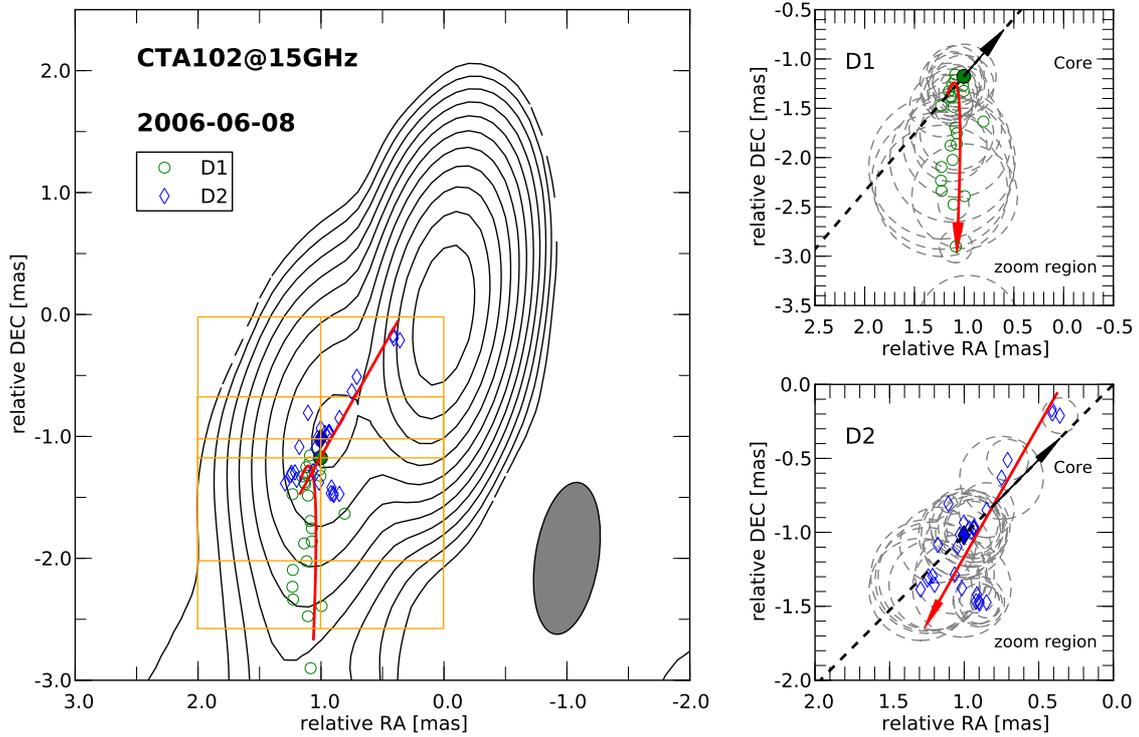} 
\caption{Same as Fig.~\ref{A1A2vec} for components D1 and D2.} 
\label{Dvec} 
\end{figure*}

The outermost component D1 is located around $3\,\mathrm{mas}$ from the core and is detected at  $8\,\mathrm{GHz}$, $15\,\mathrm{GHz}$, and $22\,\mathrm{GHz}$. We found from the long-term monitoring of \object{CTA\,102} at $15\,\mathrm{GHz}$ indications of an acceleration in the $y$ direction. This acceleration is clearly visible in the $y$-direction of the $15\,\mathrm{GHz}$ representative of D1 for $t>2005$. Since our time sampling at $8\,\mathrm{GHz}$ and $22\,\mathrm{GHz}$ spans $2005.4<t<2007.5$, the calculated values for the angular speed cannot be directly compared to the ones obtained at $15\,\mathrm{GHz}$. Therefore, we could not clearly cross-identify this feature. The angular and apparent speeds obtained at $15\,\mathrm{GHz}$ are $(0.1\pm0.02)\,\mathrm{mas/yr}$ and $(5.7\pm1.2)$\,\textit{c}. The component is moving on a highly nonradial trajectory outwards, nearly transversal to the main jet direction (see inlet in Fig.~\ref{Dvec}).

The feature labeled as D2 is detected throughout our entire multifrequency set. As in the case of D1 we found indications of an accelerating motion in both, the $x$ and $y$-component of D2 from the $15\,\mathrm{GHz}$ and $43\,\mathrm{GHz}$ observations. As mentioned before, owing to the acceleration connected to this component and the limited time sampling at $5\,\mathrm{GHz}$, $8\,\mathrm{GHz}$, and $22\,\mathrm{GHz}$ the angular speeds obtained from those observations do not necessarily correspond to the ones derived from the $15\,\mathrm{GHz}$ and $43\,\mathrm{GHz}$, but their positions are comparable within the uncertainties. We calculated an angular speed $\mu=(0.016\pm0.001)\,\mathrm{mas/yr}$, which corresponds to an apparent speed of $\beta_\mathrm{app}=(8.3\pm0.6)$\,\textit{c} using the long-term monitoring data at $15\,\mathrm{GHz}$ and $43\,\mathrm{GHz}$. The trajectory of D2 is compared to the main jet direction at $t_0$ within the uncertainties in good agreement with a radial outward motion (see inlet in Fig.~\ref{Dvec}).

At $15\,\mathrm{GHz}$ we detect one additional feature in this jet region, labeled as D3, which we could not distinguish at the other frequencies (see Appendix for all fitted components). The component D3 could be interpreted by blending effects in the core between two features that could be resolved at higher frequencies, especially $43\,\mathrm{GHz}$ (see following section).

\begin{figure}[h!]
\resizebox{\hsize}{!}{\includegraphics{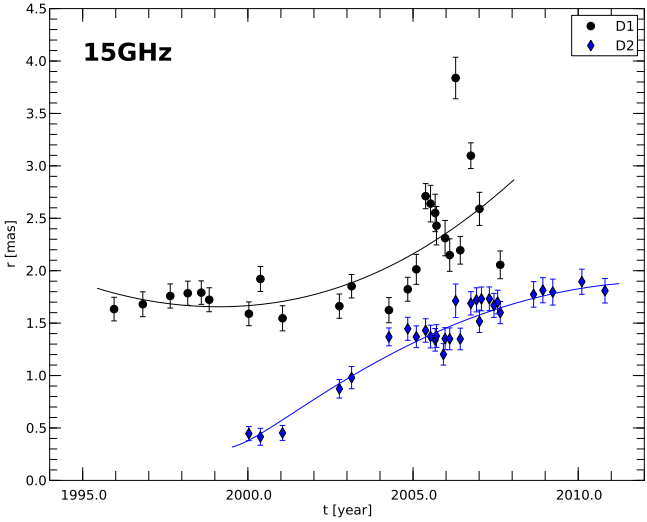}} 
\caption{Temporal separation from the core for the components D1 and D2 at $15\,\mathrm{GHz}$ taken as the representative frequency for the evolution of the cross-identified components within this region. The solid lines correspond to a first-order polynomial fit to the $x$- and $y$-positions separately.} 
\label{Dsep} 
\end{figure}

In Fig.~\ref{Dsep} we show the temporal separation from the core for D1, and D2 at $15\,\mathrm{GHz}$ and in Fig.~\ref{Dvec} we present the vector motion fits of these components  as crossidentified.

\begin{figure*}[h!]
\centering 
\includegraphics[width=17cm]{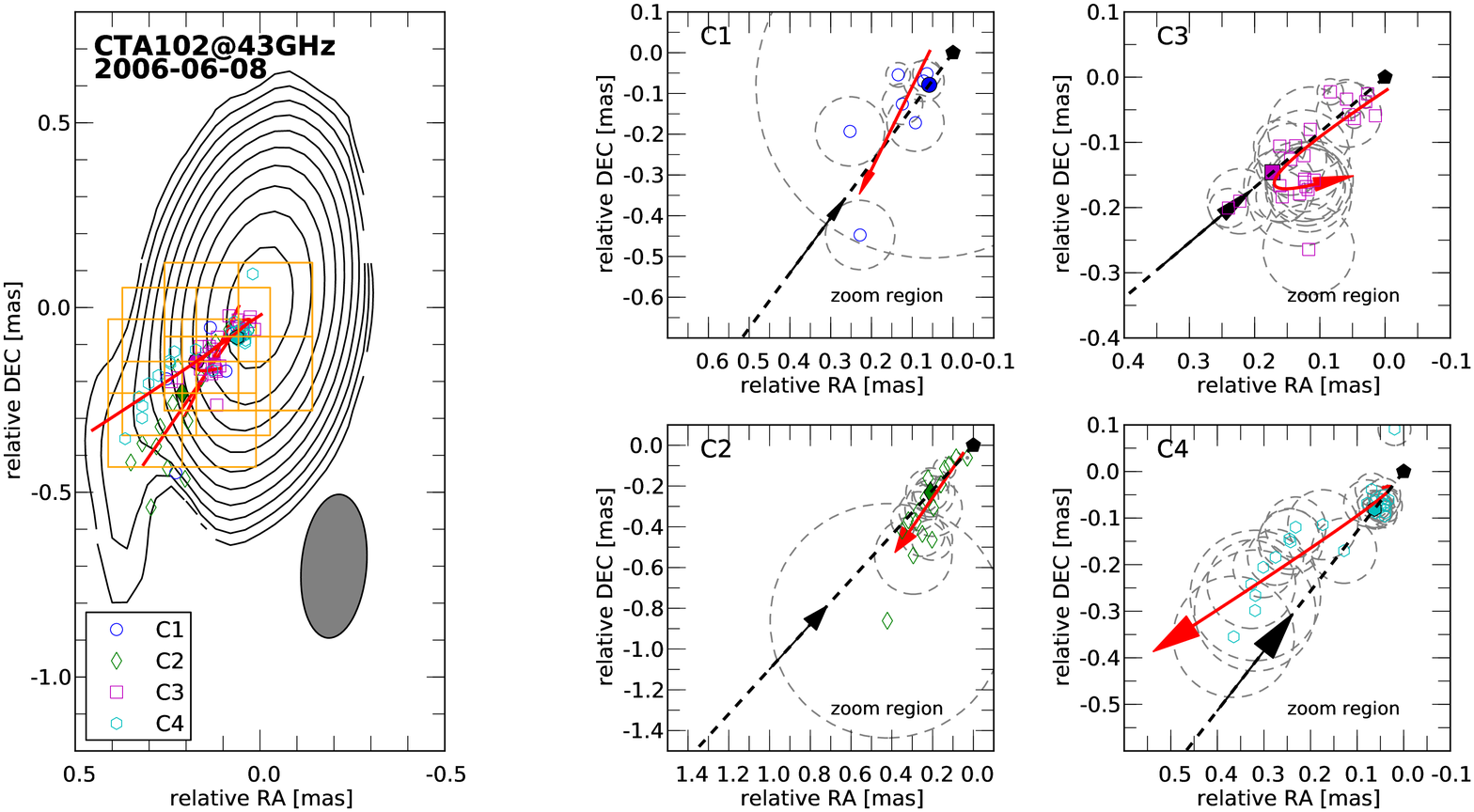} 
\caption{Same as Fig.~\ref{A1A2vec} for components C1, C2, C3, and C4, where the core is indicated here by the black pentagon marker} 
\label{Cvec} 
\end{figure*}

\subsection{Jet region C} \label{cregion}
The cross-identification {of the fitted features} at distances $r<1\,\mathrm{mas}$ from the core is a difficult task due to both the aforementioned splitting of the components and the ejection of new components, which may be associated with flares. These new components can be partially resolved at the highest frequencies but not at lower frequencies. In general such a component will appear delayed at lower frequencies after it has clearly separated from the core and can be resolved at a given frequency. Therefore, we concentrate for the core region of \object{CTA\,102} on the results of the fitting using the $43\,\mathrm{GHz}$ observations. We use six circular Gaussians, including the core, for the parameterization of the core region. Two out of these six components are within our multifrequency observations, labeled as C1 and C2.

C1 is ejected from the core around 2005.3 with an apparent speed of $\beta_\mathrm{app}=(11.6\pm3.0)$\,\textit{c} ($\mu=(0.22\pm0.06)\,\mathrm{mas/yr})$. The $y$-component of the C1 trajectory shows indication of an nonlinear motion but there was not a significant increase in the $\chi^2_\mathrm{red}$ to use a second-order polynomial fit. Within the uncertainties of our observations we classify the motion of C1 as radially outwards.

The second component is ejected from the core in $t_\mathrm{ej}=(2005.9\pm0.4)\,\mathrm{yr}$ with a comparable angular speed to C1 $\mu=(0.25\pm0.04\,\mathrm{mas/yr}$ corresponds to an apparent speed of $\beta_\mathrm{app}=(13.0\pm2.1)\,\textit{c}$. This feature is most probably connected to the strong radio outburst in \object{CTA\,102} in April 2006 and is moving radially outwards (see Sect.~\ref{disc}).

The two additional features, C3 and C4, show a clear nonlinear trajectory. In the case of C3, the component moves away from the core until a distance of $0.15\,\mathrm{mas}$ and keeps this position until the last of our observations. A second-order polynomial fit reduced the $\chi^2_\mathrm{red}$ from 5.0 to 2.3, and the apparent speed obtained is $\beta_\mathrm{app}=(3.8\pm0.6)\,\mathrm{c}$, which is three times smaller than the apparent speeds obtained for C1 and C2. Owing to the highly curved trajectory of the component, we could not compute an estimate for the ejection time. The first part of the trajectory could be classified as radially outward. However, farther downstream the component is moving nonradially inwards. 

The position of C4 remains constant at $r\sim0.1\,\mathrm{mas}$ from the core until 2009.6 and increases later. The reduced $\chi^2$ value decreased from a value of 5.5 to 2.2 by using a second-order polynomial instead of a first-order one. This fit leads to an apparent speed of $\beta_\mathrm{app}=(9.1\pm0.5)$\,\textit{c}. As in the case of C3, we could not derive an estimate for the time where the radial separation from the core is consistent with zero $(r(t_\mathrm{ej})=0)$, i. e., the ejection time. The trajectory of C3 shows clear indications of a nonradial motion. 

\begin{figure}[h!]
\resizebox{\hsize}{!}{\includegraphics{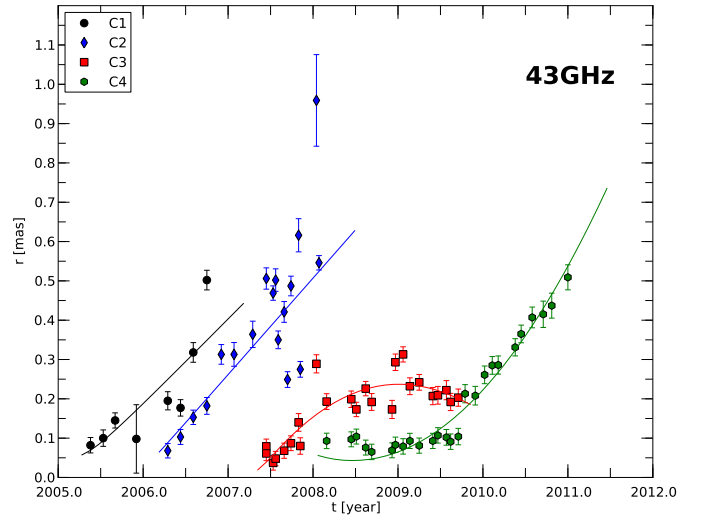}} 
\caption{Temporal separation from the core for the components C1, C2, C3, and C4 at $43\,\mathrm{GHz}$ taken as the representative frequency for the evolution of the cross-identified components within this region. The solid lines correspond to a first-order polynomial fit to the $x$- and $y$-positions separately.} 
\label{Csep} 
\end{figure}

In Fig.~\ref{Csep} we plot the radial separation from core for all components in the core region and in Fig.~\ref{Cvec} we present the vector motion fits. All fitted components in this region have the common behavior of having a nearly constant separation from the core during their early detection (See inlets in Fig.~\ref{Cvec}). 

\subsection{Summary of cross-identification}
\label{sumkin}
Here we summarize the results of the previous sections on the cross-identification of fitted components. We could clearly identify seven features throughout our entire data set. As mentioned before identification of components in the core region is a difficult task due to resolution, self-absorption effects and continuous ejection of new features. However from the $43\,\mathrm{GHz}$ observations we could detect four new features, were two of them where ejected during our multifrequency observations and one of them could be connected to the 2006 radio flare (see Sect.~\ref{disc}). In Table \ref{kinres} we present the average kinematic parameters of those components.

\begin{figure}[h!]
\resizebox{\hsize}{!}{\includegraphics{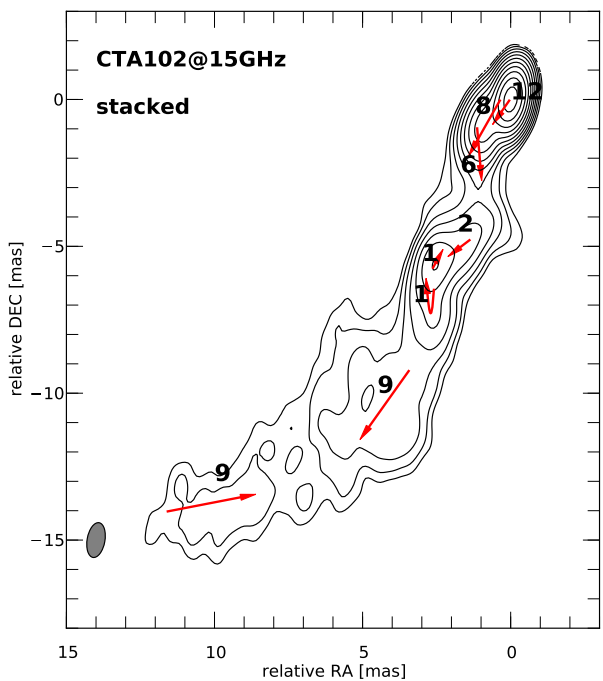}} 
\caption{Vector motion map for \object{CTA\,102}. The contour maps corresponds to the stacked $15\,\mathrm{GHz}$ observation of \object{CTA\,102}, and the arrows indicate the trajectories of each cross-identified component. The lowest contour level is drawn at 5$\times$ the average off-source rms ($4\,\mathrm{mJy}$) and the contours increase in steps of 2. The numbers above are the apparent speed of the component. 
} 
\label{vecmotion} 
\end{figure}

Only four components show a clear radial outward motion, best seen in C1 and C2. These components are detected in the core region, where the jet is straight and does not show strong bends. On the other hand, the nonradial motion of the majority of the cross-identified components may reflect the highly curved nature of \object{CTA\,102}. The inward motion of B2 could be the effect of its low pattern speed and the uncertainties in the position of the component. In Fig.~\ref{vecmotion} we show a vector motion map for all identified components.
In case of the critical angle assumption the apparent speed corresponds also the lower limit of the Doppler factor.

The detected acceleration in C3 and C4 could reflect either an acceleration of the fluid or the blending effect between a stationary feature close to core at $r\sim0.1\,\mathrm{mas}$. However, we could not detect indications of any acceleration in C1 and C2, which are located at a similar distance from core. Furthermore, \citet{Jorstad:2005p4121} find in their analysis of earlier $43\,\mathrm{GHz}$ VLBI observations a stationary feature (labeled as A1 in \citet{Jorstad:2005p4121}) at $0.1\,\mathrm{mas}$ from the core. If we assume that there is such a feature, we could split the components C3 and C4 into a stationary and traveling feature and compute the kinematic parameters for the moving ones. Such a procedure leads to an    
angular speed of $\mu=(0.20\pm0.05)\,\mathrm{mas/yr}$ ($\beta_\mathrm{app}=(10\pm2)$\,\textit{c}) for C3 and $\mu=(0.30\pm0.02)\,\mathrm{mas/yr}$ ($\beta_\mathrm{app}=(15\pm1)$\,\textit{c}) for C4. The values obtained are, within the uncertainties, in good agreement with the ones for C1 and C2, which favors the blending between a stationary and traveling components as a possible explanation for the trajectories of C3 and C4.

\begin{table*}
\caption{Results of the kinematic analysis for the cross-identified components}  
\label{kinres}
\tiny
\begin{tabular}{@{}c D{,}{\pm}{-1} D{,}{\pm}{-1} c c  D{,}{\pm}{-1} D{,}{\pm}{-1} c  c c c@{}}
\toprule
\hline
Comp	&	\multicolumn{1}{c}{$<\mu>$} &	\multicolumn{1}{c}{$\beta_\mathrm{app}$} 
& \multicolumn{1}{c}{$\delta_\mathrm{crit}$}
& \multicolumn{1}{c}{$\vartheta_\mathrm{crit}$} 
&\multicolumn{1}{c}{$t_\mathrm{ej}$}
&\multicolumn{1}{c}{$<r>$}
& \multicolumn{1}{c}{$t_\mathrm{min}$}
& \multicolumn{1}{c}{$t_\mathrm{max}$}
& \multicolumn{1}{c}{acceleration} 
& \multicolumn{1}{c}{classification}\\

& \multicolumn{1}{c}{[mas/yr]}
& \multicolumn{1}{c}{[c]}
& \multicolumn{1}{c}{[1]}
& \multicolumn{1}{c}{ [$^\circ$]}
& \multicolumn{1}{c}{ [yr]}
& \multicolumn{1}{c}{[mas]}
& \multicolumn{1}{c}{[yr]}
& \multicolumn{1}{c}{[yr]}
& \multicolumn{1}{c}{}
& \multicolumn{1}{c}{}\\
\midrule
A1 & 0.18,0.03 & 9.4,1.9       & 9   & 6   & -- & 17.0,0.5 & 1996.8 & 2010.8 & no & no-radial inward\\
A2 & 0.17,0.01 & 8.7,0.5       & 9   & 7   & 1945,6 & 11.2,0.2 & 1996.8 & 2010.8& no & radial outward\\
B1 & 0.016,0.005 & 0.85,0.25 & 1   & --    & -- & 7.6,0.3 & 1995.6 & 2010.8& yes & no-radial outward \\
B2 & 0.015,0.003 & 0.77,0.14 & 1   & --    & -- & 6.2,0.1 & 1996.8 & 2010.8 & no & radial inward\\
B3 & 0.027,0.006 & 1.41,0.32 & 2   & --    & -- & 5.1,0.2 & 1995.9 & 2010.8 & no & non-radial outward\\
D1 & 0.10,0.02 & 5.7,1.2       & 6   & 10  & -- & 2.1,0.4 & 1995.9 & 2007.6 & yes & non-radial outward\\
D2 & 0.16,0.01 & 8.3,0.6       & 8   &  7  & 1997.1,0.4 & 1.6,0.4 & 2000.0 & 2011.0 & yes & radial outward\\
C1 & 0.22,0.06 & 11.6,3.0     & 12 & 5   & 2005.1,0.2 & 0.2,0.1 & 2005.4 & 2006.8 & no & radial outward\\
C2 & 0.25,0.04 & 13.0,2.1     & 13 & 4   & 2005.9,0.2 & 0.4,0.2 & 2006.3 & 2008.0 & no & radial outward\\
C3 & 0.07,0.01 & 3.8,0.6       & 4   & 15 & -- & 0.2,0.1 & 2007.5 & 2009.7 & yes & non-radial outward\\
C4 & 0.17,0.01 & 9.1,0.5       & 9   &  6  & -- & 0.2,0.1 & 2008.2 & 2011.0 & yes & non-radial outward\\
\bottomrule
\hline
\end{tabular}
\end{table*}

\subsection{Variation in the apparent speed and Doppler factor}
\label{dopsec}
\begin{figure}[h!]
\resizebox{\hsize}{!}{\includegraphics{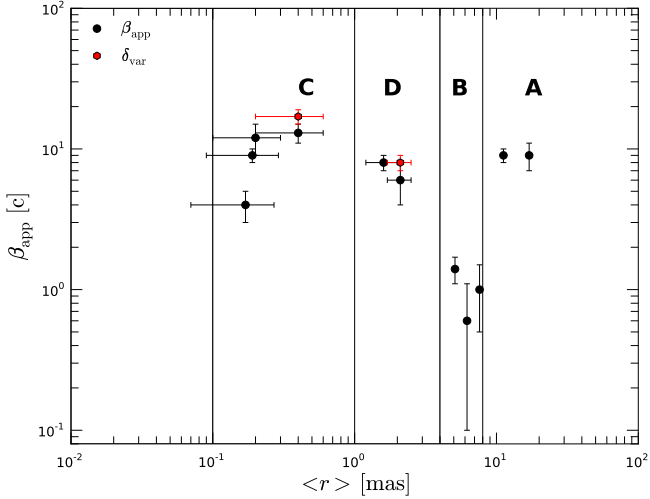}} 
\caption{Variation in the apparent speed, $\beta_\mathrm{app}$ along the jet. The two red points correspond to the variability Doppler factors, $\delta_\mathrm{var}$, computed for components C2 and D1.} 
\label{kinrad} 
\end{figure}

\begin{figure}[h!]
\resizebox{\hsize}{!}{\includegraphics{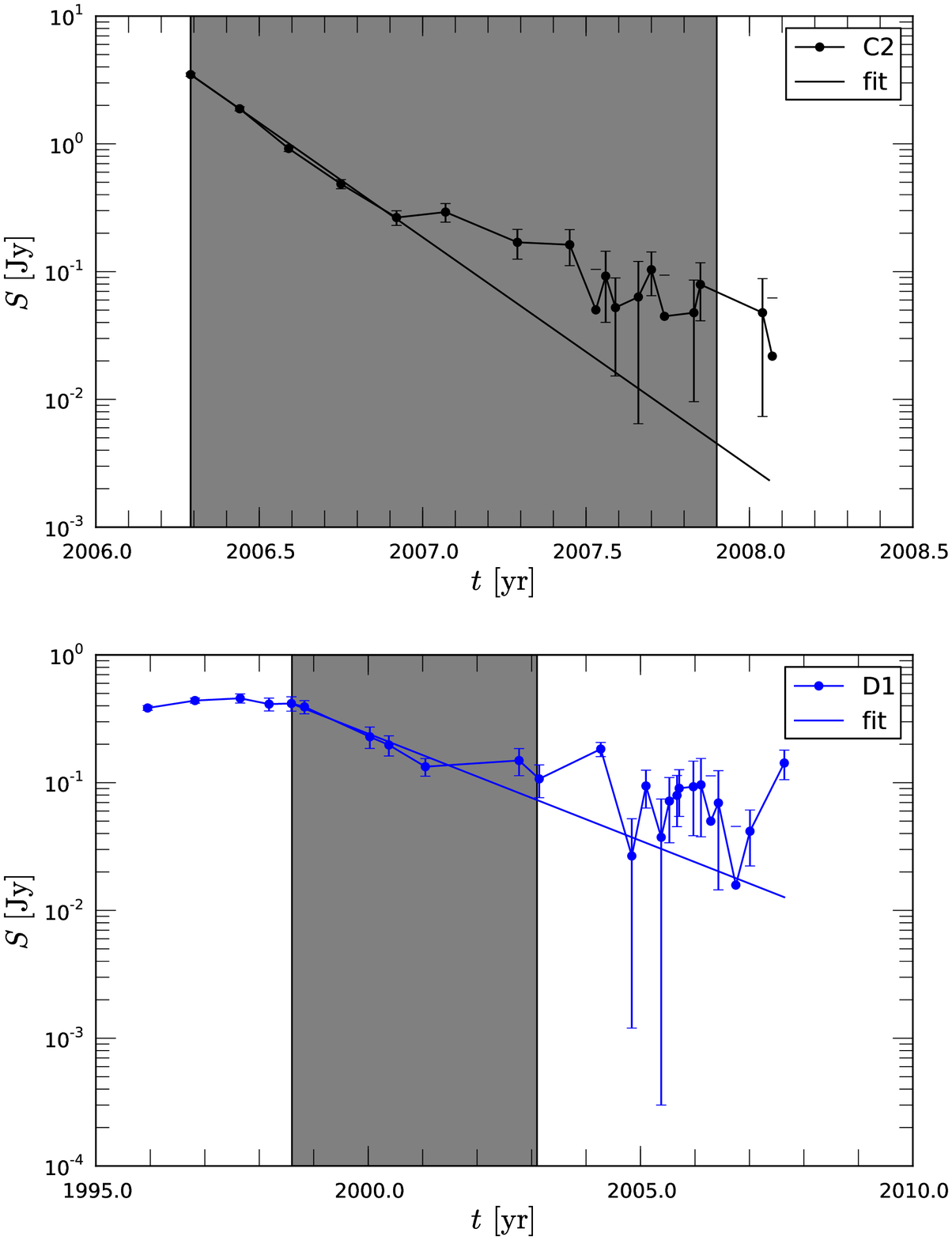}} 
\caption{Variation in the flux densities for the components C2 (left panel) and D1 (right panel). The gray shaded ares indicate the time range used for calculating the variability Doppler factor, and the solid line correspond to an exponential fit to the flux densities.} 
\label{c2d1flux} 
\end{figure}

{The results of the kinematic analysis of the cross-identified components show that the apparent speed increases between $0.2\,\mathrm{mas}<r<1.0\,\mathrm{mas}$ (region C in Fig.~\ref{allcontcore}) from $\beta_\mathrm{app}=4$ to $\beta_\mathrm{app}=13$. Farther downstream the values decrease from $\beta_\mathrm{app}=8$  to $\beta_\mathrm{app}=1$ between regions D ($1.0\,\mathrm{mas}<r<5.0\,\mathrm{mas}$) and B ($5.0\,\mathrm{mas}<r<8.0\,\mathrm{mas}$). This fall is related with the very low pattern motion of the components in the latter.
In the extended jet region A ($r>8.0\,\mathrm{mas}$), the apparent speed increases again and reaches a constant value (within uncertainties). Figure~\ref{kinrad} shows the apparent speed for all cross-identified components versus its average position, taken as a representation of its location. The apparent speed, $\beta_\mathrm{app}$, and the Doppler factor, $\delta_\mathrm{crit}$, assuming the components are traveling at the critical angle, $\vartheta_\mathrm{crit}$, are presented in Table \ref{kinres}.

The degeneracy of the apparent speed and Doppler factor on the intrinsic speed, $\beta$, and the viewing angle, $\vartheta$, does not allow any direct conclusion on the variation in the fluid speed. This degeneracy can be broken if we obtain additional information on the Doppler factor from the variability of the flux densities. One of the requirements for applying this second method in calculating of the Doppler factor is that the flux densities should have an exponential behavior for several time segments \citep{Jorstad:2005p4121}. The components C2 and D1 fulfill this requirement (see Fig.~\ref{c2d1flux}), and we computed their variability Doppler factor, $\delta_\mathrm{var}$ and the corresponding viewing angle, $\vartheta_\mathrm{var}$. The results of these calculations are presented in Table \ref{dopplertab}.   
The variability Doppler factor decreases from $\delta_\mathrm{var}=17\pm3$ at $\langle r \rangle \sim0.4\,\mathrm{mas}$ (C2) to $\delta_\mathrm{var}=8\pm2$ at $\langle r \rangle \sim2.1\,\mathrm{mas}$ (D1). The viewing angle, $\vartheta_\mathrm{var}$ is smaller than the critical angle for both components (see Tables \ref{kinres} and \ref{dopplertab}). We can then conclude that the fluid decelerates between $(0.4<r<2.1)\,\mathrm{mas}$. However, there is also a change in the viewing angle, $\vartheta$.}

\begin{table}[h!]
\caption{Used parameters for the calculation of the variability Doppler factor and viewing angle for components C2 and D1}

\label{dopplertab}
\centering
\tiny
\begin{tabular}{c c c c c}
\hline\hline
component & $d_\mathrm{eff}$  [mas]	&	$\Delta t_\mathrm{var}$ [yr]	&	$\delta_\mathrm{var}$	&	$\vartheta_\mathrm{var}$ \\
\hline
C2 (2006.3$<$t$<$2007.9)		   &$0.16\pm0.02$		&	$0.33\pm0.06$				&	$17\pm3$				&	$3.2\pm0.2$ \\
D1(1998.6$<$t$<$2003.1)		   &$0.77\pm0.05$		&	$3.3\pm0.8$				&	$8\pm2$				&	$6.8\pm0.5$ \\
\hline
\end{tabular}
\end{table}

\subsection{Evolution of the brightness temperature $T_\mathrm{b}$}\label{tbsec}
The brightness temperature is defined as
\begin{equation}
T_\mathrm{b}=1.22\times10^{12}S_{\nu}R^{-2}\nu^{-2}(1+z)\, [\mathrm{K}]
\end{equation}
where $S_{\nu}$ is the flux density in $\mathrm{Jy}$, $R$ the size of the emission zone in $\mathrm{mas}$,  and $\nu$ the observing frequency in $\mathrm{GHz}$. Some physical conditions, such as the evolution of the magnetic field in the source, can be derived from the brightness temperature \citep[e.g.,][]{Kadler:2004p2884,Schinzel:2012p5740}. We present the relations between the brightness temperature and the evolution of the physical parameters in the framework of the shock-in-jet model \citep{Marscher:1985p50}. It is known that the optically thin flux density emerging from a relativistic electron distribution $N=KE^{-s}$ is given by:
\begin{equation}
S_\nu\propto R^{2}x\delta^{(s+3)/{2}}K B^{{(s+1)}/{2}}\nu^{-(s-1)/{2}},
\label{snu}
\end{equation}
where $R$ is the width of the jet, $x$ the size of the emitting region, $B$ the magnetic field, and $s$ the spectral slope. If the shock propagates downstream, $x$ is the width of the shock, otherwise $x=R$. The width of the shock front, $x$, depends on the main energy loss mechanism, namely Compton (1), synchrotron (2), and adiabatic losses (3) \citep{Marscher:1985p50} and is given by
\begin{eqnarray}
x_1&\propto& R^{-(s+5)/8} K^{-1}B^{-3(s+1)/8}\delta^{1/2}\nu^{-1/2}\\
x_2&\propto& B^{-3/2}\delta^{1/2}\nu^{-1/2}\\
x_3&\propto& R.
\end{eqnarray}
For a more detailed derivation of the equations above, see also Paper I. Inserting the equations above into Eq. \ref{snu} leads to
\begin{eqnarray}
S_{\nu,1}&\propto& R^{-(s-11)/8}\delta^{(s+4)/2}B^{(s+1)/8}\nu^{-s/2} \label{S1}\\
S_{\nu,2}&\propto& R^2\delta^{(s+4)/{2}}K B^{{(s-2)}/{2}}\nu^{-s/2} \label{S2} \\
S_{\nu,3}&\propto& R^3\delta^{(s+3)/{2}}K B^{{(s+1)}/{2}}\nu^{-(s-1)/{2}} \label{S3}.
\end{eqnarray}
If we assume that the magnetic field, $B$, the normalization coefficient of the relativistic electron distribution, $K$, and the Doppler factor, $\delta$, follow a power law:
\begin{equation}
B\propto R^{-b} \quad K\propto R^{-k} \quad \delta\propto R^{-d},
\end{equation}
Equations~\ref{S1}$-$\ref{S3} can be written in terms of the jet radius
\begin{equation}
S_{\nu,i}\propto R^{p_i}\nu^{q_i},
\end{equation}
where the exponent $p_i$ includes the dependencies on $B$, $K$, and $\delta$, and the exponent $q_i$  includes the dependence on the frequency:
\begin{eqnarray}
p_1&=&-(s-11)/8-b(s+1)/8-d(s+4)/2\\
p_2&=&2-b(s-2)/2-d(s+4)/2-k \\
p_3&=&3-b(s+1)/2-d(s+3)/2-k \\
q_1&=&-s/2 \\
q_2&=&-s/2 \\
q_3&=&-(s-1)/2.
\end{eqnarray}

\begin{figure*}[t!]
\centering 
\includegraphics[width=17cm]{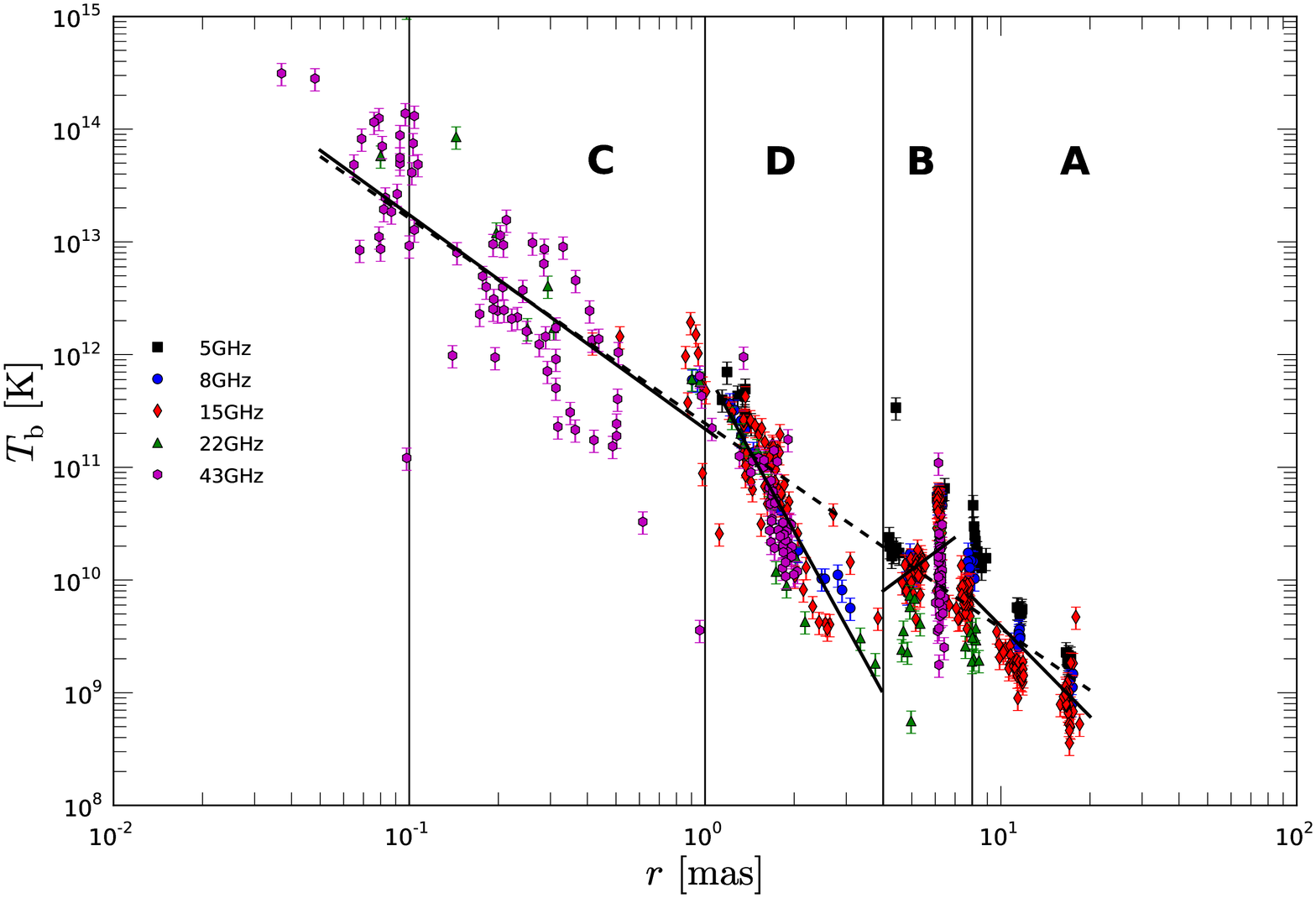}
\caption{Variation in the brightness temperature, $T_b$ with distance along the jet for all available epochs and cross-identified components (presented in Tables \ref{modelfit1} and \ref{modelfit2}). The nearly vertical variation within $6\,\mathrm{mas}<r<8\,\mathrm{mas}$ in region B reflect the temporal variation in the size of the fitted Gaussians. {The solid lines correspond to power-law fits $T_\mathrm{b}\propto r^{a}$ to the individual regions, and the dashed line to a power-law fit to the entire range $0.1\,\mathrm{mas}<r<20\,\mathrm{mas}$, which leads to an exponent $a=-1.8$ (see text).}} 
\label{tball} 
\end{figure*}

In order to generalize these equations to nonconical jets, the distance along the jet, $r$, is no longer directly proportional to the jet radius $R$. This modification is expressed by $R\propto r^{\rho}$ where $-1<\rho<1$. Finally, we can write the evolution of the brightness temperature as
\begin{equation}
T_\mathrm{b}\propto r^{\epsilon_i}\nu^{q_i-2},
\end{equation}
where the exponent $\epsilon_i$ is given by
\begin{eqnarray}
\epsilon_1&=&\rho\left[-(s+5)/8-b(s+1)/8-d(s+4)/2\right] \label{eps1} \\
\epsilon_2&=&\rho\left[-b(s-2)/2-d(s+4)/2-k\right] \label{eps2} \\
\epsilon_3&=&\rho\left[1-b(s+1)/2-d(s+3)/2-k\right] \label{eps3}.
\end{eqnarray}

The equations above can be simplified further if we assume adiabatic expansion $k=k_{\mathrm{ad}}=2(s+2)/3$ and equipartition between the magnetic energy density, $\mathcal{E}_\mathrm{mag}\propto B^2$ and the kinetic energy density, $\mathcal{E}_\mathrm{kin}\propto K$, which leads to $b=b_\mathrm{eq}=(s+2)/3$:
\begin{eqnarray}
\epsilon_1^{\mathrm{ad,eq}}&=&\rho\left[-(s^2+5s+12)/16-d(s+4)/2\right] \label{eps1} \\
\epsilon_2^{\mathrm{ad,eq}}&=&\rho\left[-(s^2+4s+8)/6-d(s+4)/2\right] \label{eps2} \\
\epsilon_3^{\mathrm{ad,eq}}&=&\rho\left[-(s^2+7s+4)/6-d(s+3)/2\right] \label{eps3}.
\end{eqnarray}

These relations can be used to derive the parameter range for exponents $\rho$, $s$, $k$, $b$, and $d$ using our calculations of the brightness temperature. Figure~\ref{tball} shows the variation in $T_b$ along the jet for several frequencies, which can be approximated by power laws, $T_b\propto r^{a}$, with changes in the exponent in different regions.

Following the convention of regions defined above, from A to D, we derived a decrease in the brightness temperature between $0.1\,\mathrm{mas}$ and $1\,\mathrm{mas}$ (region C). Region D, $1\,\mathrm{mas}$ to $4\,\mathrm{mas}$, shows a decrease in the Doppler factor (Fig.~\ref{kinrad}) and a further steepening of the brightness temperature gradient as compared to region C. In region B ($4\,\mathrm{mas}$ to $7\,\mathrm{mas}$)  the brightness temperature values increases and farther downstream ($r>7\,\mathrm{mas}$), the brightness temperature decreases again (see Fig.~\ref{tball}).
We applied power-law fits to the brightness temperature evolution as a function of the core distance to the different regions. The results are presented in Table \ref{tbevo} and are discussed in the following. The study of the transversal jet size within the analysis of the jet ridge line allowed us to constrain the estimates for the exponent $\rho$, thereby parametrizing the jet width (Paper III, in preparation). The values obtained for the parameter $\rho$ (jet opening index) are presented in Table \ref{rhotab}.

\begin{table}[h!]
\caption{Results of the power-law indices from the brightness temperature as a function of the distance to the jet basis $T_b\propto r^{a}$.}  
\label{tbevo}
\centering  
\begin{tabular}{c c c c c}
\hline\hline
region & C	&	D	&	B	&	A \\
\hline
$a$ &$-1.9\pm0.2$	&	$-4.8\pm0.3$	&	$2.0\pm0.6$	&	$-2.7\pm0.2$ \\
\hline
\end{tabular}
\end{table}

\begin{table}[h!]
\caption{Estimates for the exponent $\rho$, defining the jet geometry $R\propto r^{\rho}$ (Paper III, in preparation).}  
\label{rhotab}
\centering  
\begin{tabular}{c c c c c}
\hline\hline
region &C	&	D	&	B	&	A\\
\hline
$\rho$& $0.8\pm0.1$	&	$0.8\pm0.1$	&	$-1.0\pm0.1$	&	$1.3\pm0.2$ \\
\hline
\end{tabular}
\end{table}

We discuss the results in the following terms: i) the jet is in equipartition (magnetic energy density is equal to the kinetic energy density of the relativistic particles), ii) the magnetic field is toroidal $(b=1)$, and iii) the magnetic field is poloidal $(b=2)$. In the three cases we consider adiabatic expansion and compression. We thus assumed that, at distances $r>0.1\,\mathrm{mas}$ (deprojected $18\,\mathrm{pc}$) Compton and synchrotron losses can be neglected as the dominant energy-loss mechanism \citep{Mimica:2009p42}. The energy losses that the jet undergoes imply a decrease in the intensity of the magnetic field and in the Lorentz factors of the electrons. This makes it difficult for synchrotron radiation to dominate the losses far from the jet nozzle \citep{Mimica:2009p42}. Therefore, we considered only adiabatic losses within adequate boundaries for the parameters $s$, $b$, $k$, and $d$ in our analysis.

\subsubsection{Region C $(0.1\,\mathrm{mas}< r<1.0\,\mathrm{mas})$}
The variation in the brightness temperature in the region B ($0.1\,\mathrm{mas}<r<1.0\,\mathrm{mas}$) could be best studied at $43\,\mathrm{GHz}$. At this frequency we achieved the required resolution to resolve the jet, and the long-term monitoring provides us a large number of data points that increased the statistical significance. The power-law fit gave an exponent of $a=-1.9\pm0.2$ (Table \ref{tbevo}). In Fig.~\ref{allregions}, the different panels, labeled 1 to 3, show the range of possible parameters for the three different jet configurations proposed (indicated at the top of each panel). The panels show the variation in the exponent that determines the evolution of the Doppler factor, $\delta\propto R^{-d}$\, or in terms of distance along the jet $\delta\propto r^{-\rho d}$ using $R\propto r^\rho$ in terms of the exponents $\rho$ and $s$. 
{We find that for all three magnetic field configurations, the Doppler factor should increase with distance ($d<0$). Among the three cases studied, the slowest increase corresponds to the toroidal magnetic field ($b=1$), and it is faster for the other field configurations, with a higher value of $\mathrm{abs}(d)$. The increase in the Doppler factor could be due to an increase in the fluid speed and/or a decrease in the viewing angle. The increase in $\beta_\mathrm{app}$, together with the increasing $\delta$, suggests an acceleration of the fluid. However, our kinematic analysis of the cross-identified components did not allow us to extract any indication of fluid acceleration (see Fig.~\ref{Csep} and Table~\ref{kinres}).}

\subsubsection{Region D $(1\,\mathrm{mas}< r<4\,\mathrm{mas})$}
{The power law fitted to the evolution of the brightness temperature $(T_b\propto r^{a})$  in this region yields a value of $a=-4.8\pm0.3$. The results of the parameter space study in this case are shown in panels 4$-$6 in Fig.~\ref{allregions}. The three different configurations, using the value obtained for $\rho$ (see Table~\ref{rhotab}), lead to two clearly different results: On the one hand, if the magnetic field is purely toroidal, the Doppler factor had a decreasing exponent throughout the entire parameter space. On the other hand, an equipartition magnetic field and a purely poloidal magnetic filed permitted a decreasing or increasing Doppler factor, depending on the spectral slope. Based on the kinematics obtained, we found a slightly decreasing Doppler factor derived for components C2 and D1, which is compatible with the toroidal configuration (see panel 5 in Fig.~\ref{allregions}), but due to the lack of data within this region we could not rule out any other possibility. }

\subsubsection{Region B $(4\,\mathrm{mas}< r<7\,\mathrm{mas})$}
In contrast to the regions C, D, and A, the brightness temperature increased with distance in this region. This behavior led to a positive value for the exponent $a$ $(T_b\propto r^{a})$. We derived an exponent $a=2.0\pm0.6$ and at the same distance from the core, we found a change from an expanding jet $\rho>0$ to a collimating jet $\rho<0$ (see Table \ref{rhotab}). We calculated the possible values for the parameter $d$ for a collimating jet $\rho<0$, and the results are presented in panels 7-9 in Fig.~\ref{allregions}. 
In the case of a collimating jet, a positive value of $d$ corresponds to an increase in the Doppler factor, and the opposite is also true ($\delta\propto R^{-d}\, \&\,R\propto r^\rho \rightarrow \delta\propto r^{-\rho d}$). The parameter space study showed that the Doppler factor decreased for all jet configurations within the limits obtained for jet expansion rate, $\rho$. The major difference between the model is that the computed values for $d$ are rather extreme $0.6<d<1.6$ for nontoroidal magnetic field configurations. Assuming that the magnetic field is toroidal $\left(B\propto r^{-1}\right)$ in region D, we favor the continuation of the magnetic field configurations. 
  
\subsubsection{Region A $(r>7\,\mathrm{mas})$}
\begin{figure*}[t!]
\centering 
\includegraphics[width=15cm]{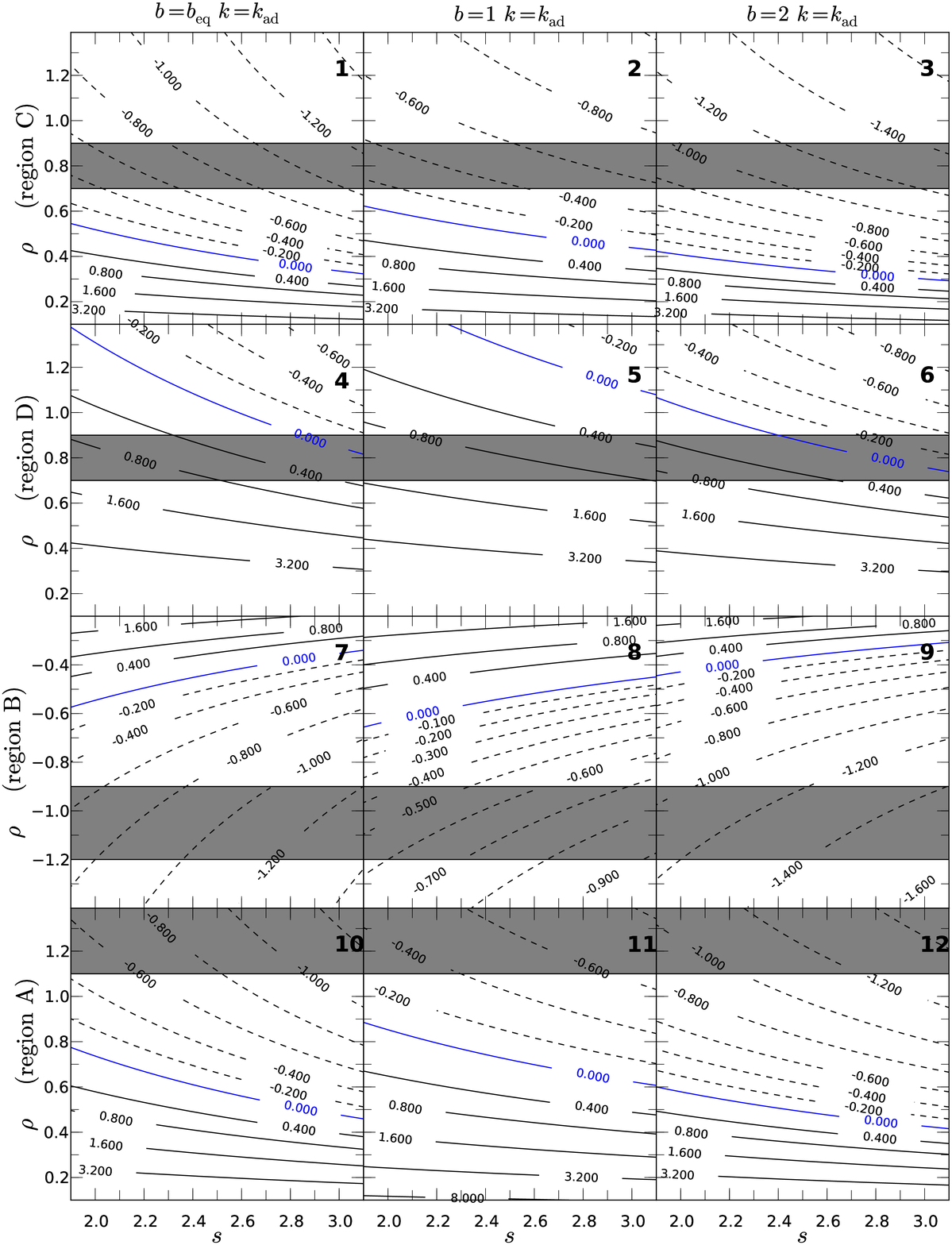} 
\caption{Parameter space for the evolution of the Doppler factor index $d$ as a function of the jet expansion index $\rho$ and the spectral slope $s$ for the different region A - D for three different jet configurations, The first column assumes a adiabatic jet ($k=k_\mathrm{ad}=2(s+2)/3$) and equipartition between the magnetic energy density and the kinetic energy of the relativistic particles, represented by $b=b_\mathrm{eq}=(s+2)/3$. The second and third columns show the evolution of $d$ for an adiabatic jet ($k=k_\mathrm{ad}$) and decreasing magnetic field $b=1$ or $b=2$, respectively.  {The iso-contours in each panel} show the allowed range for the exponent $d$ $(\delta\propto r^{-\rho d})$ and the gray shaded area the range for the parameter $\rho$ $(R\propto r^{\rho})$, with different scales for parameter $\rho$ in region B. For more details see text.} 
\label{allregions} 
\end{figure*}

{At distances over $r>7\,\mathrm{mas}$, the brightness temperature decreased again with an average exponent of $a=-2.7\pm0.2$. Taking into account that the expansion index of the jet in this region is $\rho=1.3\pm0.2$ (see Table~\ref{rhotab}), we obtained the possible values for the evolution in the Doppler factor (see panels 10 -- 12 in Fig.~\ref{allregions}). All the studied jet configurations imply an increase in the Doppler factor and the different models only differ in terms of the resulting exponent $d$. As in the case of region C, the slowest increase of the Doppler factors was obtained for $b=1$. Considering that the field geometry cannot change from toroidal to poloidal, a continuation of a toroidal magnetic field configuration is the most plausible one (see panel 11 in Fig. \ref{kinrad}).}

\section{Discussion}
\label{disc}
In the region close to the core ($r<4\,\mathrm{mas}$), we detect several
traveling features, most visible at $43\,\mathrm{GHz}$ (Figs.~\ref{Cvec} and \ref{Csep}). These features can be interpreted as traveling
shock waves, which can be generated by differential injection
pressure at the jet nozzle \citep[e.g.,][]{Perucho:2008p3885}.
Shock waves travel downstream, compressing the medium ahead of the
shock front and re-accelerating particles. The increase in radiation
induced by the shock is observed and can be traced along the jet,
until it fades due to losses and to the limited dynamical range of the
observations.
In contrast to the expected flux density evolution of a traveling
shock, we observe that C1, C3, and C4 show some local increase in their flux
densities while they propagate downstream around $r\simeq 0.1-0.2\,\mathrm{mas}$. 
This could be explained
in terms of the interaction of this shock with a standing recollimation
shock \citep[e.g.,][]{Gomez:1997p649, Mimica:2009p42},
or by an increase in the Doppler boosting of the component if it travels
along a helical jet, at the positions where it is closer to the line of
sight \citep[e.g.,][]{ Hardee:2000p5817, Aloy:2003p3889}.
Recollimation shocks may appear if the jet and the ambient medium are
not in pressure equilibrium at the nozzle. 
Our observations do not allow us to resolve any change in the
jet direction at this position, and neither the Doppler factor (Sect.~\ref{dopsec} and Fig.~\ref{kinrad}) nor the brightness temperature
(Sect.~\ref{tbsec} and Fig.~\ref{tball}) analysis permit us to obtain any clear results from this region, 
but we did find some independent evidence that favors the presence of a standing shock at this position (see Papers I and III, in prep.).

\subsection{The physical parameters along the jet}
\subsubsection{The Doppler factor $\delta$}
{In the previous sections, we have seen that Doppler factor should increase in region C $(0.1\,\mathrm{mas}<r<1\,\mathrm{mas})$, according to the shock-in-jet model applied to the observed evolution of the brightness temperature. Since our sampling at $43\,\mathrm{GHz}$ is not dense enough we could not calculate variability Doppler factors to test this conclusion.

 The variability Doppler factor, $\delta_\mathrm{var}$, computed between regions C and D shows a decrease with distance (see  Fig.~\ref{kinrad} and Sect.~\ref{dopsec}). At the same time, the apparent speed decreases and the viewing angle, $\vartheta_\mathrm{var}$, is smaller than the critical viewing angle, $\vartheta_\mathrm{crit}$. This implies that the fluid decelerates within this region. This deceleration of the fluid is in good agreement with the calculated behavior of the Doppler factor based on the gradient in the brightness temperature, $T_\mathrm{b}$ (see panel 4$-$6 in Fig.~\ref{allregions}).

For region B  ($4\,\mathrm{mas}<r<8\,\mathrm{mas}$) we could not derive Doppler factors owing to the lack of traveling components. Our calculations suggest a decrease in the Doppler factor with respect to region D, which is independent of the geometry of the magnetic field. This deceleration coincides with a region of collimation of the jet ($\rho<0$).  
Farther downstream, the jet expands again and, based on the evolution of the brightness temperature within this region, we derived an increase in $\delta$.  
The obtained evolution of the Doppler factor can be interpreted within the framework of an overpressured jet, where the fluid accelerates in the regions where the jet expands ($\rho>0$) and decelerates, on average, in the regions where the jet reconfines ($\rho<0$).}



\subsubsection{The brightness temperature $T_\mathrm{b}$}
The brightness temperature (see Sect.~\ref{tbsec} and Fig.~\ref{allregions}) decreases between $0.1\,\mathrm{mas}<r<1\,\mathrm{mas}$ (region C), and the decreasing trend continues until  $r\simeq4\,\mathrm{mas}$ (region D). However, there are hints of a change in this trend around $r\simeq1\,\mathrm{mas}$, i.e., between regions C and D. Between $r\simeq4\,\mathrm{mas}$ and  $r\simeq8\,\mathrm{mas}$ (region B), the brightness temperature increases, and decreases thereafter. 

{From the evolution of the brightness temperature in region C ($0.1\,\mathrm{mas}<r<1\,\mathrm{mas}$, see Sect.~\ref{tbsec} and Fig.~\ref{allregions}), we derive that the Doppler factor should be increasing. The value of the exponent of $\delta$ as a function of distance depends on the configuration of the magnetic field, and it increases with $b$, where $1<b<2$. The lowest range for the increase in the Doppler factor $-0.8<d<-0.2$ is obtained for a toroidal magnetic field ($b=1$), as seen in panel 2 in Fig.~\ref{allregions}}. 

In region D ($1\,\mathrm{mas}<r<4\,\mathrm{mas}$) the decrease in the Doppler factor could be related to a recollimation and deceleration process, ending up in the stationary component at $r\simeq 2\,\mathrm{mas}$. (\citet{Jorstad:2005p4121} report a standing component at the same distance.) There is an indication of a slight decrease in the growth of the jet radius with distance  that could be related to the onset of the recollimation process (Paper~III, in preparation). {The highest values for $d$, implying a steeper decrease in the Doppler factor ($\delta\propto R^{-d}$), are obtained in the case of a toroidal field and equipartition. Among these two cases, the latter implies a flatter spectral distribution, a situation that will be studied in Paper~III.} The fall in the brightness temperature could be due to the deceleration of the flow in a possible shocked region and to it's not being compensated for by the expected increase in flux density at those shocks. 


Between $r\simeq 4\,\mathrm{mas}$ and $r\simeq 8\,\mathrm{mas}$, i.e., region B, the brightness temperature increases and the jet radius decreases with distance (Paper III, in preparation, see Fig.~\ref{allregions}). According to our calculations, the Doppler factor should decrease. Thus, the increase in the brightness temperature could also be explained in terms of a new process of recollimation of the jet flow, which would generate the standing components visible in the region. It is difficult to relate the evolution of the Doppler factor with distance to the results shown in Fig.~\ref{allregions} owing to the lack of data and the large error in the determination of the exponent for the evolution of jet radius, $\rho$. For larger distances from the core the Doppler factor increases, and the brightness temperature increases again. Within this same interpretation, region A would represent a new expansion region.

\subsection{Helical versus overpressured jet}
The viewing angles calculated for the components C2 and D1 increase with distance from the core, which indicates a bend in the jet away from our line of sight. Since the viewing angles are smaller than the critical angles and the apparent speeds decrease, we know that the plasma also decelerates. The observed drop in the brightness temperature within region D, as compared to region C, is a direct consequence of the decreasing Doppler factor due to a deviation of the jet away from our line of sight, together with the deceleration of the plasma. On the one hand, the stationary behavior of components at $r\sim1.5\,\mathrm{mas}$ could indicate the presence of a recollimation shock at this position. Such recollimation shocks are expected to form in overpressured jets, and the flow should decelerate when crossing them. On the other hand, the increasing viewing angle between regions C and D could indicate the growth of helical instabilities triggered by  pressure asymmetries within the jet. 

These instabilities lead to the formation of a helical pattern with regions of higher pressure, and hence enhanced emission as compared to the rest of the jet flow. Depending on the jet properties, we would expect to see the higher pressure region tracing helical patterns at higher frequencies and the straighter flow at lower frequencies (Perucho et al. 2012). The observed morphology of \object{CTA\,102} at different frequencies confirms this aspect to the extent that the resolution at each frequency permits. In the case of a strongly over pressured jet, the subsequent expansions and collimations would dominate over the linear amplitudes of growing instabilities and we would expect a straight flow. The observed changes in the viewing angle imply that the jet is not largely over pressured with respect to its surroundings. The stability of such a flow and extensions of the regions of enhanced emission depend on the jet properties and the interaction/competition between the pinching and helical modes \citep[see, e.g.,][]{Perucho:2004p3871,Perucho:2004p3886}.


At $r>4\,\mathrm{mas}$ from the core, the source is dominated by three apparently stationary components at all the observing frequencies that allow detection. These components also show nearly constant size and flux density during the observed period. If the components are related to physical regions, this behavior can also be explained either by recollimation shocks in a pinching jet, by changes in
the viewing angle in a helical jet or a combination of both. The scenario of an overpressured jet is
supported by two facts. 
1) There are clear hints of  jet expansion and subsequent
recollimation in the innermost regions (Paper III), which is naturally explained in terms of overpressure, and 2) the position of stationary components at in region B  show that the separation between them is growing with distance along the jet. We find a separation of $\sim1.0\,\mathrm{mas}$ between B3 and
B2,  and $\sim1.5\,\mathrm{mas}$ between B2 and B1. This increase could be naturally caused by a
decrease in the ambient pressure: If the ambient medium is
homogeneous, the standing shocks appear at equidistant positions and
show similar properties, e.g., the jump in pressure and density at each
shock, whereas if the density in the ambient decreases with
distance, the shocks appear at increasing separations, while pressure jumps become smaller.

Although the increase in size and the decrease in flux density with distance of the stationary components is expected 
in the frame of an expanding jet, this is not an exclusive feature of pinching jets, but can also happen in a helical jet. 
The jet does not show any apparent kink between $r=4\,\mathrm{mas}$ and $r\simeq12\,\mathrm{mas}$, but components
B1, B2, and B3 are slightly misaligned (see Fig.~\ref{Bvec}), which leaves room for short-wavelength kinks in the jet in this region, with the changes in the jet direction occurring in the short projected distances between components in regions B and A. The increase in the 
separation between components with distance to the core could then be explained in this frame if the kinks were associated to a helical instability in an expanding jet \citep[e.g.,][]{Hardee:2003p5821, Hardee:2005p5826}. The Doppler factor seems to grow from region B to region A (see Fig.~\ref{kinrad}). However, the lack of components (notice, that Fig.~\ref{kinrad} is logarithmic) does not allow us to discard local changes on this trend, i.e., changes in scales of $1\,\mathrm{mas}$. 

It is thus possible to relate, at least partially, the changes in the Doppler factor between $r=0.1\,\mathrm{mas}$ and $r=4\,\mathrm{mas}$ (regions C and D) with changes in the viewing angle, but it is not so easy to interpret standing components along the jet within the same frame. From the visible changes in the jet direction in the radio maps, it seems that the core and the region including components B1, B2, and B3 would be viewed at short angles than the region between D1 and D2  ($2\,\mathrm{mas}<r<4\,\mathrm{mas}$) and the region beyond B3 ($r>8\,\mathrm{mas}$). These changes could be due to a helical pattern in the jet with observed wavelength $\lambda_{\rm obs}\simeq 5\,\mathrm{mas}$.
The misalignment among components B3, B2, and B1 could be explained by a small wavelength pattern ($\lambda_{\rm obs}\simeq 1\,\mathrm{mas}$) developing on top of the longer one ($\lambda_{\rm obs}\simeq 5\,\mathrm{mas}$).
A study of the variations in the ridgeline of the jet with time, such as those performed in \citet{Perucho:2012p5917}, could shed some light on possible helical patterns propagating along the jet and on their relation with the reported components.

\subsection{The connection to the 2006 radio flare}
\begin{figure*}[t!]
\centering 
\includegraphics[width=17cm]{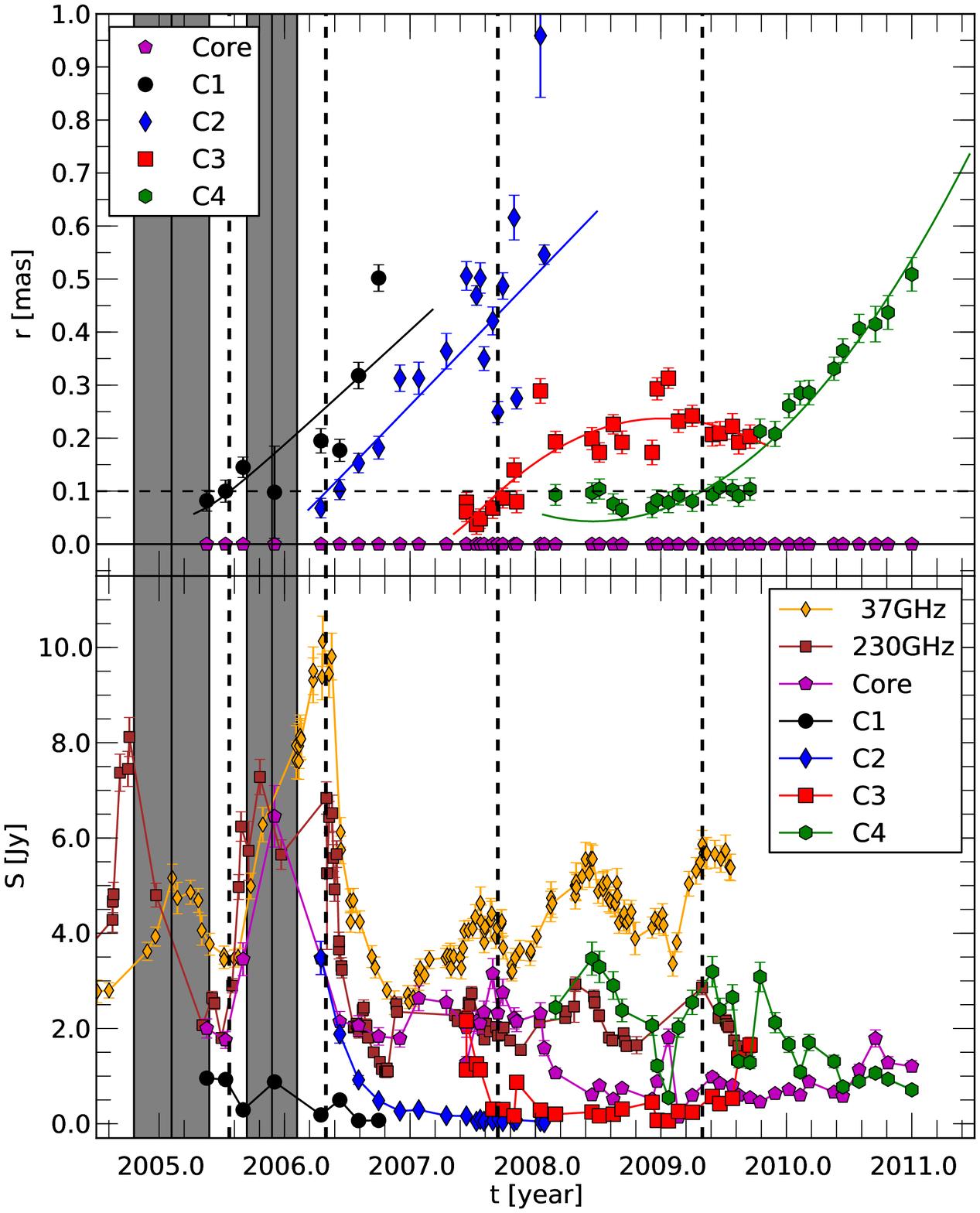} 
\caption{Evolution of the component separation from the core and the flux density for the innermost components of \object{CTA\,102}. Top: Separation from the core and fitted trajectories for C1, C2, C3, and C4. The gray shaded ares correspond to the ejection epochs for C1 and C2 and the dashed line to the crossing of a possible stationary feature at $r\sim0.1,\mathrm{mas}$ away from the core (dashed horizontal line). Bottom: Flux density evolution of the components including the core and single-dish flux density measurements at $37\,\mathrm{GHz}$ and $230\,\mathrm{GHz}$. For details see text.} 
\label{lcvlbi} 
\end{figure*}

As mentioned in Sect.~\ref{cregion} we detect several newly ejected features within the core region (see Figs.~\ref{Cvec} and \ref{Csep}). Furthermore, we proposed that the feature at a distance of $r\sim0.1\,\mathrm{mas}$ from the core is stationary. In Paper I we suggested the interaction between a traveling and a stationary  recollimation shock as a possible process behind the 2006 radio flare and the double hump structure in the turnover frequency$-$turnover flux density plane at radio frequencies in \object{CTA\,102}. In the top panel of Fig. \ref{lcvlbi} we show the trajectories of the components C1, C2, C3, and C4. The bottom panel shows the flux density evolution of C1, C2, C3, C4, and core, together with single-dish light curves at $37\,\mathrm{GHz}$ and $230\,\mathrm{GHz}$,  and the evolution of the flux density of the innermost components of \object{CTA\,102}. 

If we assume that there is a standing shock at a distance of $r\sim0.1\,\mathrm{mas}$ from the core,
and if we associate the ejected components with traveling shock waves, there will be shock-shock interaction at a given time. This crossing time, $t_\mathrm{cross}$, is indicated 
in
Fig. \ref{lcvlbi}. 
The derived ejection time for C2, $t_\mathrm{ej}=2005.9\pm0.2$ fits nicely to the first peak in the $230\,\mathrm{GHz}$ light curve and to the peak in the core flux density (see bottom panel in Fig. \ref{lcvlbi}). The value obtained for the crossing time between C2 and the proposed standing shock, $t_\mathrm{cross,C2}\sim2006.3$ is in good agreement with the peak in the $37\,\mathrm{GHz}$ and the second peak in $230\,\mathrm{GHz}$ single-dish light curve.

The ejection time for the components C3 and C4 are difficult to obtain owing to their highly non-radial trajectory.  To calculate the ejection times of these components (assuming a blending between a traveling and standing shock at $r\sim0.1\,\mathrm{mas}$), we could split the trajectory into moving and standing parts. Since this separation could lead to a strong variation in the calculated ejection time depending on the position of the separation, we only indicate the crossing time for C3 and C4 in Fig.\ref{lcvlbi}. However, the calculated crossing times $t_\mathrm{cross,C3}\sim2007.7$ and $t_\mathrm{cross,C4}\sim2009.3$ correspond to local peaks in the $37\,\mathrm{GHz}$ single dish light curves (see dashed lines in Fig. \ref{lcvlbi}). 

The peak in the $37\,\mathrm{GHz}$ single-dish light curve around 2008.6 is not connected to the ejection of a new component and seems to be connected to C4. However, this region could be highly affected by blending and/or misidentification of the components. Furthermore, if we assume that there is a standing feature at $r\sim0.1\,\mathrm{mas}$, the observed flux density behavior could reflect the highly nonlinear interaction of a the shock-shock interaction: The strong traveling shock wave associated with C2 drags the standing feature downstream. After a given time this standing shock wave is re-established at its initial position (traveling inward). During this time a new traveling shock wave is ejected from the core and collides with the inward traveling standing shock wave. This proposed scenario could not be tested within our observations due to the limited resolution within this region and sparse time sampling, and it would require high-resolution relativistic hydrodynamic simulations. 

As mentioned before, standing features arise naturally in overpressured jets. Therefore, based on the results presented above, the overpressured jet model is slightly favored, at least in the core region, as a possible jet configuration. 

\section{Summary}
\label{sum}
In this paper we present the kinematic analysis of the multifrequency VLBA observations of the blazar \object{CTA\,102} for eight observing epochs during its 2006 radio flare. We combined our data with the long-term monitoring at $15\,\mathrm{GHz}$ and at $43\,\mathrm{GHz}$ provided by the MOJAVE and the Boston Blazar Monitoring survey, respectively. The modeling of the source at different frequencies with several Gaussian features shows that the source consists of several apparently stationary components at $4\,\mathrm{mas}<r<8\,\mathrm{mas}$ and a mixture of stationary and traveling ones both in the core region and extended regions of the jet. Throughout our entire data set, we could clearly cross-identify seven components and several newly ejected components in the core region. 

Based on the observed evolution of the cross-identified component trajectories and on the brightness temperature gradients obtained, we divided the jet into four different regions. The gradients in these regions were used to estimate the possible jet configuration, e.g., orientation of the magnetic field. For our modeling we calculated the gradients for the evolution of the brightness temperature according to a generalization of the shock-in-jet model, which allows for nonconical geometry \citep{Turler:2000p1}. We applied three different configurations for the magnetic field, $b=1$, $b=b_\mathrm{eq}=(s+2)/3$, and $b=2$ ($B\propto R^{-b}$), to the observed $T_b$ gradients and derived the evolution of the Doppler factor index $d$ ($\delta\propto R^{-d}$) as a function of the spectral slope, $s$, and the jet geometry. {The result of this parameter-space study shows that i) in region C there is an increase in the Doppler factor that is independent of the magnetic field configuration, ii) the obtained decrease in $\delta$ between region C and D favors a jet with a magnetic field toroidal magnetic field and/or in equipartition, iii) the decrease in the Doppler factor in this region is due to a deceleration of the fluid and a change in the viewing angle, iv) for the recollimation region B ($\rho<0$) the Doppler factor decreases for all magnetic field geometries studied and the smallest decrease is obtained for a toroidal field, and v) farther downstream the Doppler factor rises again, with the steepnes of the acceleration depending on the geometry of the magnetic field.} 

The very detailed study of the core region at $43\,\mathrm{GHz}$ revealed four newly ejected traveling components and one possible stationary feature at $r\sim0.1\mathrm{mas}$ from the core. We could connect one of the components labeled as C2 with the April 2006 radio via a linear back-interpolation of its observed trajectory. This feature was ejected around 2005.9 from the core with an apparent speed of $(13\pm2)$\,\textit{c}. The ejection time of C2 fits nicely to the first peak in the $230\,\mathrm{GHz}$ single dish light curve and the crossing time computed for the interaction between C2, and the stationary feature corresponds to the second peak at $230\,\mathrm{GHz}$ and to the observed global maximum in the $37\,\mathrm{GHz}$ single-dish light curve.

The results were interpreted within the framework of a) an overpressured jet, b) a helical jet, and c) a combination of both. Within these models the obtained evolution in the Doppler factors and brightness temperature are caused by a) an acceleration or deceleration of the plasma generated by collimation and expansion of the jet, i.e., the pressure mismatch between the jet and the ambient medium, by b) a variation in the viewing angle and the connected Doppler boosting or deboosting and by c) nonaxial recollimation shocks that create a transversal pressure gradient and lead to a bend in the jet. Based on the analysis of the kinematics and the brightness temperature the third option was favored.

To summarize, our results on the component trajectories and the brightness temperature evolution suggest a shock-shock interaction in an overpressured jet as the driving mechanism behind the 2006 radio flare (see Paper I). However, a detailed analysis of the spectral evolution along the jet and, especially, during the 2006 radio flare could lead to further evidence of a shock-shock interaction. Our hypothesis will then be confirmed or rejected in the last part of the work described in this series of papers.

\begin{acknowledgements}
We acknowledge Dr. R. W. Porcas for careful reading and for useful comments and suggestions to the manuscript.
C.M.F. was supported for this research through a stipend from the International Max Planck Research School (IMPRS) for Astronomy and Astrophysics at the Universities of Bonn and Cologne. C.M.F. thanks the Departament d'Astronomia i Astrof\'\i sica of the Universitat de Val\`encia for its hospitality.
Part of this work was supported by the COST Action MP0905 Black Hole in a 
violent Universe through the short-term scientific missions (STSM) STSM-MP0905-140211-004292, STSM-MP0905-300711-008633, and STSM-MP0905-110711-008634.\\
E.R. acknowledges partial support from MINECO grant AYA2009-13036-C02-02 and Generalitat Valenciana's grant PROMETEO/2009/104.
M.P. acknowledges financial support from MINECO grants 
AYA2010-21322-C03-01, AYA2010-21097-C03-01, and CONSOLIDER2007-00050.
P. M. acknowledges the support from the European Research Council (grant CAMAP-259276).
Y.Y.K is partly supported by the Russian Foundation for Basic Research (project 11-02-00368), the basic research program ``Active processes in galactic and extragalactic objects'' of the Physical Sciences Division of the Russian Academy of Sciences and the Dynasty Foundation.
This work is based on observations with the VLBA, 
which is operated by the NRAO, a facility of the NSF under cooperative agreement by Associated Universities Inc.
This research made use of data from the MOJAVE database that is maintained by the MOJAVE team \citep{Lister:2009p90}. This study makes use of 43 GHz VLBA data from the Boston University gamma-ray blazar monitoring program (http://www.bu.edu/blazars/VLBAproject.html), funded by NASA through the Fermi Guest Investigator Program.
\end{acknowledgements}

\bibliographystyle{aa} 
\bibliography{biblio2}

\begin{thebibliography}{27}
\expandafter\ifx\csname natexlab\endcsname\relax\def\natexlab#1{#1}\fi

\bibitem[{Aloy {et~al.}(2003)Aloy, Mart{\'\i}, G{\'o}mez, Agudo, M{\"u}ller, \&
  Ib{\'a}{\~n}ez}]{Aloy:2003p3889}
Aloy, M.-{\'A}., Mart{\'\i}, J.-M., G{\'o}mez, J.-L., {et~al.} 2003, \apj, 585,
  L109

\bibitem[{Bj{\"o}rnsson \& Aslaksen(2000)}]{Bjornsson:2000p32}
Bj{\"o}rnsson, C.-I. \& Aslaksen, T. 2000, \apj, 533, 787

\bibitem[{Fromm {et~al.}(2011)Fromm, Perucho, Ros, Savolainen, Lobanov, Zensus,
  Aller, Aller, Gurwell, \& L{\"a}hteenm{\"a}ki}]{Fromm:2011p4088}
Fromm, C.~M., Perucho, M., Ros, E., {et~al.} 2011, \aap, 531, 95, Paper I

\bibitem[{G{\'o}mez {et~al.}(1997)G{\'o}mez, Mart{\'\i}, Marscher,
  Ib{\'a}{\~n}ez, \& Alberdi}]{Gomez:1997p649}
G{\'o}mez, J.~L., Mart{\'\i}, J.~M., Marscher, A.~P., Ib{\'a}{\~n}ez, J.~M., \&
  Alberdi, A. 1997, \apjl, 482, L33

\bibitem[{Hardee(2000)}]{Hardee:2000p5817}
Hardee, P.~E. 2000, \apj, 533, 176

\bibitem[{Hardee \& Hughes(2003)}]{Hardee:2003p5821}
Hardee, P.~E. \& Hughes, P.~A. 2003, \apj, 583, 116

\bibitem[{Hardee {et~al.}(2005)Hardee, Walker, \& G{\'o}mez}]{Hardee:2005p5826}
Hardee, P.~E., Walker, R.~C., \& G{\'o}mez, J.~L. 2005, \apj, 620, 646

\bibitem[{Jorstad {et~al.}(2005)Jorstad, Marscher, Lister, Stirling, Cawthorne,
  Gear, G{\'o}mez, Stevens, Smith, Forster, \& Robson}]{Jorstad:2005p4121}
Jorstad, S.~G., Marscher, A.~P., Lister, M.~L., {et~al.} 2005, \apj, 130, 1418

\bibitem[{Kadler {et~al.}(2004)Kadler, Ros, Lobanov, Falcke, \&
  Zensus}]{Kadler:2004p2884}
Kadler, M., Ros, E., Lobanov, A.~P., Falcke, H., \& Zensus, J.~A. 2004, \aap,
  426, 481

\bibitem[{Kellermann {et~al.}(1998)Kellermann, Vermeulen, Zensus, \&
  Cohen}]{Kellermann:1998p5022}
Kellermann, K.~I., Vermeulen, R.~C., Zensus, J.~A., \& Cohen, M.~H. 1998, \aj,
  115, 1295

\bibitem[{Lister {et~al.}(2009{\natexlab{a}})Lister, Aller, Aller, Cohen,
  Homan, Kadler, Kellermann, Kovalev, Ros, Savolainen, Zensus, \&
  Vermeulen}]{Lister:2009p90}
Lister, M.~L., Aller, H.~D., Aller, M.~F., {et~al.} 2009{\natexlab{a}}, \apj,
  137, 3718

\bibitem[{Lister {et~al.}(2009{\natexlab{b}})Lister, Cohen, Homan, Kadler,
  Kellermann, Kovalev, Ros, Savolainen, \& Zensus}]{Lister:2009p8}
Lister, M.~L., Cohen, M.~H., Homan, D.~C., {et~al.} 2009{\natexlab{b}}, \apj,
  138, 1874

\bibitem[{Lister \& Homan(2005)}]{Lister:2005p5497}
Lister, M.~L. \& Homan, D.~C. 2005, \apj, 130, 1389

\bibitem[{Lobanov \& Zensus(1999)}]{Lobanov:1999p2299}
Lobanov, A.~P. \& Zensus, J.~A. 1999, \apj, 521, 509

\bibitem[{Marscher \& Gear(1985)}]{Marscher:1985p50}
Marscher, A.~P. \& Gear, W.~K. 1985, \apj, 298, 114

\bibitem[{Mimica {et~al.}(2009)Mimica, Aloy, Agudo, Mart{\'\i}, G{\'o}mez, \&
  Miralles}]{Mimica:2009p42}
Mimica, P., Aloy, M.-A., Agudo, I., {et~al.} 2009, \apj, 696, 1142

\bibitem[{Perucho {et~al.}(2008)Perucho, Agudo, G{\'o}mez, Kadler, Ros, \&
  Kovalev}]{Perucho:2008p3885}
Perucho, M., Agudo, I., G{\'o}mez, J.~L., {et~al.} 2008, \aap, 489, L29

\bibitem[{Perucho {et~al.}(2004{\natexlab{a}})Perucho, Hanasz, Mart{\'\i}, \&
  Sol}]{Perucho:2004p3871}
Perucho, M., Hanasz, M., Mart{\'\i}, J.~M., \& Sol, H. 2004{\natexlab{a}},
  \aap, 427, 415

\bibitem[{Perucho {et~al.}(2012)Perucho, Kovalev, Lobanov, Hardee, \&
  Agudo}]{Perucho:2012p5917}
Perucho, M., Kovalev, Y.~Y., Lobanov, A.~P., Hardee, P.~E., \& Agudo, I. 2012,
  \apj, 749, 55

\bibitem[{Perucho {et~al.}(2004{\natexlab{b}})Perucho, Mart{\'\i}, \&
  Hanasz}]{Perucho:2004p3886}
Perucho, M., Mart{\'\i}, J.~M., \& Hanasz, M. 2004{\natexlab{b}}, \aap, 427,
  431

\bibitem[{Piner {et~al.}(2007)Piner, Mahmud, Fey, \&
  Gospodinova}]{Piner:2007p5743}
Piner, B.~G., Mahmud, M., Fey, A.~L., \& Gospodinova, K. 2007, \apj, 133, 2357

\bibitem[{Savolainen {et~al.}(2002)Savolainen, Wiik, Valtaoja, Jorstad, \&
  Marscher}]{Savolainen:2002p2410}
Savolainen, T., Wiik, K., Valtaoja, E., Jorstad, S.~G., \& Marscher, A.~P.
  2002, \aap, 394, 851

\bibitem[{Savolainen {et~al.}(2008)Savolainen, Wiik, Valtaoja, \&
  Tornikoski}]{Savolainen:2008p2958}
Savolainen, T., Wiik, K., Valtaoja, E., \& Tornikoski, M. 2008, Extragalactic
  Jets: Theory and Observation from Radio to Gamma Ray ASP Conference Series,
  386, 451

\bibitem[{Schinzel {et~al.}(2012)Schinzel, Lobanov, Taylor, Jorstad, Marscher,
  \& Zensus}]{Schinzel:2012p5740}
Schinzel, F.~K., Lobanov, A.~P., Taylor, G.~B., {et~al.} 2012, \aap, 537, 70

\bibitem[{Shepherd(1997)}]{Shepherd:1997p2298}
Shepherd, M.~C. 1997, Astronomical Data Analysis Software and Systems VI, 125,
  77

\bibitem[{T{\"u}rler {et~al.}(2000)T{\"u}rler, Courvoisier, \&
  Paltani}]{Turler:2000p1}
T{\"u}rler, M., Courvoisier, T. J.-L., \& Paltani, S. 2000, \aap, 361, 850

\bibitem[{Zensus {et~al.}(2002)Zensus, Ros, Kellermann, Cohen, Vermeulen, \&
  Kadler}]{Zensus:2002p2202}
Zensus, J.~A., Ros, E., Kellermann, K.~I., {et~al.} 2002, \aj, 124, 662

\end{thebibliography}
\section*{Appendix}
Here we show results of modeling for the individual frequencies $(5\,\mathrm{GHz}-43\,\mathrm{GHz})$ during our multifrequency campaign (May 2005 until April 2007). For each frequency we present contour plots of the source, including the position of the fitted circular Gaussian components and its structural evolution during our observations. Furthermore, we show the temporal evolution of the component parameters, e.g., position, flux density and size, and the results of a polynomial fit to the cross identified features are summarized in Tables \ref{5restab}  - \ref{43restab} . The component parameters for all frequencies and epochs can be found at the end of the Appendix  in Tables \ref{modelfit1} and \ref{modelfit2}. 

\subsection*{$5\,\mathrm{GHz}$ VLBI observations of \object{CTA\,102}}

\begin{table*}[b!]
\caption{Results of the kinematic analysis for the fitted components at $5\,\mathrm{GHz}$}  
\label{5restab}
\tiny
\begin{tabular}{@{}c D{,}{\pm}{-1} D{,}{\pm}{-1} c c c D{,}{\pm}{-1} D{,}{\pm}{-1} c  c c c@{}}
\toprule
\hline
Comp	&	\multicolumn{1}{c}{$\mu$} &	\multicolumn{1}{c}{$\beta_\mathrm{app}$} 
& \multicolumn{1}{c}{$\delta_\mathrm{min}$}
& \multicolumn{1}{c}{$\vartheta_\mathrm{max}$} 
&\multicolumn{1}{c}{$\Gamma_\mathrm{min}$}
&\multicolumn{1}{c}{$t_\mathrm{ej}$}
&\multicolumn{1}{c}{$<r>$}
& \multicolumn{1}{c}{$t_\mathrm{min}$}
& \multicolumn{1}{c}{$t_\mathrm{max}$}\\

& \multicolumn{1}{c}{[mas/yr]}
& \multicolumn{1}{c}{[c]}
& \multicolumn{1}{c}{[1]}
& \multicolumn{1}{c}{ [$^\circ$]}
& \multicolumn{1}{c}{[1]}
& \multicolumn{1}{c}{ [yr]}
& \multicolumn{1}{c}{[mas]}
& \multicolumn{1}{c}{[yr]}
& \multicolumn{1}{c}{[yr]}
& \multicolumn{1}{c}{acceleration}
& \multicolumn{1}{c}{classification}\\
\midrule
A1 & 0.2,0.1 & 8,4 & 8 & 7 & 8 & -- & 16.9,0.2 & 2005.39 & 2007.32 &no& non-radial outward\\
A2 & 0.20,0.05& 11,2 & 11& 5& 11 & 1947,15 & 11.6,0.2 & 2005.39 & 2007.32&no& radial outward\\
B1 & 0.07,0.07 & 4,4 & 4 & 16 & 4 & -- & 8,0.3 & 2005.39 & 2007.32 &no& non-radial outward\\
B2 & 0.05,0.04 & 2,2 & 2 & 21 & 2 & -- & 6.2,0.1 & 2005.39 & 2007.32 &no& non-radial outward \\
B3 & 0.07,0.06 & 4,3 & 4 & 14 & 4 & -- & 4.3,0.1 & 2005.39 & 2007.32 &no &non-radial outward\\
D2 & 0.14,0.01 & 8,1 & 8 & 8 & 8 & 1997.3,0.7 & 1.3,0.1 & 2005.39 & 2007.32 &no& radial outward\\
\bottomrule
\hline
\end{tabular}
\end{table*}

\begin{figure}[h!]
\resizebox{\hsize}{!}{\includegraphics{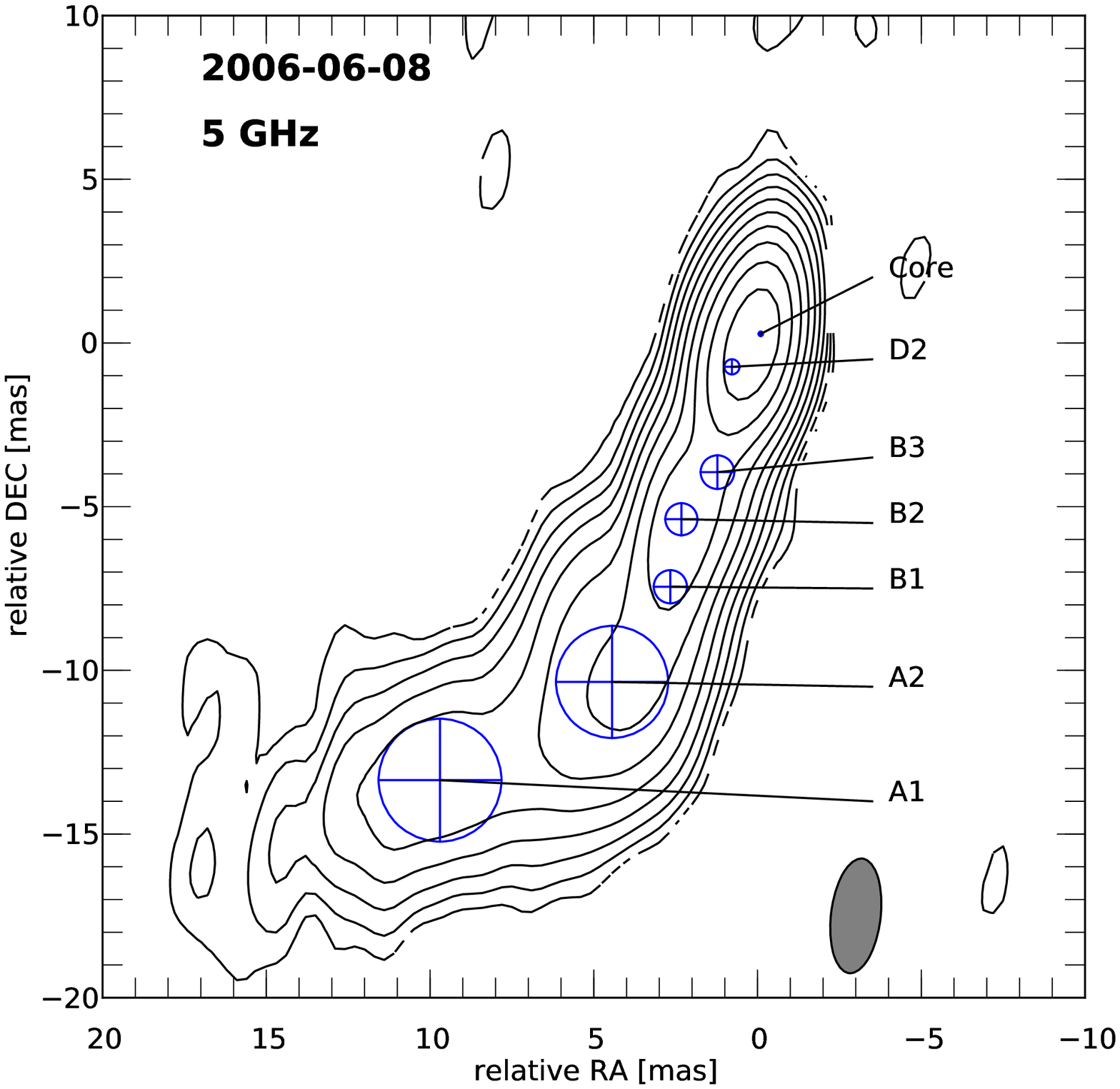}} 
\caption{$5\,\mathrm{GHz}$ uniformly weighted VLBA CLEAN image of \object{CTA\,102} observed on 8 June 2006 with fitted circular Gaussian components overlaid. The map peak flux density was $2.0\,\mathrm{Jy/beam}$, where the convolving beam was $3.5\times1.5\,\mathrm{mas}$ at P.A. $-5.3$. The lowest contour is plotted at $10\times$ the off-source $\mathrm{rms}$ and increases in steps of 2.} 
\label{5cont} 
\end{figure}

\begin{figure}[h!]
\resizebox{\hsize}{!}{\includegraphics{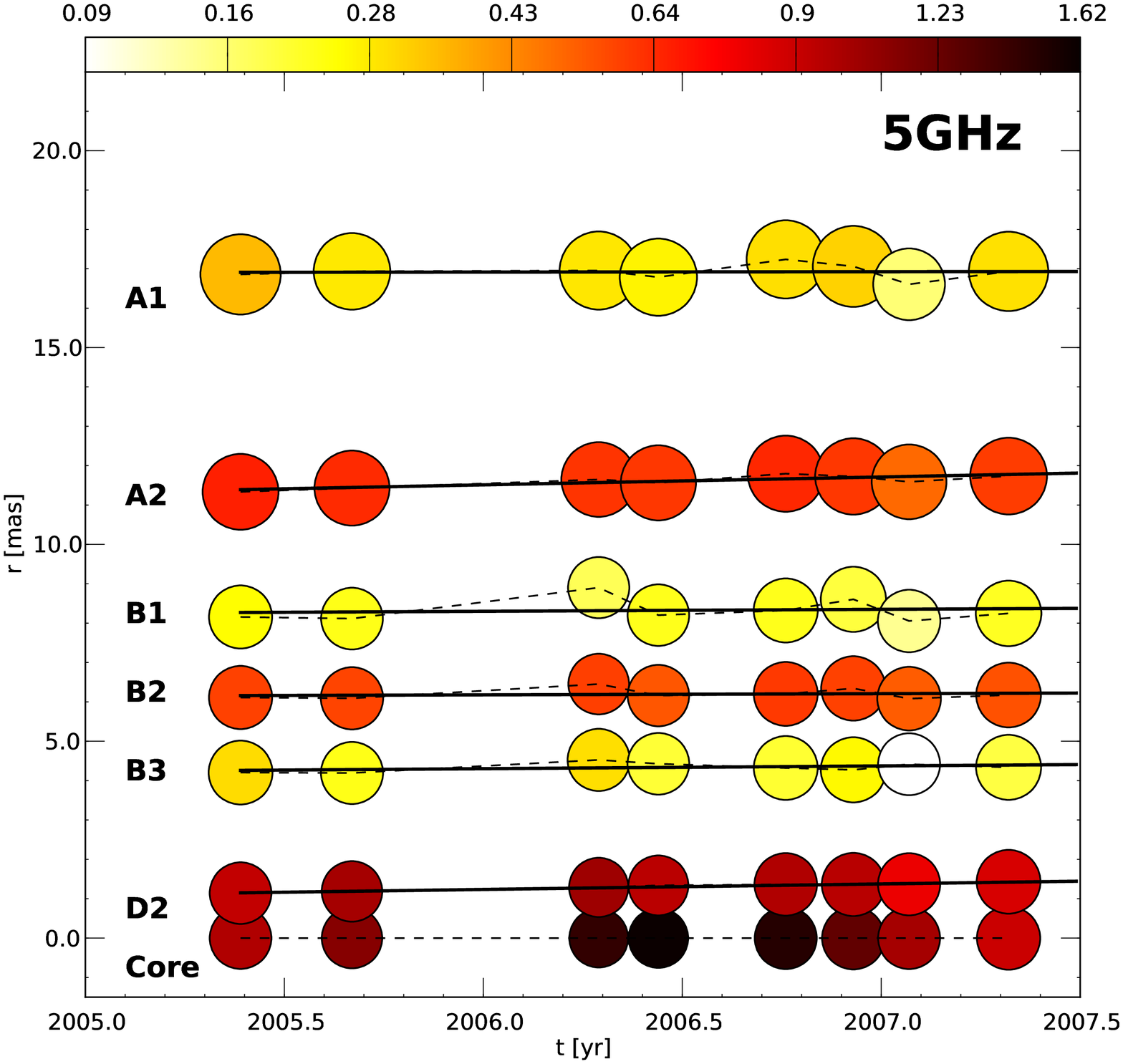}} 
\caption{Temporal evolution of the separation from the core for the $5\,\mathrm{GHz}$ components. The color scale corresponds to the flux density and the size of the circles to the relative size (FWHM) of the components. The solid black lines correspond to a linear fits of the component trajectory.} 
\label{5kin} 
\end{figure}

\clearpage
\begin{landscape}
\begin{figure}
\resizebox{\hsize}{!}{\includegraphics{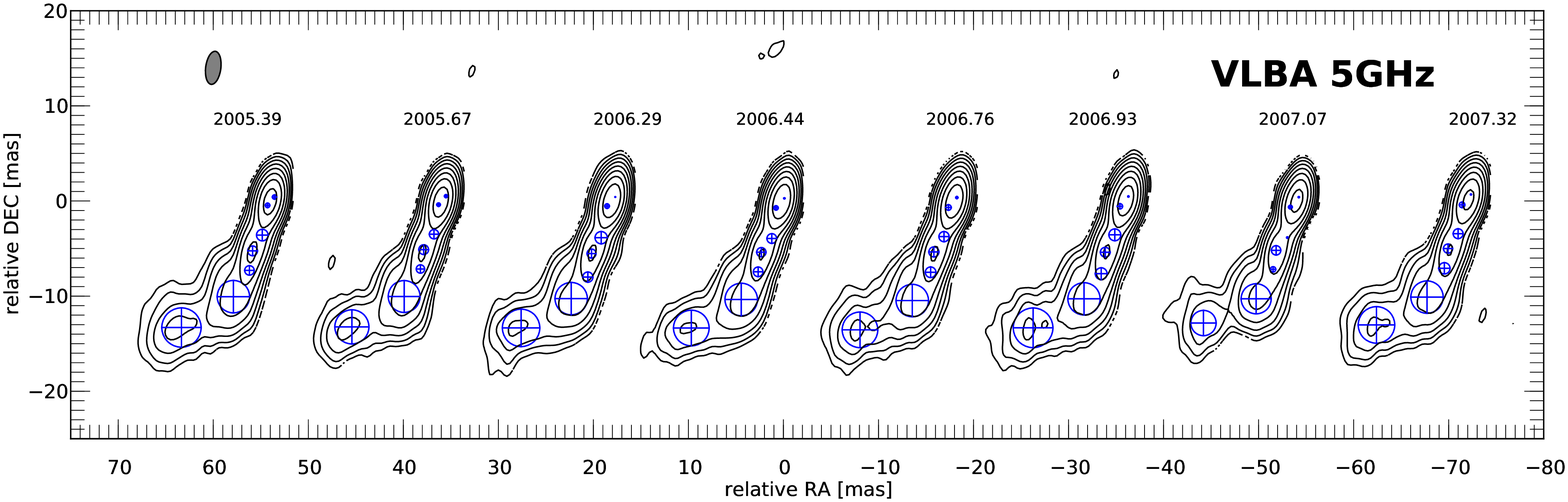}} 
\caption{$5\,\mathrm{GHz}$ uniform weighted VLBA images of \object{CTA\,102} with fitted circular Gaussian components. For better comparison all maps are convolved with a common beam of $3.5\times1.6\,\mathrm{mas}$ at P.A. $-7.4$ and the epoch of the observations is  indicated above each contour map. The lowest contour levels is plotted $10\times$ of the maximum off-source $\mathrm{rms}$ and increases in steps of 2.} 
\label{contall5}
\end{figure}
\end{landscape}


\subsection*{$8\,\mathrm{GHz}$ VLBI observations of \object{CTA\,102}}

\begin{table*}
\caption{Results of the kinematic analysis for the fitted components at $8\,\mathrm{GHz}$}  
\label{8restab}
\tiny
\begin{tabular}{@{}c D{,}{\pm}{-1} D{,}{\pm}{-1} c c c D{,}{\pm}{-1} D{,}{\pm}{-1} c  c c c@{}}
\toprule
\hline
Comp	&	\multicolumn{1}{c}{$\mu$} &	\multicolumn{1}{c}{$\beta_\mathrm{app}$} 
& \multicolumn{1}{c}{$\delta_\mathrm{min}$}
& \multicolumn{1}{c}{$\vartheta_\mathrm{max}$} 
&\multicolumn{1}{c}{$\Gamma_\mathrm{min}$}
&\multicolumn{1}{c}{$t_\mathrm{ej}$}
&\multicolumn{1}{c}{$<r>$}
& \multicolumn{1}{c}{$t_\mathrm{min}$}
& \multicolumn{1}{c}{$t_\mathrm{max}$}\\

& \multicolumn{1}{c}{[mas/yr]}
& \multicolumn{1}{c}{[c]}
& \multicolumn{1}{c}{[1]}
& \multicolumn{1}{c}{ [$^\circ$]}
& \multicolumn{1}{c}{[1]}
& \multicolumn{1}{c}{ [yr]}
& \multicolumn{1}{c}{[mas]}
& \multicolumn{1}{c}{[yr]}
& \multicolumn{1}{c}{[yr]}
& \multicolumn{1}{c}{acceleration}
& \multicolumn{1}{c}{classification}\\
\midrule
A1 & 0.30,0.20 & 15,8 & 15 & 4 & 15 & -- & 17.2,0.3 & 2005.39 & 2007.32 &no & non-radial outward\\
A2 & 0.10,0.06 & 6,3 & 6& 10 & 6 & -- & 11.5,0.1 & 2005.39 & 2007.32 &no & radial outward \\
B1 & 0.08,0.08 & 4,4 & 4 & 13 & 4 & -- & 7.9,0.1 & 2005.39 & 2007.32 &no & radial inward \\
B2 & 0.04,0.03 & 2,2 & 2 & 26 & 2 & -- & 6.2,0.1 & 2005.39 & 2007.32 &no& radial inward \\
B3 & 0.14,0.07 & 7,4 & 7 & 8 & 7 & 1970,20 & 5.0,0.2 & 2005.39 & 2007.32 &no& radial outward\\
D1 & 0.8,0.1 & 43,7 & 43 & 1 & 43 & -- & 2.4,0.5 & 2005.39 & 2007.32 &no& non-radial outward \\
D2 & 0.30,0.02 & 16,1 & 16 & 4 & 16 & 2002.3,0.2 & 1.2,0.2 & 2005.39 & 2007.32 &no& radial outward \\
\bottomrule
\hline
\end{tabular}
\end{table*}

\begin{figure}[h!]
\resizebox{\hsize}{!}{\includegraphics{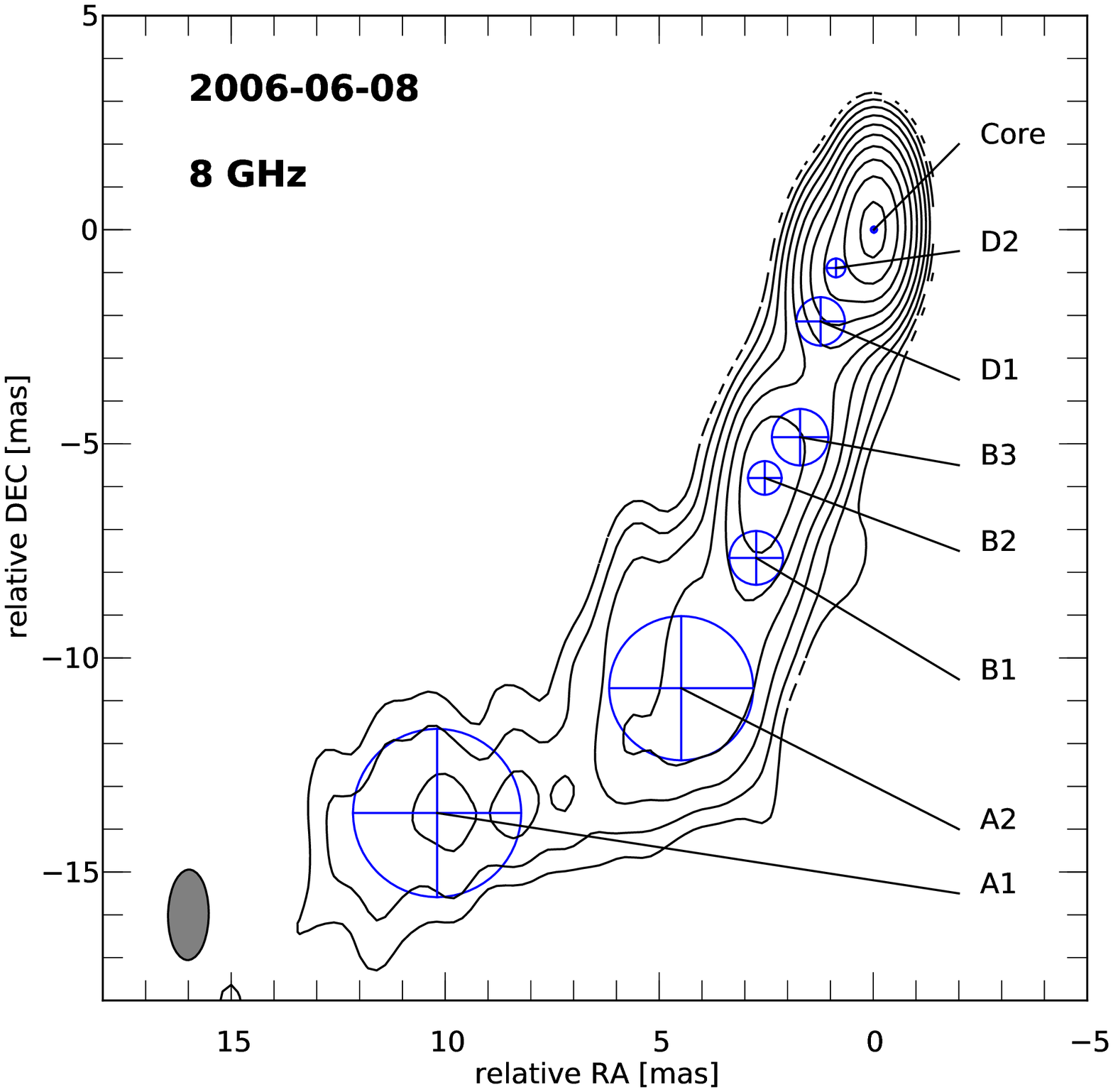}} 
\caption{$8\,\mathrm{GHz}$ uniformly weighted VLBA CLEAN image of \object{CTA\,102} observed on 8 June 2006 with fitted circular Gaussian components overlaid. The map peak flux density was $2.8\,\mathrm{Jy/beam}$, where the convolving beam was $2.1\times0.9\,\mathrm{mas}$ at P.A. $-0.9$. The lowest contour is plotted at $10\times$ the off-source $\mathrm{rms}$ and increases in steps of 2.} 
\label{8cont} 
\end{figure}

\begin{figure}[b!]
\resizebox{\hsize}{!}{\includegraphics{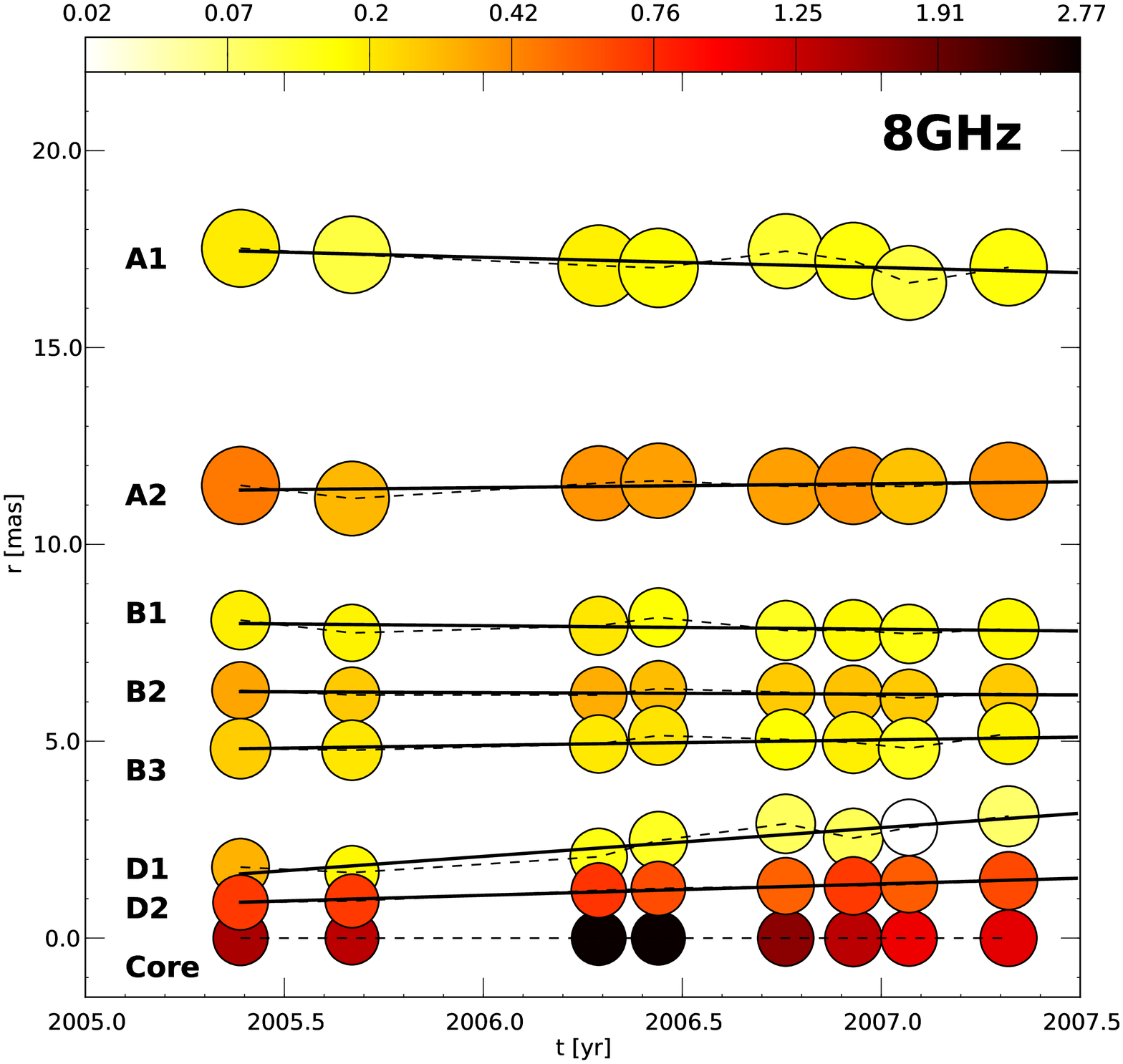}} 
\caption{Temporal evolution of the separation from the core for the $8\,\mathrm{GHz}$ modeled features. The color scale corresponds to the flux density and the size of the circles to the relative size (FWHM) of the components. The solid black lines correspond to a linear fit of the component trajectory.} 
\label{8kin} 
\end{figure}


\begin{landscape}
\begin{figure}
\centering
\resizebox{\hsize}{!}{\includegraphics{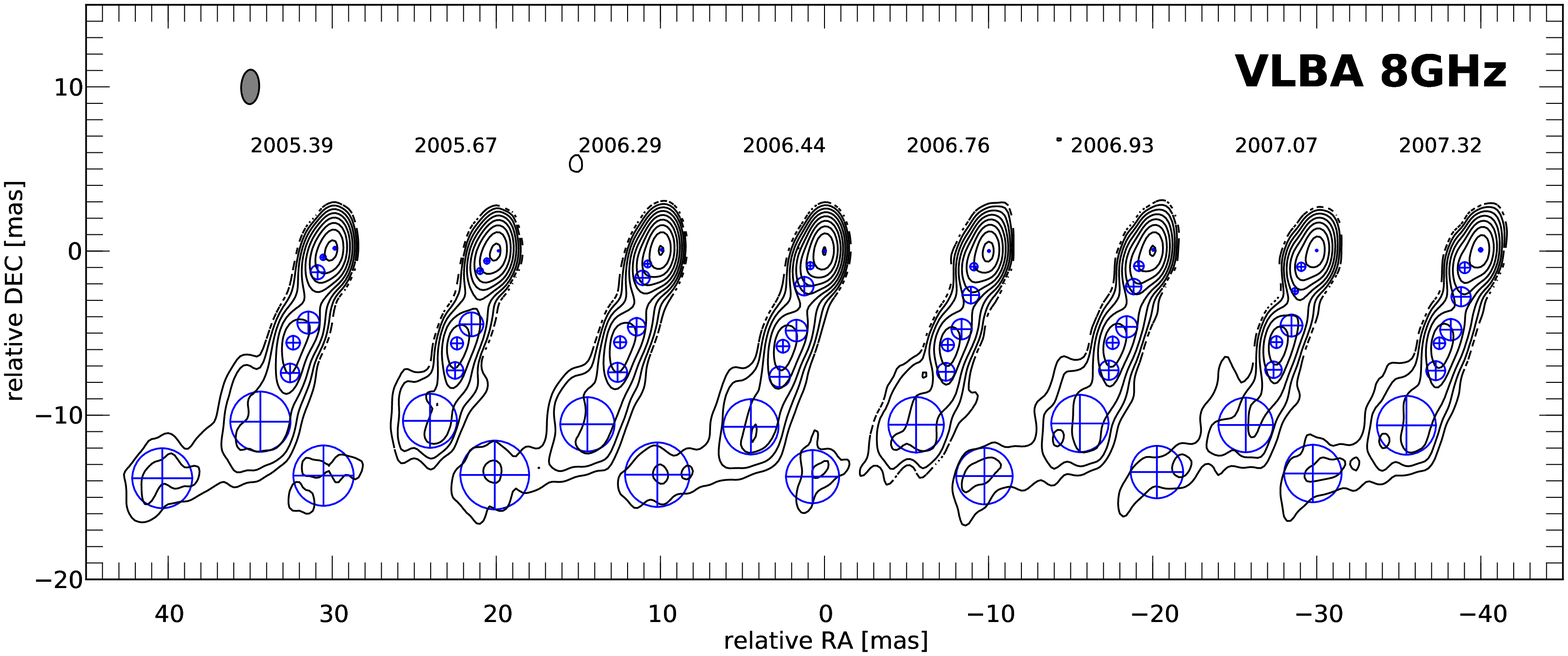}} 
\caption{$8\,\mathrm{GHz}$ uniform weighted VLBA images of \object{CTA\,102} with fitted circular Gaussian components. For better comparison all maps are convolved with a common beam of $2.1\times1.1\,\mathrm{mas}$ at P.A. $-2.6$ and the epoch of the observations is  indicated above each contour map. The lowest contour levels is plotted $10\times$ of the maximum off-source $\mathrm{rms}$ and increases in steps of 2.} 
\label{contall8} 
\end{figure}
\end{landscape}

\subsection*{$15\,\mathrm{GHz}$ VLBI observations of \object{CTA\,102}}

\begin{figure}[h!]
\resizebox{\hsize}{!}{\includegraphics{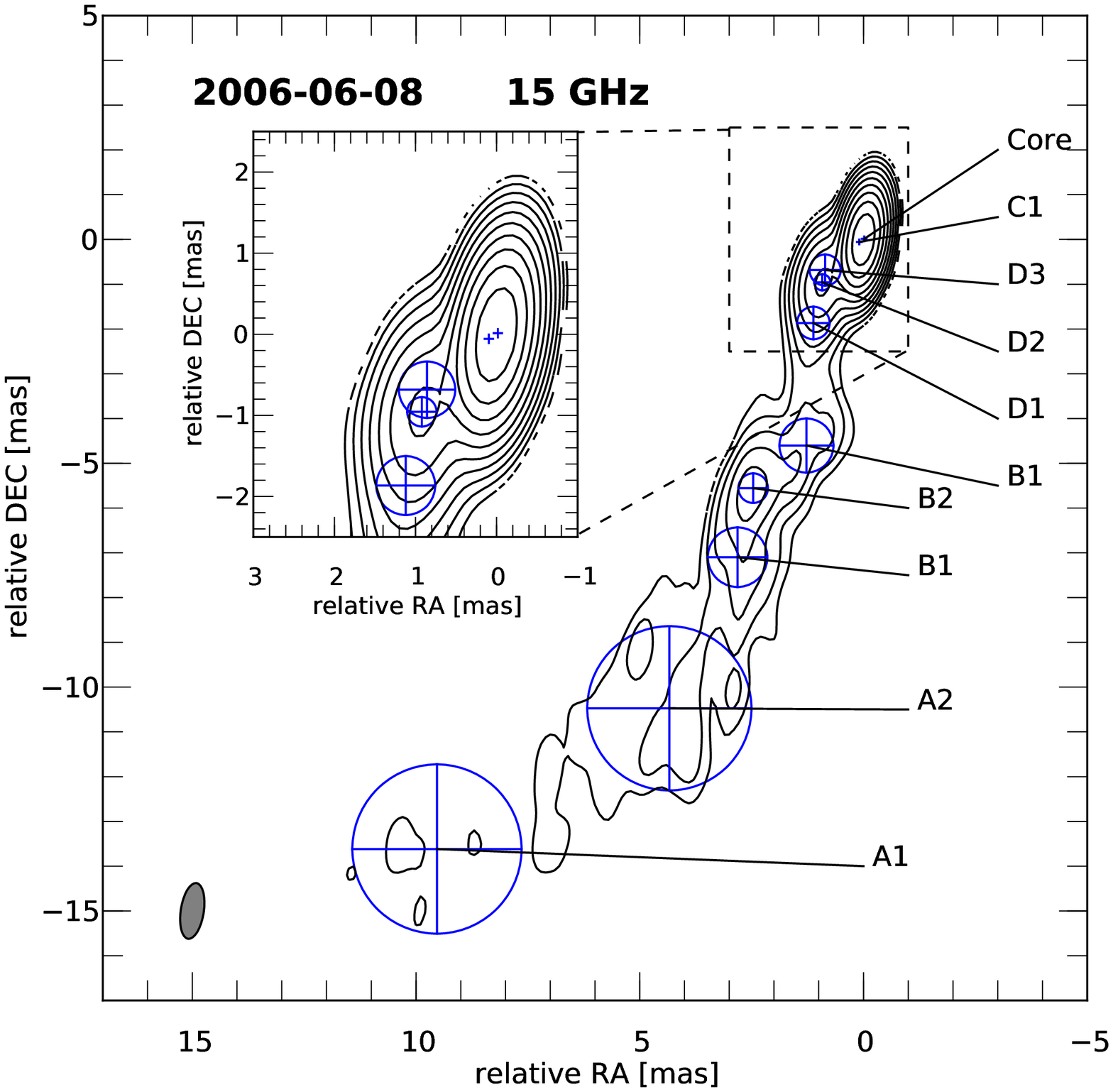}} 
\caption{$15\,\mathrm{GHz}$ uniformly weighted VLBA CLEAN image of \object{CTA\,102} observed on 8 June 2006 with fitted circular Gaussian components overlaid. The map peak flux density was $4.13\,\mathrm{Jy/beam}$, where the convolving beam was $1.3\times0.5\,\mathrm{mas}$ at P.A. $-6.2$. The lowest contour is plotted at $10\times$ the off-source $\mathrm{rms}$ and increases in steps of 2.} 
\label{15cont} 
\end{figure}

\begin{figure}[h!]
\resizebox{\hsize}{!}{\includegraphics{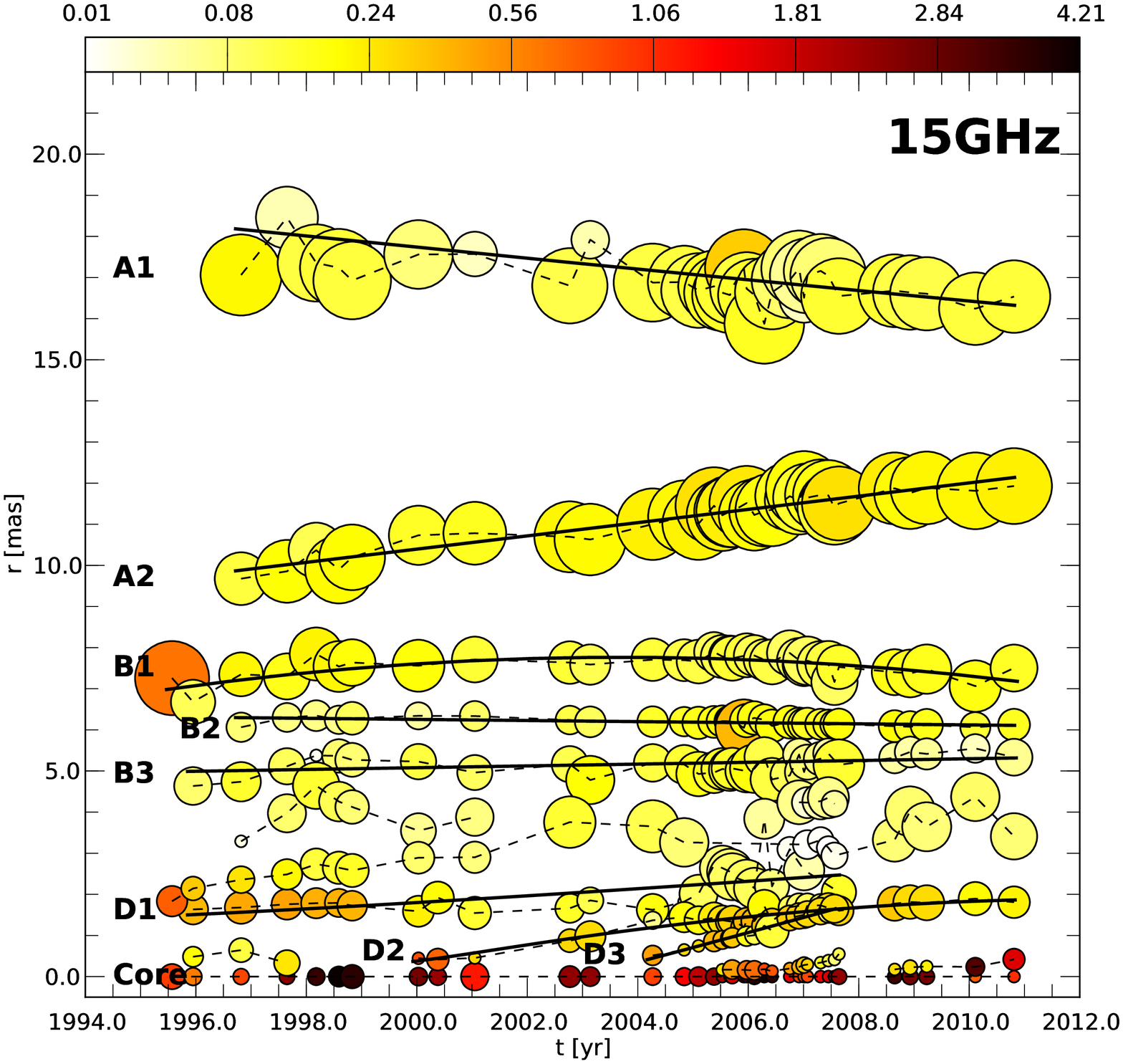}} 
\caption{Temporal evolution of the separation from the core for the $15\,\mathrm{GHz}$ components from \citet{Lister:2009p90} and combined with our $15\,\mathrm{GHz}$ data. The color scale corresponds to the flux density and the size of the circles to the relative size (FWHM) of the components. The solid black lines correspond to a polynomial fit of the component trajectory.} 
\label{15kin} 
\end{figure}

\begin{table*}
\caption{Results of the kinematic analysis for the fitted components at $15\,\mathrm{GHz}$}
\label{15restab} 
\tiny
\begin{tabular}{@{}c D{,}{\pm}{-1} D{,}{\pm}{-1} c c c D{,}{\pm}{-1} D{,}{\pm}{-1} c  c c c@{}}
\toprule
\hline
Comp	&	\multicolumn{1}{c}{$\mu$} &	\multicolumn{1}{c}{$\beta_\mathrm{app}$} 
& \multicolumn{1}{c}{$\delta_\mathrm{min}$}
& \multicolumn{1}{c}{$\vartheta_\mathrm{max}$} 
&\multicolumn{1}{c}{$\Gamma_\mathrm{min}$}
&\multicolumn{1}{c}{$t_\mathrm{ej}$}
&\multicolumn{1}{c}{$<r>$}
& \multicolumn{1}{c}{$t_\mathrm{min}$}
& \multicolumn{1}{c}{$t_\mathrm{max}$}\\

& \multicolumn{1}{c}{[mas/yr]}
& \multicolumn{1}{c}{[c]}
& \multicolumn{1}{c}{[1]}
& \multicolumn{1}{c}{ [$^\circ$]}
& \multicolumn{1}{c}{[1]}
& \multicolumn{1}{c}{ [yr]}
& \multicolumn{1}{c}{[mas]}
& \multicolumn{1}{c}{[yr]}
& \multicolumn{1}{c}{[yr]}
& \multicolumn{1}{c}{acceleration}
& \multicolumn{1}{c}{classification}\\
\midrule
A1 & 0.20,0.03 & 9,2       & 9   & 6   & 9  & -- & 17.0,0.5 & 1996.8 & 2010.8 & no & no-radial inward\\
A2 & 0.20,0.01 & 9,1       & 9   & 7   & 9  & 1945,6 & 11.2,0.2 & 1996.8 & 2010.8& no & radial outward\\
B1 & 0.02,0.01 & 1.0,0.5 & 1   & --   & -- & -- & 7.6,0.3 & 1995.6 & 2010.8& yes & no-radial outward \\
B2 & 0.01,0.01 & 0.6,0.5 & 1   & --   & -- & -- & 6.2,0.1 & 1996.8 & 2010.8 & no & radial inward\\
B3 & 0.03,0.01 & 1.4,0.3 & 2   & --   & -- & -- & 5.1,0.2 & 1995.9 & 2010.8 & no & non-radial outward\\
D1 & 0.10,0.02 & 6,2       & 6   & 10 & 6  & -- & 2.1,0.4 & 1995.9 & 2007.6 & no & non-radial outward\\
D2 & 0.20,0.01 & 8,1       & 8   &  7  & 7  & 1997.1,0.4 & 1.6,0.4 & 2000.0 & 2011.0 & yes & radial outward\\
D3 & 0.37,0.02 & 20,1     & 20 & 3   & 20 & -- & 1.1,0.4 & 2004.3 & 2007.6 & no & non-radial outward\\
\bottomrule
\hline
\end{tabular}
\end{table*} 

\clearpage
\begin{landscape}
\begin{figure}
\centering
\resizebox{\hsize}{!}{\includegraphics{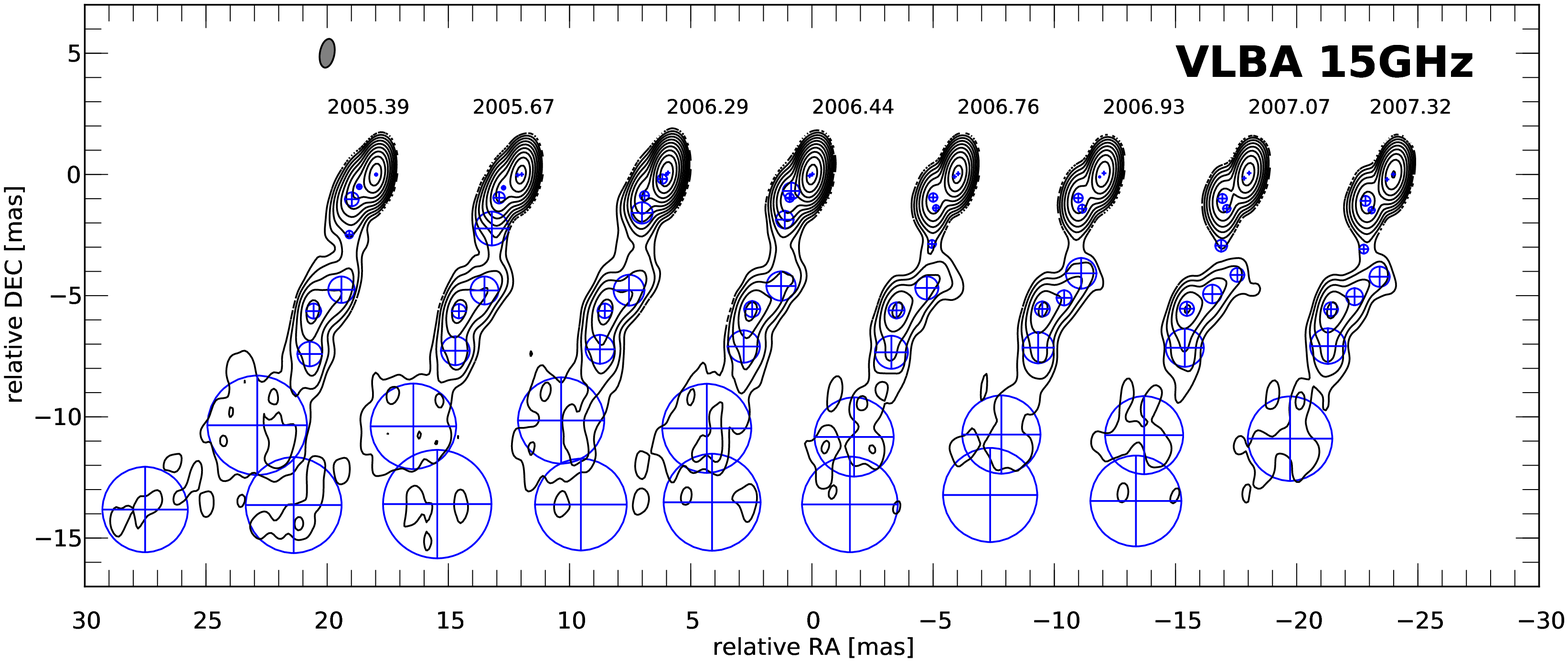}} 
\caption{$15\,\mathrm{GHz}$ uniform weighted VLBA images of \object{CTA\,102} with fitted circular Gaussian components. For better comparison all maps are convolved with a common beam of $1.2\times0.6\,\mathrm{mas}$ at P.A. $-9.9$ and the epoch of the observations is  indicated above each contour map. The lowest contour levels is plotted $10\times$ of the maximum off-source $\mathrm{rms}$ and increases in steps of 2.Notice that the outermost component (A1) for each epoch is plotted as almost overlapping with the jet of the earlier epoch (to its left) in this representation.} 
\label{contall15} 
\end{figure}
\end{landscape}

\subsection*{$22\,\mathrm{GHz}$ VLBI observations of \object{CTA\,102}}

\begin{figure}[h!]
\resizebox{\hsize}{!}{\includegraphics{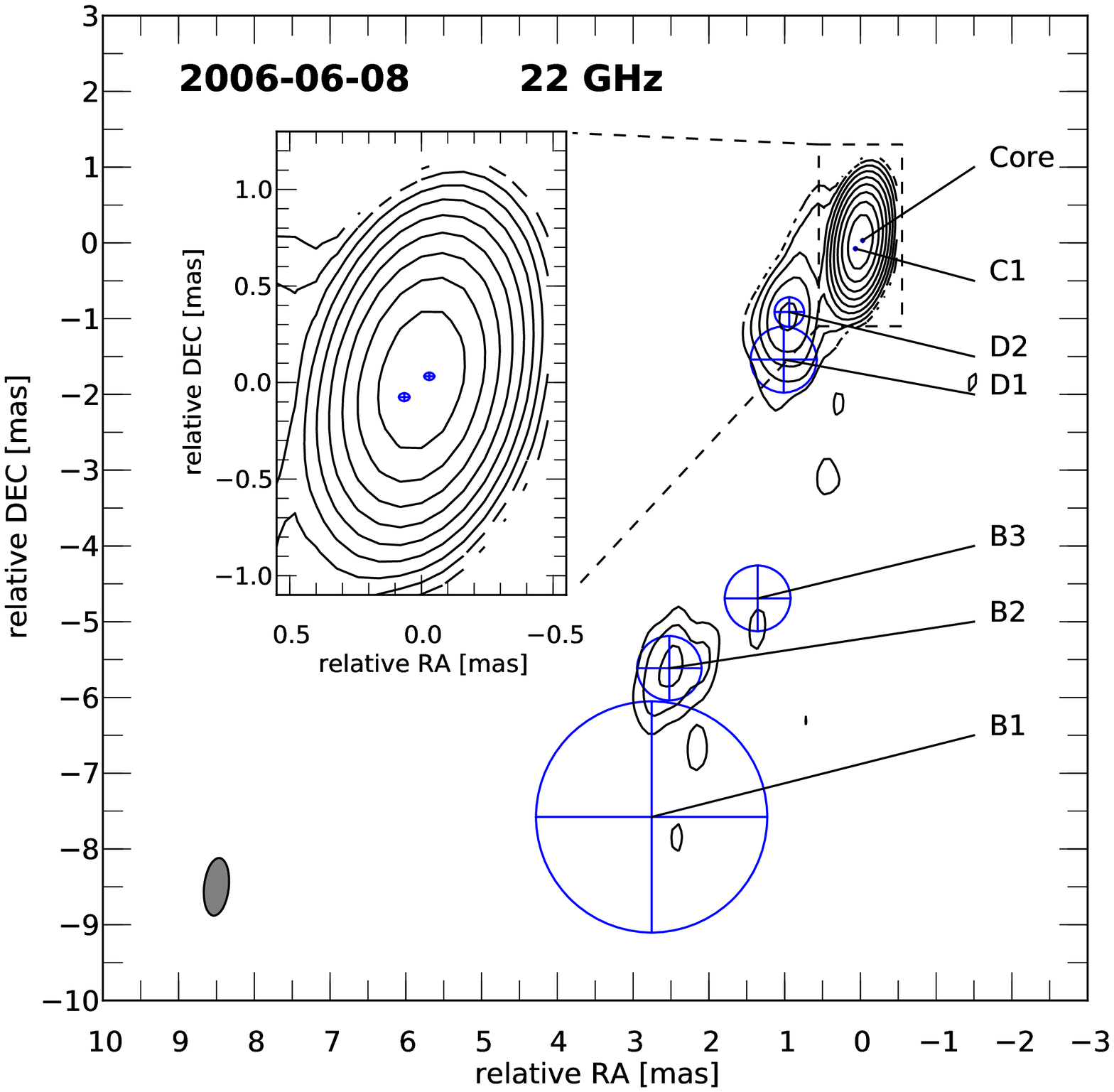}} 
\caption{$22\,\mathrm{GHz}$ uniformly weighted VLBA CLEAN image of \object{CTA\,102} with fitted circular Gaussian components observed 8th of June 2006 overlaid. The map peak flux density was $3.8\,\mathrm{Jy/beam}$, where the convolving beam was $0.76\times0.33\,\mathrm{mas}$ at P.A. $-6.2$. he lowest contour is plotted at $10\times$ the off-source $\mathrm{rms}$ and increases in steps of 2.} 
\label{22cont} 
\end{figure}

\begin{figure}[h!]
\resizebox{\hsize}{!}{\includegraphics{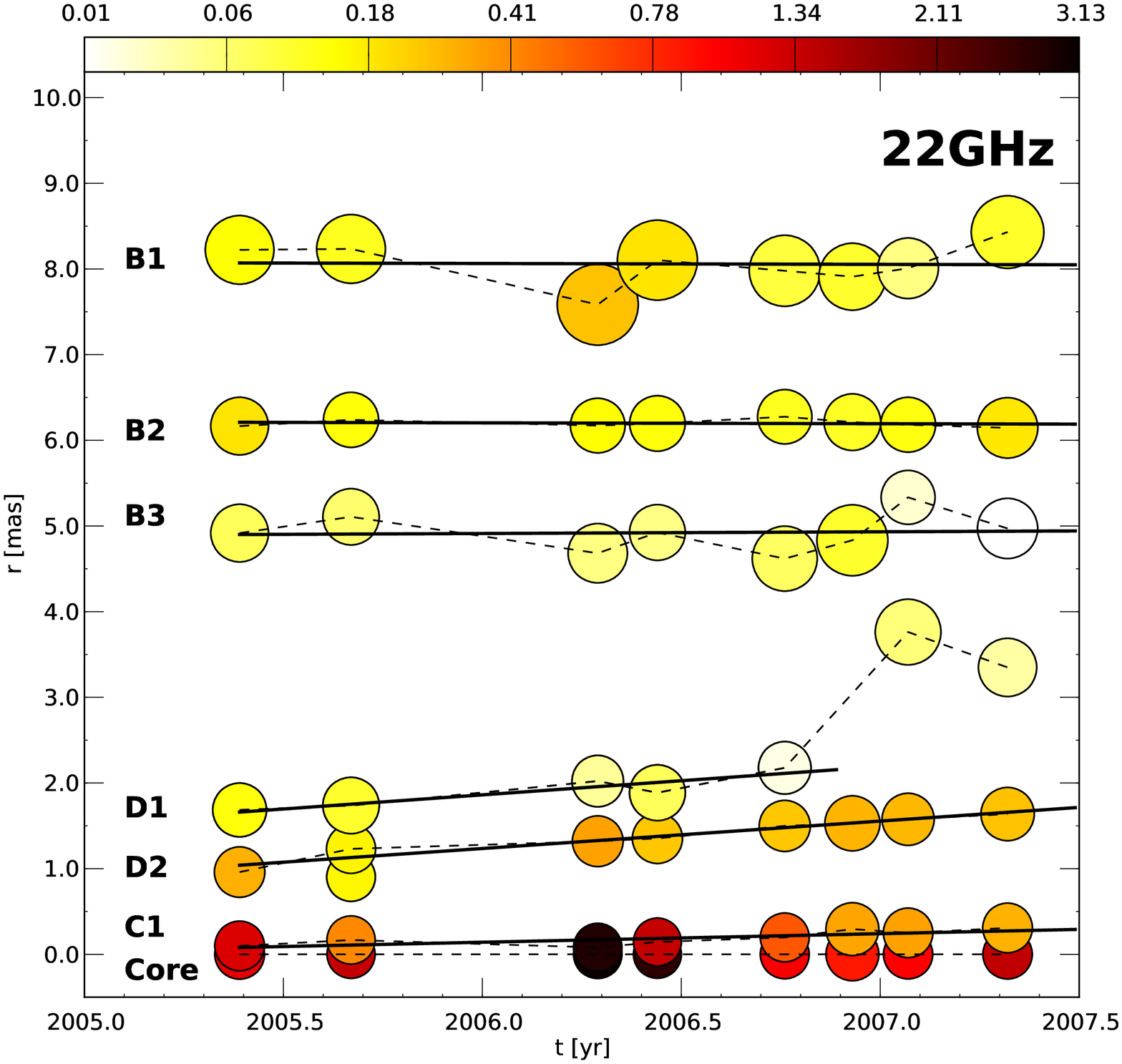}} 
\caption{Temporal evolution of the separation from the core for the $22\,\mathrm{GHz}$ components. The color scale corresponds to the flux density and the size of the circles to the relative size (FWHM) of the components. The solid black lines correspond to a linear fit of the component trajectory.} 
\label{22kin} 
\end{figure}

\begin{table*}
\caption{Results of the kinematic analysis for the fitted components at $22\,\mathrm{GHz}$}  
\label{22restab}
\tiny
\begin{tabular}{@{}c D{,}{\pm}{-1} D{,}{\pm}{-1} c c c D{,}{\pm}{-1} D{,}{\pm}{-1} c  c c c@{}}
\toprule
\hline
Comp	&	\multicolumn{1}{c}{$\mu$} &	\multicolumn{1}{c}{$\beta_\mathrm{app}$} 
& \multicolumn{1}{c}{$\delta_\mathrm{min}$}
& \multicolumn{1}{c}{$\vartheta_\mathrm{max}$} 
&\multicolumn{1}{c}{$\Gamma_\mathrm{min}$}
&\multicolumn{1}{c}{$t_\mathrm{ej}$}
&\multicolumn{1}{c}{$<r>$}
& \multicolumn{1}{c}{$t_\mathrm{min}$}
& \multicolumn{1}{c}{$t_\mathrm{max}$}\\

& \multicolumn{1}{c}{[mas/yr]}
& \multicolumn{1}{c}{[c]}
& \multicolumn{1}{c}{[1]}
& \multicolumn{1}{c}{ [$^\circ$]}
& \multicolumn{1}{c}{[1]}
& \multicolumn{1}{c}{ [yr]}
& \multicolumn{1}{c}{[mas]}
& \multicolumn{1}{c}{[yr]}
& \multicolumn{1}{c}{[yr]}
& \multicolumn{1}{c}{acceleration}
& \multicolumn{1}{c}{classification}\\
\midrule
B1 & 0.06,0.11 & 3,6 & -- & -- & -- & -- & 8.1,0.2 & 2005.39 & 2007.32 & no& non-radial outward\\ 
B2 & 0.03,0.02 & 2,1 & 2 & 32 & 2 & -- & 6.2,0.1 & 2005.39 & 2007.32 &no& radial inward \\
B3 & 0.1,0.1 & 7,7 & 7 & 8 & 7 & -- & 4.9,0.2 & 2005.39 & 2007.32 & no & non-radial outward\\
D1 & 0.6,0.1 & 31,6 & 31 & 2,& 31 & -- & 1.9,0.2 & 2005.39 & 2006.76 &no& non-radial outward \\
D2 & 0.30,0.03 & 17,2 & 17 & 3 & 17 & -- & 1.4,0.2 & 2005.39 & 2007.32 &no & non-radial outward \\
C1 & 0.10,0.03 & 5 & 5 & 12 & 5 & 2004.4,0.5 & 0.2,0.1 & 2005.39 & 2007.32 &no & radial outward \\
\bottomrule
\hline
\end{tabular}
\end{table*}

\clearpage
\begin{landscape}
\begin{figure}
\centering
\resizebox{\hsize}{!}{\includegraphics{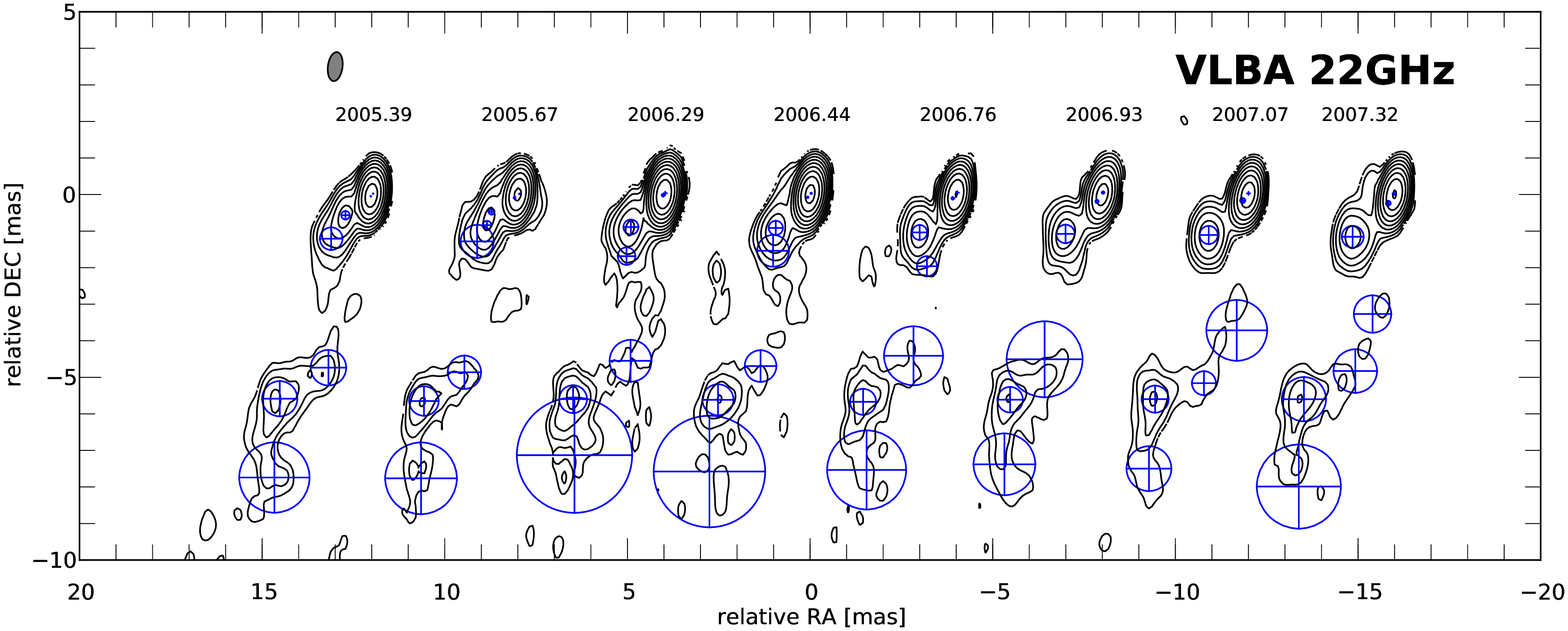}} 
 \caption{$22\,\mathrm{GHz}$ uniform weighted VLBA images of \object{CTA\,102} with fitted circular Gaussian components. For better comparison all maps are convolved with a common beam of $0.8\times0.4\,\mathrm{mas}$ at P.A. $-7.6$ and the epoch of the observations is  indicated above each contour map. The lowest contour levels is plotted $10\times$ of the maximum off-source $\mathrm{rms}$ and increases in steps of 2.} 
\label{contall22} 
\end{figure}
\end{landscape}

\subsection*{$43\,\mathrm{GHz}$ VLBI observations of \object{CTA\,102}}

\begin{figure}[h!]
\resizebox{\hsize}{!}{\includegraphics{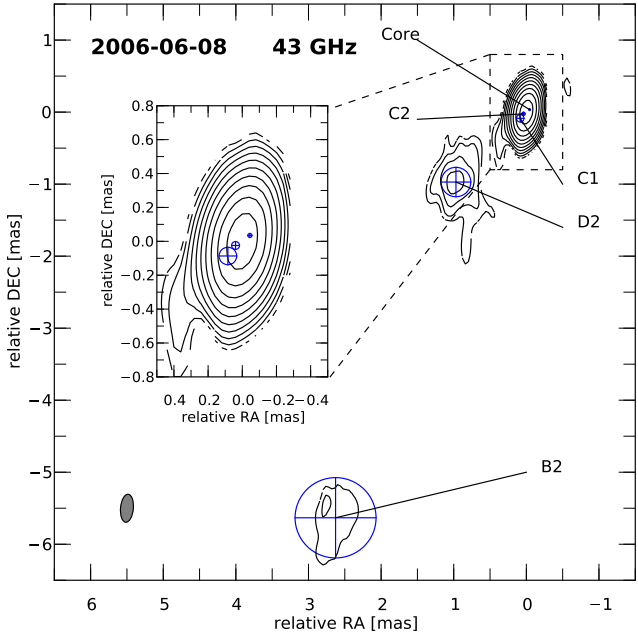}} 
\caption{$43\,\mathrm{GHz}$ uniformly weighted VLBA image of \object{CTA\,102} with fitted circular Gaussian components observed 8th of June 2006. The map peak flux density was $3.61\,\mathrm{Jy/beam}$, where the convolving beam was $0.39\times0.19\,\mathrm{mas}$ at P.A. $-5.0$. The lowest contour is plotted at $5\times$ the off-source $\mathrm{rms}$ and increases in steps of 2.} 
\label{43cont} 
\end{figure}

\begin{figure}[h!]
\resizebox{\hsize}{!}{\includegraphics{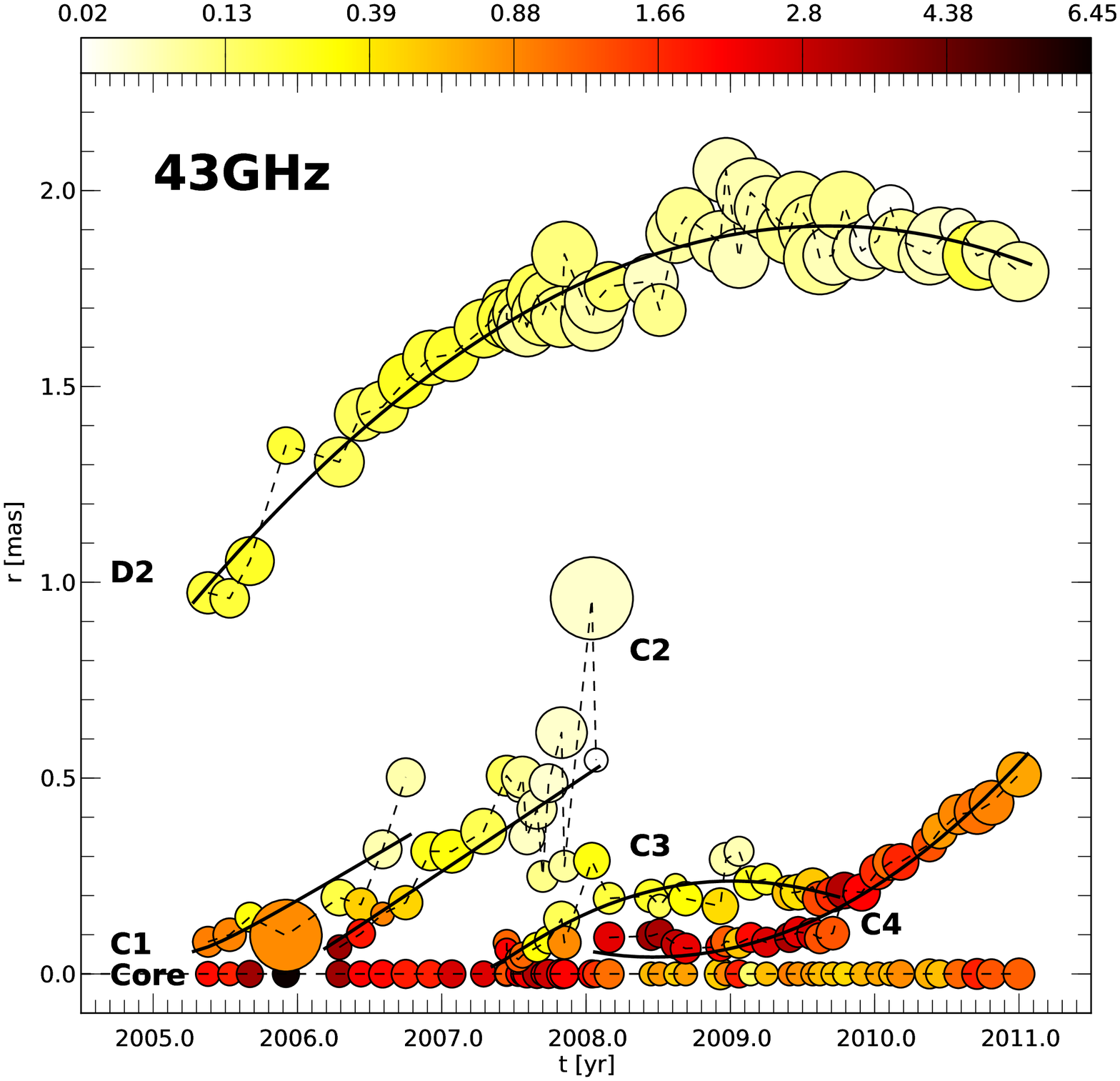}} 
\caption{Temporal evolution of the separation from the core for the $43\,\mathrm{GHz}$ components. The color scale corresponds to the flux density and the size of the circles to the relative size (FWHM) of the components. The solid black lines correspond to a polynomial fit of the component trajectory.} 
\label{43kin} 
\end{figure}

\begin{table*}
\caption{Results of the kinematic analysis for the fitted components at $43\,\mathrm{GHz}$}  
\label{43restab}
\tiny
\begin{tabular}{@{}c D{,}{\pm}{-1} D{,}{\pm}{-1} c c c D{,}{\pm}{-1} D{,}{\pm}{-1} c  c c c@{}}
\toprule
\hline
Comp	&	\multicolumn{1}{c}{$\mu$} &	\multicolumn{1}{c}{$\beta_\mathrm{app}$} 
& \multicolumn{1}{c}{$\delta_\mathrm{min}$}
& \multicolumn{1}{c}{$\vartheta_\mathrm{max}$} 
&\multicolumn{1}{c}{$\Gamma_\mathrm{min}$}
&\multicolumn{1}{c}{$t_\mathrm{ej}$}
&\multicolumn{1}{c}{$<r>$}
& \multicolumn{1}{c}{$t_\mathrm{min}$}
& \multicolumn{1}{c}{$t_\mathrm{max}$}\\

& \multicolumn{1}{c}{[mas/yr]}
& \multicolumn{1}{c}{[c]}
& \multicolumn{1}{c}{[1]}
& \multicolumn{1}{c}{ [$^\circ$]}
& \multicolumn{1}{c}{[1]}
& \multicolumn{1}{c}{ [yr]}
& \multicolumn{1}{c}{[mas]}
& \multicolumn{1}{c}{[yr]}
& \multicolumn{1}{c}{[yr]}
& \multicolumn{1}{c}{acceleration}
& \multicolumn{1}{c}{classification}\\
\midrule
C1 & 0.20,0.10 & 12,3     & 12 & 5   & 12& 2005.1,0.2 & 0.2,0.1 & 2005.4 & 2006.8 & no & radial outward\\
C2 & 0.25,0.04 & 13,2     & 13 & 4   & 13& 2005.9,0.2 & 0.4,0.2 & 2006.3 & 2008.0 & no & radial outward\\
C3 & 0.07,0.01 & 4,1       & 4   & 15 & 4  & -- & 0.2,0.1 & 2007.5 & 2009.7 & yes & radial/non-radial outward\\
C4 & 0.17,0.01 & 9,1       & 9   &  6  & 9  & -- & 0.2,0.1 & 2008.2 & 2011.0 & yes & non-radial outward\\
\bottomrule
\hline
\end{tabular}
\end{table*} 

\clearpage
\begin{landscape}
\begin{figure}
\centering
\resizebox{\hsize}{!}{\includegraphics{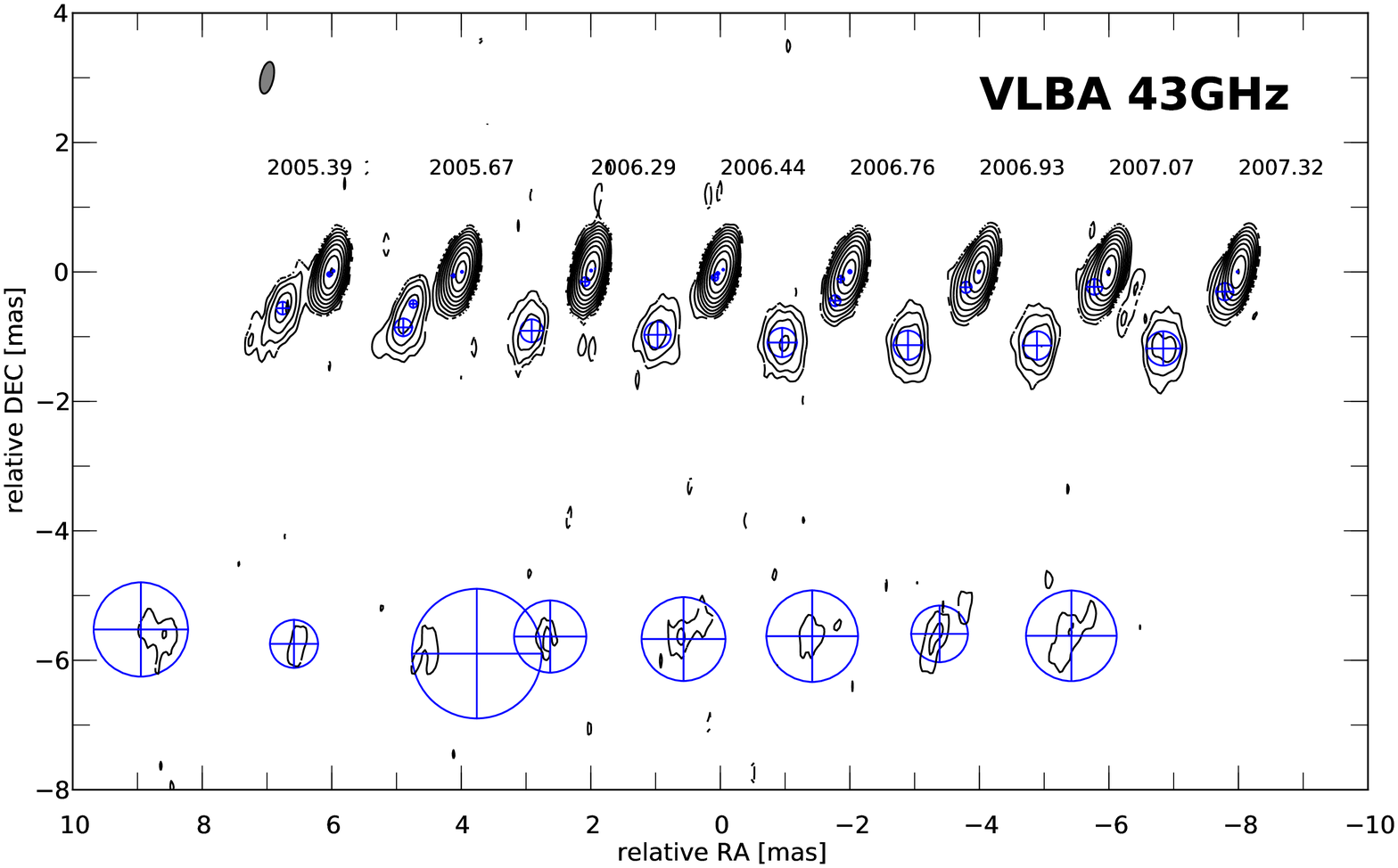}} 
\caption{$43\,\mathrm{GHz}$ uniform weighted VLBA images of \object{CTA\,102} with fitted circular Gaussian components. For better comparison all maps are convolved with a common beam of $0.5\times0.2\,\mathrm{mas}$ at P.A. $-11.9$ and the epoch of the observations is  indicated above each contour map. The lowest contour levels is plotted $10\times$ of the maximum off-source $\mathrm{rms}$ and increases in steps of 2.} 
\label{contall43} 
\end{figure}
\end{landscape}

\begin{landscape}
\begin{table}
\caption{Fitted components for multifrequency VLBI observations}  
\label{modelfit1}
\begin{tabular}{c c c c c c c c c c c c c c c c c c c c} 
\hline\hline
\multicolumn{5}{c}{2005-05-19}& \multicolumn{5}{c}{2005-09-01} & \multicolumn{5}{c}{2006-04-14} &\multicolumn{5}{c}{2006-06-08}\\
\hline
ID & $S$ & $R$ & $\vartheta$ & FWHM  &    ID & $S$ & $R$ & $\vartheta$ & FWHM &  ID & $S$ & $R$ & $\vartheta$ & FWHM & ID & $S$ & $R$ & $\vartheta$ & FWHM  \\
     & [Jy] & [mas] &$^\circ$ & [mas] &      & [Jy] & [mas] &$^\circ$ & [mas] &     & [Jy] & [mas] &$^\circ$ & [mas] &      & [Jy] & [mas] &$^\circ$ & [mas] \\
\hline
\multicolumn{18}{l}{$\nu=5\,\mathrm{GHz}$}\\
core & 0.98 & 0.00 & 0.00 & 0.45 & core & 1.12 & 0.00 & 0.00 & 0.33 & core & 1.45 & 0.00 & 0.00 & 0.07 & core & 1.62 & 0.00 & 0.00 & 0.14 \\
D2 & 0.92 & 1.14 & $-$50.18 & 0.48 & D2 & 1.02 & 1.19 & $-$49.08 & 0.38 & D2 & 1.04 & 1.29 & $-$47.82 & 0.49 & D2 & 0.94 & 1.34 & $-$48.87 & 0.46 \\
B3 & 0.30 & 4.21 & $-$72.20 & 1.22 & B3 & 0.23 & 4.19 & $-$72.57 & 0.98 & B3 & 0.29 & 4.53 & $-$70.87 & 1.30 & B3 & 0.21 & 4.43 & $-$72.62 & 1.03 \\
B2 & 0.58 & 6.11 & $-$67.85 & 1.03 & B2 & 0.58 & 6.09 & $-$67.39 & 1.03 & B2 & 0.59 & 6.45 & $-$67.22 & 0.95 & B2 & 0.54 & 6.16 & $-$66.83 & 0.98 \\
B1 & 0.25 & 8.16 & $-$71.13 & 1.00 & B1 & 0.23 & 8.12 & $-$70.67 & 0.88 & B1 & 0.18 & 8.90 & $-$71.08 & 1.08 & B1 & 0.23 & 8.20 & $-$70.34 & 1.01 \\
A2 & 0.67 & 11.34 & $-$67.56 & 3.41 & A2 & 0.64 & 11.43 & $-$67.38 & 3.35 & A2 & 0.61 & 11.65 & $-$66.53 & 3.41 & A2 & 0.61 & 11.56 & $-$66.88 & 3.42 \\
A1 & 0.35 & 16.86 & $-$54.49 & 4.14 & A1 & 0.28 & 16.93 & $-$54.24 & 3.59 & A1 & 0.28 & 16.96 & $-$54.24 & 3.94 & A1 & 0.26 & 16.79 & $-$54.31 & 3.75 \\
\multicolumn{18}{l}{$\nu=8\,\mathrm{GHz}$}\\
core & 1.44 & 0.00 & 0.00 & 0.18 & core & 1.33 & 0.00 & 0.00 & 0.08 & core & 2.76 & 0.00 & 0.00 & 0.13 & core & 2.77 & 0.00 & 0.00 & 0.14 \\
D2 & 0.69 & 0.90 & $-$37.79 & 0.34 & D2 & 0.70 & 0.94 & $-$41.33 & 0.34 & D2 & 0.72 & 1.21 & $-$45.04 & 0.44 & D2 & 0.63 & 1.26 & $-$45.33 & 0.44 \\
D1 & 0.31 & 1.80 & $-$54.26 & 0.87 & D1 & 0.18 & 1.67 & $-$47.67 & 0.38 & D1 & 0.15 & 2.06 & $-$55.53 & 0.89 & D1 & 0.13 & 2.48 & $-$59.85 & 1.13 \\
B3 & 0.26 & 4.81 & $-$70.38 & 1.35 & B3 & 0.21 & 4.77 & $-$69.82 & 1.46 & B3 & 0.20 & 4.93 & $-$71.93 & 1.08 & B3 & 0.21 & 5.14 & $-$70.42 & 1.32 \\
B2 & 0.35 & 6.29 & $-$66.17 & 0.84 & B2 & 0.26 & 6.18 & $-$65.87 & 0.75 & B2 & 0.33 & 6.17 & $-$65.74 & 0.75 & B2 & 0.29 & 6.33 & $-$66.27 & 0.79 \\
B1 & 0.19 & 8.07 & $-$70.21 & 1.14 & B1 & 0.18 & 7.75 & $-$70.06 & 1.02 & B1 & 0.20 & 7.93 & $-$70.17 & 1.17 & B1 & 0.16 & 8.14 & $-$70.26 & 1.26 \\
A2 & 0.47 & 11.50 & $-$66.76 & 3.67 & A2 & 0.30 & 11.16 & $-$68.04 & 3.29 & A2 & 0.40 & 11.55 & $-$66.91 & 3.30 & A2 & 0.36 & 11.62 & $-$67.20 & 3.37 \\
A1 & 0.20 & 17.52 & $-$53.09 & 3.65 & A1 & 0.11 & 17.36 & $-$52.10 & 3.68 & A1 & 0.19 & 17.08 & $-$53.42 & 4.19 & A1 & 0.17 & 17.02 & $-$53.17 & 3.93 \\
\multicolumn{18}{l}{$\nu=15\,\mathrm{GHz}$}\\
core & 2.35 & 0.00 & 0.00 & 0.10 & core & 1.77 & 0.00 & 0.00 & 0.00 & core & 4.07 & 0.00 & 0.00 & 0.00 & core & 3.62 & 0.00 & 0.00 & 0.00 \\
C1 & -- & -- & -- & -- & C1 & 0.23 & 0.17 & $-$9.95 & 0.00 & C1 & 0.91 & 0.18 & $-$38.88 & 0.00 & C1 & 0.82 & 0.13 & $-$31.94 & 0.00 \\
D3 & 0.39 & 0.86 & $-$35.85 & 0.20 & D3 & 0.34 & 0.93 & $-$36.36 & 0.15 & D3 & 0.20 & 1.71 & $-$53.60 & 0.66 & D3 & 0.13 & 1.12 & $-$38.58 & 0.70 \\
D2 & 0.24 & 1.43 & $-$45.52 & 0.57 & D2 & 0.30 & 1.34 & $-$45.86 & 0.48 & D2 & 0.36 & 1.24 & $-$42.07 & 0.34 & D2 & 0.34 & 1.35 & $-$45.97 & 0.36 \\
D1 & 0.04 & 2.71 & $-$65.94 & 0.31 & D1 & 0.08 & 2.55 & $-$61.06 & 1.39 & D1 & 0.05 & 3.84 & $-$75.49 & 1.04 & D1 & 0.07 & 2.19 & $-$58.83 & 0.73 \\
B3 & 0.14 & 4.96 & $-$73.25 & 1.09 & B3 & 0.15 & 5.03 & $-$72.20 & 1.16 & B3 & 0.11 & 5.34 & $-$71.22 & 1.01 & B3 & 0.12 & 4.79 & $-$74.38 & 1.21 \\
B2 & 0.18 & 6.19 & $-$65.40 & 0.62 & B2 & 0.19 & 6.21 & $-$65.31 & 0.59 & B2 & 0.18 & 6.25 & $-$65.27 & 0.58 & B2 & 0.18 & 6.10 & $-$65.98 & 0.66 \\
B1 & 0.10 & 7.89 & $-$69.68 & 1.02 & B1 & 0.11 & 7.77 & $-$69.37 & 1.19 & B1 & 0.11 & 7.71 & $-$69.12 & 1.11 & B1 & 0.12 & 7.65 & $-$68.30 & 1.33 \\
A2 & 0.26 & 11.45 & $-$64.62 & 4.10 & A2 & 0.22 & 11.31 & $-$66.72 & 3.53 & A2 & 0.23 & 11.32 & $-$66.79 & 3.39 & A2 & 0.19 & 11.36 & $-$67.45 & 3.67 \\
A1 & 0.15 & 16.79 & $-$55.41 & 3.52 & A1 & 0.15 & 16.57 & $-$55.40 & 3.96 & A1 & 0.15 & 15.87 & $-$55.25 & 4.38 & A1 & 0.14 & 16.64 & $-$54.99 & 3.78 \\
\multicolumn{18}{l}{$\nu=22\,\mathrm{GHz}$}\\ 
core & 1.16 & 0.00 & 0.00 & 0.01 & core & 1.37 & 0.00 & 0.00 & 0.01 & core & 3.13 & 0.00 & 0.00 & 0.01 & core & 2.72 & 0.00 & 0.00 & 0.04 \\
-- & -- & -- & -- & -- & -- & 0.16 & 0.90 & $-$32.83 & 0.16 & -- & -- & -- & -- & -- & -- & -- & -- & -- & -- \\
C1 & 1.22 & 0.10 & $-$48.10 & 0.01 & C1 & 0.42 & 0.17 & $-$40.03 & 0.05 & C1 & 2.86 & 0.08 & $-$41.50 & 0.07 & C1 & 1.38 & 0.14 & $-$48.53 & 0.04 \\
D2 & 0.30 & 0.96 & $-$38.01 & 0.23 & D1 & 0.10 & 1.74 & $-$48.44 & 0.91 & D2 & 0.34 & 1.32 & $-$44.48 & 0.41 & D1 & 0.07 & 1.89 & $-$56.45 & 0.87 \\
D1 & 0.13 & 1.69 & $-$47.03 & 0.62 & D2 & 0.16 & 1.23 & $-$44.15 & 0.24 & D1 & 0.04 & 2.02 & $-$58.35 & 0.48 & D2 & 0.25 & 1.35 & $-$44.31 & 0.39 \\
B3 & 0.07 & 4.92 & $-$75.59 & 0.97 & B3 & 0.06 & 5.11 & $-$72.90 & 0.91 & B3 & 0.05 & 4.68 & $-$78.27 & 1.15 & B3 & 0.04 & 4.92 & $-$73.65 & 0.87 \\
B2 & 0.19 & 6.17 & $-$65.50 & 0.97 & B2 & 0.14 & 6.24 & $-$65.27 & 0.81 & B2 & 0.14 & 6.17 & $-$65.86 & 0.76 & B2 & 0.13 & 6.20 & $-$65.68 & 0.85 \\
B1 & 0.14 & 8.22 & $-$70.81 & 1.93 & B1 & 0.11 & 8.23 & $-$70.93 & 1.96 & B1 & 0.26 & 7.58 & $-$70.87 & 3.16 & B1 & 0.19 & 8.10 & $-$69.90 & 3.05 \\
\multicolumn{18}{l}{$\nu=43\,\mathrm{GHz}$}\\
core & 1.84 & 0.00 & 0.00 & 0.03 & core & 3.45 & 0.00 & 0.00 & 0.03 & core & 3.48 & 0.00 & 0.00 & 0.03 & core & 2.14 & 0.00 & 0.00 & 0.02 \\
C2 & -- & -- & -- & -- & C2 & -- & -- & -- & -- & C2 & 3.48 & 0.07 & 154.45 & 0.01 & C2 & 1.88 & 0.10 & $-$35.12 & 0.05 \\
C1 & 0.19 & 0.97 & $-$36.19 & 0.20 & C1 & 0.29 & 0.14 & $-$22.02 & 0.06 & C1 & 0.19 & 0.20 & $-$61.77 & 0.14 & C1 & 0.50 & 0.18 & $-$43.26 & 0.10 \\
D2 & 0.93 & 0.08 & $-$38.94 & 0.07 & D2 & 0.24 & 1.05 & $-$39.11 & 0.33 & D2 & 0.15 & 1.31 & $-$45.41 & 0.35 & D2 & 0.15 & 1.43 & $-$44.77 & 0.41 \\
B2 & 0.16 & 6.29 & $-$61.75 & 1.46 & B2 & 0.05 & 6.25 & $-$65.31 & 0.48 & B2 & 0.07 & 6.18 & $-$73.39 & 2.00 & B2 & 0.10 & 6.26 & $-$64.76 & 1.12 \\
\hline
\end{tabular} 
\end{table}
\end{landscape}

\begin{landscape}
\begin{table}
\caption{Fitted components for multifrequency VLBI observations}  
\label{modelfit2}

\begin{tabular}{c c c c c c c c c c c c c c c c c c c c} 
\hline\hline
\multicolumn{5}{c}{2006-10-02}& \multicolumn{5}{c}{2006-12-04} & \multicolumn{5}{c}{2007-01-26} &\multicolumn{5}{c}{2007-04-26}\\
\hline
ID & $S$ & $R$ & $\vartheta$ & FWHM  &    ID & $S$ & $R$ & $\vartheta$ & FWHM &  ID & $S$ & $R$ & $\vartheta$ & FWHM & ID & $S$ & $R$ & $\vartheta$ & FWHM  \\
     & [Jy] & [mas] &$^\circ$ & [mas] &      & [Jy] & [mas] &$^\circ$ & [mas] &     & [Jy] & [mas] &$^\circ$ & [mas] &      & [Jy] & [mas] &$^\circ$ & [mas] \\
\hline
\multicolumn{18}{l}{$\nu=5\,\mathrm{GHz}$}\\
core & 1.50 & 0.00 & 0.00 & 0.24 & core & 1.26 & 0.00 & 0.00 & 0.14 & core & 1.02 & 0.00 & 0.00 & 0.14 & core & 0.90 & 0.00 & 0.00 & 0.12 \\
D2 & 0.98 & 1.36 & $-$49.67 & 0.59 & D2 & 0.96 & 1.36 & $-$49.23 & 0.57 & D2 & 0.80 & 1.37 & $-$49.75 & 0.40 & D2 & 0.86 & 1.43 & $-$49.51 & 0.59 \\
B3 & 0.21 & 4.32 & $-$71.81 & 1.08 & B3 & 0.26 & 4.27 & $-$70.51 & 1.25 & B3 & 0.09 & 4.42 & $-$74.15 & 0.16 & B3 & 0.20 & 4.34 & $-$71.89 & 1.04 \\
B2 & 0.60 & 6.20 & $-$67.09 & 1.06 & B2 & 0.59 & 6.34 & $-$67.42 & 1.03 & B2 & 0.53 & 6.08 & $-$66.88 & 0.99 & B2 & 0.55 & 6.17 & $-$67.06 & 0.97 \\
B1 & 0.23 & 8.32 & $-$70.65 & 1.12 & B1 & 0.20 & 8.61 & $-$70.59 & 1.25 & B1 & 0.14 & 8.05 & $-$70.38 & 0.56 & B1 & 0.22 & 8.25 & $-$70.21 & 1.19 \\
A2 & 0.65 & 11.80 & $-$66.58 & 3.42 & A2 & 0.61 & 11.73 & $-$66.53 & 3.40 & A2 & 0.50 & 11.59 & $-$67.22 & 3.17 & A2 & 0.59 & 11.73 & $-$66.81 & 3.41 \\
A1 & 0.29 & 17.24 & $-$53.77 & 3.75 & A1 & 0.31 & 17.06 & $-$54.01 & 4.17 & A1 & 0.16 & 16.61 & $-$52.84 & 2.66 & A1 & 0.29 & 16.93 & $-$54.07 & 3.85 \\
\multicolumn{18}{l}{$\nu=8\,\mathrm{GHz}$}\\
core & 1.64 & 0.00 & 0.00 & 0.12 & core & 1.33 & 0.00 & 0.00 & 0.13 & core & 1.03 & 0.00 & 0.00 & 0.10 & core & 1.10 & 0.00 & 0.00 & 0.22 \\
D2 & 0.55 & 1.32 & $-$46.27 & 0.46 & D2 & 0.69 & 1.33 & $-$48.44 & 0.60 & D2 & 0.56 & 1.37 & $-$46.55 & 0.51 & D2 & 0.63 & 1.46 & $-$48.33 & 0.68 \\
D1 & 0.09 & 2.90 & $-$67.69 & 1.03 & D1 & 0.09 & 2.54 & $-$61.80 & 0.95 & D1 & 0.02 & 2.81 & $-$61.84 & 0.39 & D1 & 0.08 & 3.10 & $-$67.31 & 1.18 \\
B3 & 0.17 & 5.04 & $-$70.73 & 1.24 & B3 & 0.19 & 4.97 & $-$70.86 & 1.30 & B3 & 0.14 & 4.82 & $-$71.45 & 1.34 & B3 & 0.18 & 5.19 & $-$69.53 & 1.31 \\
B2 & 0.27 & 6.24 & $-$66.34 & 0.76 & B2 & 0.28 & 6.18 & $-$66.32 & 0.77 & B2 & 0.26 & 6.09 & $-$66.29 & 0.73 & B2 & 0.26 & 6.22 & $-$66.16 & 0.71 \\
B1 & 0.14 & 7.82 & $-$70.36 & 1.09 & B1 & 0.18 & 7.82 & $-$69.69 & 1.19 & B1 & 0.15 & 7.73 & $-$70.10 & 1.01 & B1 & 0.17 & 7.85 & $-$69.60 & 1.16 \\
A2 & 0.36 & 11.48 & $-$67.25 & 3.38 & A2 & 0.40 & 11.49 & $-$67.06 & 3.49 & A2 & 0.28 & 11.47 & $-$67.87 & 3.32 & A2 & 0.39 & 11.61 & $-$67.06 & 3.59 \\
A1 & 0.12 & 17.45 & $-$51.98 & 3.24 & A1 & 0.16 & 17.20 & $-$53.25 & 3.44 & A1 & 0.11 & 16.64 & $-$54.20 & 3.20 & A1 & 0.16 & 17.04 & $-$53.11 & 3.50 \\
\multicolumn{18}{l}{$\nu=15\,\mathrm{GHz}$}\\
core & 1.53 & 0.00 & 0.00 & 0.00 & core & 0.92 & 0.00 & 0.00 & 0.00 & core & 0.89 & 0.00 & 0.00 & 0.00 & core & 1.47 & 0.00 & 0.00 & 0.00 \\
C1& 0.47 & 0.20 & $-$39.77 & 0.00 &  C1 & 0.41 & 0.24 & $-$40.72 & 0.05 & C1 & 0.30 & 0.29 & $-$46.86 & 0.00 & -- & 0.15 & 0.34 & $-$39.12 & 0.00 \\
D3 & 0.10 & 1.69 & $-$57.23 & 0.25 & D3 & 0.09 & 1.72 & $-$58.46 & 0.34 & D3 & 0.09 & 1.73 & $-$57.88 & 0.31 & D3 & 0.09 & 1.73 & $-$58.79 & 0.27 \\
D2 & 0.32 & 1.43 & $-$43.67 & 0.35 & D2 & 0.32 & 1.48 & $-$44.16 & 0.37 & D2 & 0.28 & 1.52 & $-$43.66 & 0.38 & D2 & 0.28 & 1.59 & $-$43.70 & 0.41 \\
D1 & 0.02 & 3.10 & $-$69.54 & 0.33 & A13 & 0.06 & 4.24 & $-$77.13 & 1.26 & D1 & 0.01 & 3.21 & $-$69.00 & 0.49 & D1 & 0.01 & 3.32 & $-$68.39 & 0.37 \\
D0 & -- & -- & -- & -- & D0 & -- & -- & -- & -- & D0 & 0.03 & 4.23 & $-$83.43 & 0.58 & D0 & 0.04 & 4.27 & $-$82.33 & 0.84 \\
B3 & 0.10 & 4.89 & $-$74.66 & 0.95 & B3 & 0.06 & 5.40 & $-$72.27 & 0.63 & B3 & 0.07 & 5.22 & $-$73.09 & 0.77 & B3 & 0.07 & 5.30 & $-$72.40 & 0.70 \\
B2 & 0.18 & 6.19 & $-$65.74 & 0.66 & B2 & 0.16 & 6.16 & $-$65.58 & 0.62 & B2 & 0.15 & 6.16 & $-$65.38 & 0.59 & B2 & 0.16 & 6.13 & $-$65.20 & 0.58 \\
B1 & 0.09 & 7.87 & $-$69.50 & 1.36 & B1 & 0.10 & 7.70 & $-$69.33 & 1.29 & B1 & 0.09 & 7.68 & $-$69.79 & 1.57 & B1 & 0.11 & 7.59 & $-$69.15 & 1.47 \\
A2 & 0.16 & 11.68 & $-$68.43 & 3.26 & A2 & 0.15 & 11.59 & $-$68.56 & 3.23 & A2 & 0.14 & 11.65 & $-$68.18 & 3.22 & A2 & 0.16 & 11.71 & $-$68.65 & 3.48 \\
A1 & 0.09 & 16.94 & $-$53.17 & 4.00 & A1 & 0.08 & 17.22 & $-$52.51 & 3.94 & A1 & 0.05 & 17.04 & $-$51.19 & 3.88 & A1 & 0.07 & 17.16 & $-$51.76 & 3.75 \\
\multicolumn{18}{l}{$\nu=22\,\mathrm{GHz}$}\\ 
core & 1.04 & 0.00 & 0.00 & 0.00 & core & 0.88 & 0.00 & 0.00 & 0.07 & core & 1.02 & 0.00 & 0.00 & 0.00 & core & 1.40 & 0.00 & 0.00 & 0.00 \\
C1 & 0.59 & 0.20 & $-$50.59 & 0.07 & C1 & 0.33 & 0.29 & $-$55.68 & 0.09 & C1 & 0.34 & 0.25 & $-$50.84 & 0.14 & C1 & 0.29 & 0.31 & $-$59.64 & 0.13 \\
D2 & 0.24 & 1.50 & $-$46.43 & 0.41 & D2 & 0.29 & 1.53 & $-$47.75 & 0.51 & D2 & 0.28 & 1.58 & $-$46.27 & 0.50 & D2 & 0.26 & 1.64 & $-$46.29 & 0.60 \\
D1 & 0.01 & 2.18 & $-$67.46 & 0.55 & D1 & -- & -- & -- & -- & D1 & 0.05 & 3.76 & $-$85.10 & 1.67 & D1 & 0.03 & 3.35 & $-$79.88 & 1.02 \\
B3 & 0.06 & 4.62 & $-$74.93 & 1.61 & B3 & 0.10 & 4.83 & $-$70.66 & 2.08 & B3 & 0.02 & 5.33 & $-$76.78 & 0.66 & B3 & 0.01 & 4.97 & $-$77.68 & 1.19 \\
B2 & 0.11 & 6.28 & $-$65.71 & 0.71 & B2 & 0.11 & 6.21 & $-$65.83 & 0.70 & B2 & 0.13 & 6.18 & $-$65.58 & 0.74 & B2 & 0.18 & 6.14 & $-$66.33 & 1.21 \\
B1 & 0.09 & 7.98 & $-$71.86 & 2.16 & B1 & 0.10 & 7.91 & $-$70.04 & 1.69 & B1 & 0.05 & 8.01 & $-$70.10 & 1.23 & B1 & 0.10 & 8.43 & $-$72.00 & 2.30 \\
\multicolumn{18}{l}{$\nu=43\,\mathrm{GHz}$}\\
core & 1.83 & 0.00 & 0.00 & 0.05 & core & 1.77 & 0.00 & 0.00 & 0.04 & core & 2.64 & 0.00 & 0.00 & 0.04 & core & 2.54 & 0.00 & 0.00 & 0.02 \\
C2 & 0.48 & 0.18 & $-$39.01 & 0.11 & C2 & 0.30 & 0.31 & $-$50.25 & 0.18 & C2 & 0.29 & 0.31 & $-$47.75 & 0.24 & C2 & 0.18 & 0.36 & $-$57.45 & 0.27 \\
C1 & 0.07 & 0.50 & $-$62.96 & 0.17 & C1 & -- & -- & -- & -- & C1 & -- & -- & -- & -- & C1 & -- & -- & -- & -- \\
D2 & 0.25 & 1.51 & $-$46.27 & 0.45 & D2 & 0.20 & 1.57 & $-$46.01 & 0.45 & D2 & 0.23 & 1.58 & $-$46.00 & 0.44 & D2 & 0.18 & 1.65 & $-$45.94 & 0.53 \\
B2 & 0.15 & 6.23 & $-$65.64 & 1.29 & B2 & 0.14 & 6.19 & $-$65.43 & 1.41 & B2 & 0.13 & 6.17 & $-$65.05 & 0.87 & B2 & 0.16 & 6.18 & $-$65.49 & 1.40 \\
\hline
\end{tabular} 
\end{table}
\end{landscape}

\end{document}